\documentclass[aps, prd, amsmath, amssymb, twocolumn, 10pt, longbibliography, superscriptaddress, nofootinbib]{revtex4-2}

\usepackage[utf8]{inputenc}
\usepackage{lmodern}
\usepackage[T1]{fontenc}
\usepackage{textcomp}
\usepackage{amsmath}
\usepackage{amsfonts}
\usepackage{calc}
\usepackage{mathrsfs}
\usepackage{array}
\usepackage{xcolor}
\usepackage{bm}
\usepackage{graphicx}
\usepackage{hyperref}
\usepackage{tensor}
\usepackage{mathtools}
\usepackage[caption=false]{subfig}
\usepackage{float}
\usepackage{orcidlink}
\usepackage[normalem]{ulem}
\usepackage{placeins}
\usepackage{siunitx}
\usepackage{upgreek}
\usepackage[capitalise]{cleveref}
\usepackage{comment}
\usepackage{cancel}

\graphicspath{{Figures/}}

\newcommand{\res}{\mathcal{R}}
\newcommand{\TeukO}{\mathcal{O}}
\renewcommand{\S}{{\rm S}}
\newcommand{\R}{\mathrm{R}}
\newcommand{\e}{\varepsilon}

\newcommand{\scri}{{\mathscr I}}
\newcommand{\hor}{{\mathscr H}}

\newcommand{\diff}[2]  {\frac{d #1}{d #2}}

\newcommand{\pdiff}[2]  {\frac{\partial #1}{\partial #2}}

\newcommand{\eqn}[1] {Eq.~\eqref{eq:#1}}
\newcommand{\fig}[1] {Fig.~\ref{fig:#1}}

\newcommand{\secref}[1] {Sec.~\ref{sec:#1}}

\newcommand{\nn}{\nonumber}

\renewcommand{\l}{\left(}
\renewcommand{\r}{\right)}

\newcommand{\lm}  {\ell m}
\newcommand{\m}  {m}

\newcommand{\rplus} {r_{+}}
\newcommand{\rminus} {r_{-}}
\newcommand{\psifour} {\psi_{4}}

\newcommand{\rstar} {r_{*}}
\newcommand{\rsharp} {r_{\sharp}}

\newcommand{\barh} {\bar{h}}
\newcommand{\rp} {r_{p}}

\newcommand{\ellmax} {\ell_{\rm max}}

\DeclarePairedDelimiterX\braket[1]{\langle}{\rangle}{#1}

\newcommand{\beq}{\begin{equation}}
\newcommand{\eeq}{\end{equation}}
\newcommand{\bec}{\begin{cases}}
\newcommand{\eec}{\end{cases}}

\newcommand{\calP}{\mathcal{P}}

\newcommand{\order}{\mathcal{O}}
\newcommand{\calR}{\mathcal{R}}
\newcommand{\calW}{\mathcal{W}}

\newcommand{\phihat}{\hat{\varphi}}
\newcommand{\barm}{\bar{\mtetrad}}
\newcommand{\mass}{\mu}
\newcommand{\sY}[1]{{}_{#1}Y}

\newcommand{\hs}[1]{\mathring{h}_{#1}}

\newcommand{\dd}{d}
\newcommand{\ee}{e}
\newcommand{\half}{\tfrac{1}{2}}

\newcommand{\ethb}{\bar{\eth}}

\newcommand{\Lz}{\mathcal{L}_{z}}
\newcommand{\En}{\mathcal{E}}
\newcommand{\rcov}{\mathsf{r}}
\newcommand{\scov}{\mathsf{s}}

\newcommand{\ltetrad}{{\rm l}}
\newcommand{\ntetrad}{{\rm n}}
\newcommand{\mtetrad}{{\rm m}}
\newcommand{\mb}{{\rm \bar m}}

\DeclareMathOperator{\diag}{diag}

\definecolor{colour1}{HTML}{0571b0} 
\definecolor{colour2}{HTML}{92c5de} 
\definecolor{colour3}{HTML}{f4a582} 
\definecolor{colour4}{HTML}{ca0020} 
\definecolor{colour5}{HTML}{fe4a49} 
\definecolor{colour6}{HTML}{004F0E} 
\definecolor{colour7}{HTML}{35978f} 

\hypersetup{colorlinks=true, linkcolor=colour1, citecolor=colour1,
filecolor=colour1, urlcolor=colour1, pdfencoding=auto}

\begin{document}
\title{Kerr metric perturbations for second-order self-force theory}

\author{Benjamin Leather\,\orcidlink{0000-0001-6186-7271}} 
\email{bjl1u25@soton.ac.uk}
\affiliation{School of Mathematical Sciences and STAG Research Centre, University of Southampton, Southampton, United Kingdom, SO17 1BJ}

\author{Samuel D.\ Upton\,\orcidlink{0000-0003-2965-7674}}
\email{sam.upton@ucd.ie}
\affiliation{School of Mathematics and Statistics, University College Dublin, Belfield, Dublin 4, Ireland}
\affiliation{School of Mathematical Sciences and STAG Research Centre, University of Southampton, Southampton, United Kingdom, SO17 1BJ}

\author{Patrick Bourg\,\orcidlink{0000-0003-0015-0861}}
\affiliation{School of Mathematics and Statistics, University College Dublin, Belfield, Dublin 4, Ireland}
\affiliation{Institute for Mathematics, Astrophysics and Particle Physics, Radboud University, Heyendaalseweg 135, 6525 AJ Nijmegen, The Netherlands}

\author{Kevin O. Cunningham\,\orcidlink{0000-0002-1094-2267}}
\affiliation{School of Mathematical Sciences and STAG Research Centre, University of Southampton, Southampton, United Kingdom, SO17 1BJ}
\affiliation{School of Mathematics and Statistics, University College Dublin, Belfield, Dublin 4, Ireland}

\author{Sam R.\ Dolan\, \orcidlink{0000-0002-4672-6523}}
\affiliation{Consortium for Fundamental Physics, School of Mathematics and Statistics, University of Sheffield, Hicks Building, Hounsfield Road, Sheffield S3 7RH, United Kingdom}

\author{Conor Dyson\,\orcidlink{0000-0002-9742-9422}}
\affiliation{School of Physics and Astronomy \& Institute for Gravitational Wave Astronomy, University of Birmingham, Birmingham, B15 2TT, UK}
\affiliation{Center of Gravity, Niels Bohr Institute, Blegdamsvej 17, 2100 Copenhagen, Denmark}

\author{Zachary Nasipak\,\orcidlink{0000-0002-5109-9704}}
\affiliation{School of Mathematical Sciences and STAG Research Centre, University of Southampton, Southampton, United Kingdom, SO17 1BJ}

\author{Nami Nishimura\,\orcidlink{0009-0002-4398-9130}}
\affiliation{Max Planck Institute for Gravitational Physics (Albert Einstein Institute), Am M\" uhlenberg 1, Potsdam 14476, Germany}
\affiliation{Department of Physics, University of Maryland, College Park, Maryland 20742, USA}

\author{Andrew Spiers\,\orcidlink{0000-0003-0222-7578}}
\affiliation{School of Mathematics and Statistics, University College Dublin, Belfield, Dublin 4, Ireland}

\author{Maarten van de Meent\,\orcidlink{0000-0002-0242-2464}}
\affiliation{Center of Gravity, Niels Bohr Institute, Blegdamsvej 17, 2100 Copenhagen, Denmark}
\affiliation{Max Planck Institute for Gravitational Physics (Albert Einstein Institute), Am M\" uhlenberg 1, Potsdam 14476, Germany}

\author{Barry Wardell\,\orcidlink{0000-0001-6176-9006}}
\affiliation{School of Mathematics and Statistics, University College Dublin, Belfield, Dublin 4, Ireland}

\date{\today}

\begin{abstract}
 We present a calculation of the first-order metric perturbation arising from a point mass moving on a circular, equatorial orbit in Kerr spacetime. The metric perturbation is specifically aimed at meeting the requirements of second-order self-force theory: it is highly accurate; incorporates all contributions, including those associated with perturbations to the black hole's mass and angular momentum; is in Lorenz gauge near the worldline and in (near) Bondi--Sachs gauge at large radius; and can be efficiently evaluated across the entire exterior spacetime. The Lorenz gauge gives a well-understood singular structure at the particle, while the (near) Bondi--Sachs gauge ensures the asymptotic fall-off needed for an asymptotically flat second-order source. Our formalism reconstructs the metric from separable Teukolsky-type equations, decomposed into spin-weighted spheroidal harmonic modes and solved spectrally on hyperboloidal slices. The resulting $(\ell,m)$-mode sums converge rapidly across most of the domain, but slowly near the point mass; we overcome this with a novel use of a puncture to extract accurate $m$-mode data, yielding output accurate to $\sim 10^{-10}$ per $m$-mode and suitable for use at second order. Our metric-perturbation results cover a range of Kerr spins $-0.9 \le a/M \le 0.9$ and orbital radii out to $50M$. In addition to its immediate application in second-order self-force calculations, our results can also be used as input to other calculations involving Kerr metric perturbations, including studies of environmental effects and quasinormal modes.
\end{abstract}

\maketitle
\tableofcontents


\section{Introduction}

The third generation of gravitational-wave detectors, including LISA \cite{LISA:2024hlh}, the Einstein Telescope \cite{ET:2025xjr} and Cosmic Explorer \cite{Hall:2022dik}, are now under active development. When they come online in the mid-to-late 2030s, these instruments will usher in an era of unprecedented sensitivity to sources across the gravitational-wave spectrum \cite{Gupta:2023lga}. However, the full scientific gain from this dramatic increase in sensitivity will only be realised if there is commensurate improvement in the accuracy of waveform models. 
Existing models in use by the LIGO-Virgo-KAGRA collaboration are far from meeting the accuracy requirements of third-generation detectors. In addition, the new low-frequency sensitivity of third-generation detectors (in particular the LISA mission, which will be sensitive to millihertz gravitational-wave sources) will open gravitational-wave astronomy up to entirely new classes of sources not observable by LIGO-Virgo-KAGRA \cite{LISAConsortiumWaveformWorkingGroup:2023arg}. Key among these are Extreme Mass-Ratio Inspirals (EMRIs), binary systems consisting of a $\sim 10^6 M_\odot$ black hole (the primary) with a $\sim 10 M_\odot$ compact object in orbit around it \cite{Babak:2017tow} (the secondary). Whereas comparable-mass inspirals are nearly circular at the time of detection, EMRIs are expected to remain eccentric and inclined up to the plunge, and EMRIs are among the most scientifically rich sources that LISA can expect to observe.

The development of EMRI waveform models for LISA fundamentally relies on solving the equations of black hole perturbation theory to develop a model based on gravitational self-force theory \cite{Poisson:2011nh, Barack:2018yvs}. In this perturbative scheme, the secondary (`the particle') sources a metric perturbation on the background spacetime of the primary. It has long been established that the accuracy requirements of LISA can only be met by developing models accurate to second order (in the mass ratio) in perturbation theory \cite{Hinderer:2008dm}. The challenge of second-order self-force calculations stems from the fact that they are large, multifaceted problems requiring high precision at every stage \cite{Pound:2019lzj, Warburton:2021kwk, Wardell:2021fyy, Leather:2026zhl}. An essential ingredient in the second-order calculation is the first-order metric perturbation. This paper is part of a larger effort to develop second-order self-force-based models of \emph{spinning} EMRIs, and it addresses the fundamental challenge of obtaining a first-order metric perturbation that is sufficiently accurate and in a form suitable for second-order self-force calculations. This amounts to solving the equations for metric perturbations of Kerr spacetime --- a problem that has been studied for over half a century \cite{Teukolsky:1973ha, chrzanowski1975vector, Wald:1978vm}, and yet which still requires new refinements to meet the requirements of second-order self-force theory. These requirements guide the setup of our calculation:
\begin{enumerate}
    \item High accuracy: required so that numerical error at first order does not contaminate highly-sensitive second-order calculations;
    \item Complete: all contributions to the metric perturbation must be included, including radiative and zero-frequency modes, gauge pieces, and completion contributions arising from the perturbation to the mass and angular momentum of the spacetime;
    \item Comprehensive: second-order self-force calculations require the metric perturbation across the whole spacetime exterior to the horizon;
    \item Gauge: the metric perturbation must be in a suitable globally-regular gauge; 
    \item Coordinates: the metric perturbation must be expressed in terms of background spacetime coordinates free of spurious coordinate singularities.
\end{enumerate}
The gauge requirements motivate the use of Lorenz gauge, due to its well-understood symmetric singular structure near the worldline and absence of singular features elsewhere in the domain at first order. We introduce a (near) Bondi--Sachs gauge at large radius to eliminate issues related to the asymptotic behavior of Lorenz gauge perturbations in the far-field. The accuracy requirements favour metric-reconstruction schemes in which the solution can be constructed from solutions to ordinary, rather than partial, differential equations.

A variety of approaches to computing Lorenz gauge perturbations for circular orbits on Kerr spacetime have been adopted, though each comes with certain limitations that have impeded its use in second-order self-force calculations:
\begin{itemize}
    \item Time-domain evolutions in 2+1D \cite{Barack:2007we, Dolan:2010mt, Dolan:2011dx, Dolan:2012jg, Thornburg:2016msc} --- the existing (finite-difference) implementations are insufficiently accurate for second-order self-force, and are hampered by low-multipole linear instabilities (though see Ref.~\cite{Thornburg:2026eeb} for recent progress);
    \item Solutions to elliptic PDEs in the frequency domain with custom solvers \cite{Osburn:2022bby, Osburn:2026uct}, or with spectral methods from numerical relativity \cite{Vu:2026ypc} --- these methods show promise, but implementations are still in the early stages of development;
    \item Reconstruction of the metric perturbation from (differential operators acting on) solutions of separable Teukolsky equations \cite{Dolan:2021ijg, Dolan:2023enf, Wardell:2024yoi} --- the first implementation \cite{Dolan:2023enf} of a fully separated $(\ell,m)$-mode approach yielded accurate results for individual modes, but came with (expected) accuracy issues relating to poor convergence of the (truncated) infinite sum over $\ell$ in the vicinity of the worldline, and this directly feeds into well-known issues in the infinite mode coupling required to produce a second order source \cite{Miller:2016hjv}.
\end{itemize}
In this work, we build on the last of these three approaches in order to meet our requirements. The method is based on a high-accuracy implementation of the fully separated $(\ell,m)$-mode reconstruction approach combined with a puncture scheme designed to handle the issue of poor convergence of the sum over $\ell$ near the worldline. By working in hyperboloidal coordinates that are regular at the future horizon and null infinity, we avoid coordinate singularities; by working in Lorenz gauge near the worldline, we have a well-understood singular behavior near the worldline and a well-defined second-order source~\cite{Detweiler:2011tt,Upton:2021oxf,Upton:2025bja,Musaeus:2026kgp}; and by applying a transformation to (near) Bondi--Sachs gauge at large radius, we circumvent the issues of asymptotic divergences in the second-order problem in Lorenz gauge. 

The end result is a metric perturbation expressed in terms of azimuthal modes ($m$-modes). An $m$-mode decomposition has major advantages for second-order calculations, because it largely circumvents the problem of computing the second-order source through complicated and slowly-converging $(\ell,m)$-mode coupling sums \cite{spiers2023second,Miller:2016hjv}. (The coupling of $m$-modes remains, but it is comparatively simple to express in terms of products of Fourier series.) There remain the linked issues of convergence in the sum over $m$, and the logarithmic divergence of the $m$-modes of the metric perturbation at the particle. Both of these issues are readily handled by employing a suitable puncture scheme (similar to what was previously done for $(\ell,m)$ modes \cite{Miller:2016hjv,Upton:2025bja} but more immediately applicable to Kerr perturbations), as is shown here. Overall, we aim to show that an $ m$-mode decomposition offers a good compromise between the simplicity of constructing the second-order source, the accuracy of the result, and the numerical handling of divergences.

Though the focus here is on second-order self-force, the first-order metric perturbation may also serve as an ingredient for other calculations. First, the Lorenz gauge is well suited to calculations of the back-reaction on EMRIs of non-vacuum black hole environments, such as boson clouds \cite{Dyson:2025dlj, Li:2025ffh, Li:2026foy} and accretion discs \cite{Dyson:2026ddd}, and scalar-charged secondary black holes~\cite{maselli2020detecting,Spiers:2023cva, Barsanti:2026ulr}. Second, in black hole spectroscopy, there is much interest in calculating the effect of metric perturbations on the quasinormal mode spectrum of the Kerr black hole \cite{Berti:2025hly}. Finally, the results can serve as a benchmark for comparison with complementary methods for reconstructing the metric in other gauges, such as the GHZ method \cite{Green:2019nam, Hollands:2024iqp} in which the radiation-gauge CCK procedure \cite{chrzanowski1975vector} is extended to non-vacuum with the introduction of a corrector tensor and a shadowless gauge \cite{Toomani:2021jlo}. Other approaches include Refs.~\cite{Loutrel:2020wbw, Ripley:2020xby, Casals:2024ynr, Li:2026rkf}.

\subsection{Conventions and notation}

We use a mostly positive metric signature, $(-,+,+,+)$, and geometric units where $c=G=1$.
Greek letters indicate spacetime indices and Latin letters from the beginning of the alphabet $(a,b,c)$ are tetrad indices.
Individual tetrad legs are written in a serif font $(\ltetrad,\ntetrad,\mtetrad,\mb)$ to distinguish from harmonic modes where $s$, $\ell$, $j$, and $m$ are the spin weight, spherical mode degree, spheroidal mode degree, and azimuthal mode number, respectively.

\section{Perturbations of Kerr spacetime}
\label{sec:perturbations}

\subsection{Background motion in Kerr spacetime}
\label{sec:perturbations/background}
We consider a point particle of mass $\mass$ moving on a circular, equatorial geodesic of a Kerr spacetime with mass $M$ and spin parameter $a$, where $a > 0$ ($a < 0$) describes a prograde (retrograde) orbit.
In Boyer--Lindquist (BL) coordinates $(t,r,\theta,\varphi)$, the Kerr metric reads
\begin{multline}
    ds^{2} = -\left(1 - \frac{2Mr}{\Sigma}\right)dt^{2} + \frac{\Sigma}{\Delta}dr^{2} + \Sigma\, d\theta^{2} \\
    + \left(r^{2} + a^{2} + \frac{2Ma^{2}r}{\Sigma}\sin^{2}\theta \right) \sin^{2}\theta\, d\varphi^{2} \\
    - \frac{4Mar\sin^{2}\theta}{\Sigma}\, dt\, d\varphi,
    \label{eq:kerr_line_element}
\end{multline}
where $\Sigma \coloneqq r^{2} + a^{2}\cos^{2}\theta$ and $\Delta \coloneqq r^{2} - 2Mr + a^{2} = (r - \rplus)(r - \rminus)$, with
\begin{align}
    \kappa &\coloneqq \sqrt{M^{2} - a^{2}}, 
    &
    r_{\pm} &\coloneqq M \pm \kappa.
\end{align}
Hence $\rplus\rminus = a^{2}$, $\rplus + \rminus = 2M$ and $\rplus - \rminus = 2\kappa$.
The horizon $r=\rplus$ then has angular velocity $a/(2M\rplus)$ and surface gravity $\kappa/(2M\rplus)$.

We will need the tortoise coordinate $\rstar$ and its angular counterpart $\rsharp$, defined by
\begin{equation}
    \frac{d\rstar}{dr} = \frac{r^{2} + a^{2}}{\Delta}, \qquad \frac{d\rsharp}{dr} = \frac{a}{\Delta}.
    \label{eq:tortoise_coordinates}
\end{equation}
We fix the integration constants such that
\begin{align}
    \rstar ={}& r + \frac{M\rplus}{\kappa}\ln \left(\frac{r - \rplus}{2M}\right) - \frac{M\rminus}{\kappa}\ln \left(\frac{r - \rminus}{2M}\right), \\
    \rsharp ={}& \frac{a}{2\kappa}\ln \left(\frac{r - \rplus}{r - \rminus} \right).
    \label{eq:tortoise_explicit}
\end{align}
From these, we form ingoing Kerr coordinates $(v, r, \theta, \varphi_{+})$ and outgoing Kerr coordinates $(u, r, \theta, \varphi_{-})$, where $v \coloneqq t + \rstar$, $u \coloneqq t - \rstar$ and $\varphi_{\pm} \coloneqq \varphi \pm \rsharp$.
In the former, the metric is regular across the future horizon $\hor^{+}$; in the latter, after conformal compactification, it is regular at future null infinity $\scri^{+}$.

The particle then orbits at a fixed BL radius $\rp$ in the equatorial plane, $\theta_p = \pi/2$, with azimuthal angle $\varphi_p(t) = \Omega t$ and orbital frequency
\begin{equation}
    \Omega = \frac{\sqrt{M}}{\rp^{3/2}+a\sqrt{M}}.
    \label{eq:Omega}
\end{equation}
Its four-velocity is $u^\alpha = u^t(1,0,0,\Omega)$, with
\begin{equation}
    u^t = \frac{1+a\sqrt{M}\rp^{-3/2}}{\sqrt{1-3M/\rp+2a\sqrt{M}\rp^{-3/2}}},
    \label{eq:orbit}
\end{equation}
and $1/u^t$ is the redshift.
The conserved specific energy $\En \coloneqq -u_\mu(\partial_t)^\mu$ and specific azimuthal angular momentum $\Lz \coloneqq u_\mu(\partial_\varphi)^\mu$, associated with the Killing vectors $\partial_t$ and $\partial_\varphi$, are
\begin{align} \label{eq:En}
    \En &= \frac{1-2M/\rp+a\sqrt{M}\rp^{-3/2}}{\sqrt{1-3M/\rp+2a\sqrt{M}\rp^{-3/2}}},
    \\ \label{eq:Lz}
    \Lz &= \frac{\sqrt{M\rp}\,(1-2a\sqrt{M}\rp^{-3/2}+a^2/\rp^2)}{\sqrt{1-3M/\rp+2a\sqrt{M}\rp^{-3/2}}}.
\end{align}
They satisfy $\En - \Omega\Lz = 1/u^t$, which follows from the normalisation $u_\mu u^\mu = -1$.

\subsection{Metric perturbations}
\label{sec:perturbations/metric-perturbations}

We expand the exact metric as
\begin{equation}
    g_{\mu\nu}^\text{exact} = g_{\mu\nu} + \varepsilon h^{(1)}_{\mu\nu} + \varepsilon^{2} h^{(2)}_{\mu\nu} + \mathcal{O}(\varepsilon^{3}),
\end{equation}
where $\varepsilon = \mu/M$ is the small mass-ratio. 
The perturbations $h^{(n)}_{\mu\nu}$ are therefore independent of $\mu$.
Throughout this paper we work at first order in $\varepsilon$; the second order enters only through the regularity it requires of the first-order field, discussed at the end of this section.

The first-order perturbation obeys the linearised Einstein equation,
\begin{equation}
    G^{(1)}_{\mu\nu}[h^{(1)}] = 8\pi T_{\mu\nu},
    \label{eq:linEinstein}
\end{equation}
sourced by the stress-energy of a point mass,
\begin{equation}
    T^{\mu\nu} = M \int \frac{u^{\mu} u^{\nu}}{\sqrt{-g}}\, \delta^{(4)}[x - x_p(\tau)]\, d\tau.
\end{equation}
Here $G^{(n)}_{\mu\nu}$ denotes the term of $n$th order in the expansion of the Einstein tensor of $g_{\mu\nu} + h_{\mu\nu}$ about the background,
\begin{equation}
    G_{\mu\nu}[g + h] = G^{(1)}_{\mu\nu}[h] + G^{(2)}_{\mu\nu}[h, h] + {\cal O}(h^{3}),
    \label{eq:einstein_tensor_expansion}
\end{equation}
where we have used $G_{\mu\nu}[g] = 0$ for the vacuum Kerr background; $G^{(1)}_{\mu\nu}$ is linear in its argument and $G^{(2)}_{\mu\nu}$ is quadratic.

For a trace-reversed perturbation $\bar h_{\mu\nu} \coloneqq h_{\mu\nu} - \half g_{\mu\nu} g^{\alpha\beta} h_{\alpha\beta}$, we define its divergence and the wave operator as,
\begin{align}
    Z_\mu[\bar h] \coloneqq{}& \nabla^\nu \bar h_{\mu\nu}, \\
    E_{\mu\nu}[\bar h] \coloneqq{}& \Box\bar h_{\mu\nu} + 2R\indices{_\mu^\alpha_\nu^\beta}\bar h_{\alpha\beta},
\end{align}
with $\Box \coloneqq g^{\alpha\beta}\nabla_\alpha\nabla_\beta$.
In the terms of these, the linearised Einstein tensor is \cite{Pound:2021qin}
\begin{equation}
    G^{(1)}_{\mu\nu}[h] = -\frac{1}{2}E_{\mu\nu}[{\bar h}] + \nabla_{(\mu}Z_{\nu)}[{\bar h}] - \frac{1}{2}g_{\mu\nu}\nabla^{\alpha}Z_{\alpha}[{\bar h}].
\end{equation}
In the Lorenz gauge,
\begin{equation}
    Z_\mu[\bar h^{(1)}] = 0,
    \label{eq:lorenz_condition}
\end{equation}
so that \eqn{linEinstein} reduces to the hyperbolic system
\begin{equation}
    E_{\mu\nu}[\bar h^{(1)}] = -16\pi T_{\mu\nu}.
    \label{eq:lorenz}
\end{equation}
Because the background is Ricci-flat, the Riemann term drops out of the trace of \eqn{lorenz}.
The trace $h^{(1)} \coloneqq g^{\mu\nu}h^{(1)}_{\mu\nu}$ therefore obeys the scalar wave equation $\Box h^{(1)} = 16\pi T$, with $T \coloneqq g^{\mu\nu}T_{\mu\nu}$.

For the circular equatorial orbit, the stress-energy is
\begin{equation}
    T_{\mu\nu} = M \frac{\,u_\mu u_\nu}{u^t\,\rp^2}\,\delta(r-\rp)\,\delta(\theta-\tfrac{\pi}{2})\,\delta(\varphi-\Omega t),
    \label{eq:Tmunu}
\end{equation}
where we have used $\sqrt{-g} = \Sigma\sin\theta$, which equals $\rp^2$ on the orbit.

Following Detweiler and Whiting~\cite{Detweiler:2002mi}, we split the retarded solution of \eqn{lorenz} as
\begin{equation}
    h^{(1)}_{\mu\nu} = h^{\S1}_{\mu\nu} + h^{\R1}_{\mu\nu}.
\end{equation}
The singular field $h^{\S1}_{\mu\nu}$ is a particular solution of the sourced equation that exerts no force on the particle.
The regular field $h^{\R1}_{\mu\nu}$ is a smooth vacuum solution in a neighbourhood of the worldline.
The particle moves on a geodesic of the effective metric $g_{\mu\nu} + \varepsilon h^{\R1}_{\mu\nu} + \order(\e^2)$.
Equivalently, it follows a forced geodesic in the background $g_{\mu\nu}$,
\begin{equation}
   \mass u^\beta \nabla_\beta u^\alpha = F^\alpha 
\end{equation}
with the first-order self-force~\cite{Mino:1996nk,Quinn:1996am,Detweiler:2002mi}
\begin{equation}
    F^\alpha = -\frac{\mass^2}{2 M}\bigl(g^{\alpha\delta}+u^\alpha u^\delta\bigr)\bigl(2\nabla_\beta h^{\R1}_{\delta\gamma}-\nabla_\delta h^{\R1}_{\beta\gamma}\bigr)u^\beta u^\gamma .
    \label{eq:GSF}
\end{equation}
In practice, we work with a puncture $h^{\calP1}_{\mu\nu}$, a truncated local expansion of $h^{\S1}_{\mu\nu}$ about the worldline, and the residual field $h^{\res1}_{\mu\nu} \coloneqq h^{(1)}_{\mu\nu} - h^{\calP1}_{\mu\nu}$.
On the worldline, $h^{\res1}_{\mu\nu}$ agrees with $h^{\R1}_{\mu\nu}$ through as many derivatives as the order of the puncture allows~\cite{Barack:2007we,Vega:2007mc,Wardell:2015ada}.
The puncture we use, described in \secref{punctures}, reproduces the singular field through order $\lambda^2$ in the distance to the worldline, so $h^{\res1}_{\mu\nu}$ is $C^2$ there.
This regularity is needed at second order, where the source contains second derivatives of $h^{\res1}_{\mu\nu}$.
Indeed, away from the worldline, the second-order perturbation satisfies
\begin{equation}
    G^{(1)}_{\mu\nu}[h^{(2)}] = -G^{(2)}_{\mu\nu}[h^{(1)},h^{(1)}],
    \label{eq:second_order}
\end{equation}
with $G^{(2)}_{\mu\nu}$ defined in \eqn{einstein_tensor_expansion}.
In the Lorenz gauge, $Z_\mu[\bar h^{(2)}]=0$ and this becomes $E_{\mu\nu}[\bar h^{(2)}] = 2\,G^{(2)}_{\mu\nu}[h^{(1)},h^{(1)}]$.

The source $G^{(2)}_{\mu\nu}$ is a sum of products of the form $h\,\partial^2h$ and $\partial h\,\partial h$.
Near the particle, $G^{(2)}_{\mu\nu}[h^{\S1},h^{\S1}]\sim\lambda^{-4}$, which is not locally integrable.
The second-order field is therefore also split into a puncture $h^{\calP2}_{\mu\nu}$ and a residual field $h^{\res2}_{\mu\nu}$, which off the worldline obeys
\begin{equation}
    E_{\mu\nu}[\bar h^{\res2}] = 2\,G^{(2)}_{\mu\nu}[h^{(1)},h^{(1)}] - E_{\mu\nu}[\bar h^{\calP2}].
    \label{eq:second_order_residual}
\end{equation}
The second-order puncture is constructed so that $E_{\mu\nu}[\bar h^{\calP2}]$ cancels the singular part of the first term.
Substituting $h^{(1)}_{\mu\nu}=h^{\calP1}_{\mu\nu}+h^{\res1}_{\mu\nu}$, the source in \eqn{second_order_residual} splits into terms that involve the puncture and the term $2G^{(2)}_{\mu\nu}[h^{\res1},h^{\res1}]$, which involves only the residual field.
Since $G^{(2)}_{\mu\nu}$ contains at most two derivatives of each factor and $h^{\res1}_{\mu\nu}$ is $C^2$, this residual--residual term is continuous at the worldline.

From here on we suppress the label ``$1$" on first-order quantities, writing, for example, $h_{\mu\nu}$ for $h^{(1)}_{\mu\nu}$, except where second-order quantities also appear.

\subsection{Field variables and mode decompositions}
\label{sec:perturbations/reconstruction-in-lm-modes}
We work with the components of the perturbation on the unnormalised tetrad legs of Ref.~\cite{Dolan:2023enf},
\begin{subequations}\label{eq:ddkwlegs}
\begin{align}
    \ltetrad^{\mu}_{+} &= \ltetrad^{\mu},
    &
    \ltetrad^{\mu}_{-} &= -\frac{2\Sigma}{\Delta}\,\ntetrad^{\mu}, \\
    \mtetrad^{\mu}_{+} &= \sqrt{2}\,\bar{\zeta}\,\mtetrad^{\mu},
    &
    \mtetrad^{\mu}_{-} &= \sqrt{2}\,\zeta\,\barm^{\mu},
\end{align}
\end{subequations}
where $(\ltetrad^\mu,\ntetrad^\mu,\mtetrad^\mu,\barm^\mu)$ is the Kinnersley tetrad, and $\zeta = r - i a \cos \theta$; see Appendix~\ref{sec:newman_penrose_conventions} for further details.

We rescale these components by the factors that clear the denominators $\bar\zeta$, $\zeta$ and $\Sigma$ in the reconstruction formulas of Ref.~\cite{Dolan:2023enf} (their Eq.~(84)), multiply $h_{\ltetrad_+\ltetrad_-}$ by $\Delta$ as in that equation, and use the trace in place of $h_{\mtetrad_+\mtetrad_-}$:
\begin{subequations}\label{eq:slots}
\begin{align}
    \hs{\ltetrad_\pm \ltetrad_\pm} ={}& h_{\ltetrad_\pm \ltetrad_\pm}, & \hs{\mtetrad_\pm \mtetrad_\pm} ={}& h_{\mtetrad_\pm \mtetrad_\pm},\\
    \hs{\ltetrad_+\mtetrad_+} ={}& \bar\zeta\,h_{\ltetrad_+\mtetrad_+}, {}& \hs{\ltetrad_+\mtetrad_-} ={}& \zeta\,h_{\ltetrad_+\mtetrad_-},\\
    \hs{\ltetrad_-\mtetrad_+} ={}& \zeta\,h_{\ltetrad_-\mtetrad_+}, {}& \hs{\ltetrad_-\mtetrad_-} ={}& \bar\zeta\,h_{\ltetrad_-\mtetrad_-},\\
    \hs{\ltetrad_+\ltetrad_-} ={}& \Sigma\Delta\,h_{\ltetrad_+\ltetrad_-}, {}& \hs{\rm tr} ={}& g^{\mu\nu}h_{\mu\nu}.
\end{align}
\end{subequations}
The Kinnersley components follow algebraically; for example, $h_{\ltetrad\ltetrad} = \hs{\ltetrad_{+} \ltetrad_{+}}$ and $h_{\ntetrad\ntetrad} = \Delta^{2} \hs{\ltetrad_{-} \ltetrad_{-}}/(4\Sigma^{2})$.

For the circular orbit, the field depends on $t$ and $\varphi$ only through $\varphi-\Omega t$.
We therefore decompose each component $\hs{A}$ of \eqn{slots} into $m$-modes,
\begin{equation}
\begin{aligned}
    \hs{A} &= \sum_{m=-\infty}^{\infty}\hs{A}^{m}(r,\theta)\,e^{im(\varphi-\Omega t)},
\end{aligned}
\label{eq:mmode-def}
\end{equation}
so that the $m$-mode has frequency $\omega \coloneqq m\Omega$ and time dependence $e^{-i\omega t}$.
We expand each $m$-mode in spin-weighted spherical harmonics,
\begin{equation}
    \hs{A}^{m}(r, \theta) = \sum^{\infty}_{\ell = \ell_{\rm min}} \hs{A}^{\ell m}(r)\, {}_{s_{A}}Y_{\ell m}(\theta, 0),
    \label{eq:lmmode-def}
\end{equation}
where $s_A$ is the spin weight of $\hs{A}$, the number of its $m_+$ indices minus the number of its $m_-$ indices (Appendix~\ref{sec:newman_penrose_conventions}), and $\ell_{\rm min} \coloneqq \max(|m|, |s_{A}|)$.
We call the coefficients $\hs{A}^{\lm}(r)$ the $(\ell, m)$-modes of the field.
Only $m \geq 0$ need be computed, since the modes with $m<0$ follow from the reality of $h_{\mu\nu}$.

Perturbation equations on a Kerr background, such as the Teukolsky equation of Sec.~\ref{sec:perturbations/teukolsky-equation}, do not separate in the spherical harmonics of \eqn{lmmode-def} but in spin-weighted spheroidal harmonics ${}_{s}S^{a\omega}_{jm}(\theta,\varphi)$, so we work with both bases and convert between them.
Throughout, $\ell$ labels spherical and $j$ spheroidal multipoles.
The spheroidal harmonics are normalised to unity on the sphere, their eigenvalues ${}_{s}\lambda^{a\omega}_{jm}$ are those of Teukolsky~\cite{Teukolsky:1973ha}, reducing to $(j-s)(j+s+1)$ for $a\omega=0$ (for $s>0$, Ref.~\cite{Dolan:2023enf} uses instead ${}_{-s}\lambda^{a\omega}_{jm}={}_{s}\lambda^{a\omega}_{jm}+2s$), and they are expanded in the spherical basis as
\begin{equation}
    {}_{s}S^{a\omega}_{jm}(\theta,\varphi) = \sum_{\ell = \max(|m|,|s|)}^{\infty} {}_{s}b^{(m , a\omega)}_{j \ell}\, {}_{s}Y_{\ell m}(\theta, \varphi),
    \label{eq:spheroidal_to_spherical}
\end{equation}
for $j\geq\max(|m|,|s|)$, with the coefficients ${}_{s}b^{(m , a\omega)}_{j \ell}$ computed as in Appendix~A of Ref.~\cite{Hughes:1999bq}.
Products of spherical harmonics with powers of $\cos\theta$ and $\sin\theta$, which arise from the $\theta$-dependence of the Kerr background, are re-expanded in spherical harmonics using Clebsch--Gordan coefficients, as in Eqs.~(132)--(133) of Ref.~\cite{Dolan:2023enf}.

\subsection{The Teukolsky equation}
\label{sec:perturbations/teukolsky-equation}

The fields from which we build the metric perturbation obey Teukolsky equations~\cite{Teukolsky:1973ha} of spin weight $s\in\{0,\pm1,\pm2\}$, or equations with the same radial operator, each with its own source. For the circular orbit with $\omega = m \Omega$, a field of spin weight $s$, separates as $\sum_{jm}{}_{s}R_{jm}(r)\,{}_{s}S^{a\omega}_{jm}(\theta,\varphi)\,e^{-i\omega t}$, and its radial functions obey
\begin{multline}
    \biggl[\Delta^{-s}\diff{}{r}\Bigl(\Delta^{s+1} \diff{}{r}\Bigr)+\frac{K^2-2is(r-M)K}{\Delta}
    \\
    +4is\omega r-{}_s\lambda^{a\omega}_{jm}\biggr]{}_{s}R_{jm}={}_{s}T_{jm},
    \label{eq:teukolsky_radial}
\end{multline}
where $K\coloneqq(r^2+a^2)\omega-ma$ and ${}_{s}T_{jm}$ is the source.

For $\omega\neq0$, the two causal homogeneous solutions are defined by
\begin{equation}
\begin{aligned}
    {}_{s}R^{\rm in}_{jm}(r\to\rplus)&\sim\Delta^{-s}\ee^{-i(\omega-m\Omega_\mathcal{H})\rstar},
    \\
    {}_{s}R^{\rm up}_{jm}(r\to\infty)&\sim r^{-(2s+1)}\ee^{i\omega\rstar},
\end{aligned}
\label{eq:bcs}
\end{equation}
that is, purely ingoing at $\hor^+$ and purely outgoing at $\scri^+$.
A field sourced only at the particle is therefore, for $r\neq r_p$, a piecewise combination of the two,
\begin{multline}
    {}_{s}R_{jm}(r) = {}_{s}Z^{\rm in}_{jm}{}_{s}R^{\rm in}_{jm}(r)\,\Theta(r_p-r)\\
    +{}_{s}Z^{\rm up}_{jm}{}_{s}R^{\rm up}_{jm}(r)\,\Theta(r-r_p).
    \label{eq:inup}
\end{multline}
The amplitudes ${}_{s}Z^{\rm in/up}_{jm}$ follow from variation of parameters, as integrals of the source against the homogeneous solutions, or equivalently from the jumps of the field and its radial derivative across $r=r_p$, obtained by integrating \eqn{teukolsky_radial} across the orbit.

\subsection{Overview of the method}
\label{sec:perturbations/overview}
On a Kerr background, the Lorenz gauge field equations~\eqref{eq:lorenz} do not separate.
Dolan \emph{et al.}~\cite{Dolan:2021ijg} showed, however, that their solutions can be built from solutions of separable equations.
This construction was applied to circular equatorial orbits in Ref.~\cite{Dolan:2023enf} and has since been implemented in Refs.~\cite{Dyson:2025dlj,Li:2025ffh,Li:2026foy,Dyson:2026ddd}. It was later extended to arbitrary sources in Ref.~\cite{Wardell:2024yoi}, with implementations for generic bound Kerr geodesics in Refs.~\cite{Cunningham_2026, Cunningham_inprep}.

Our starting point is to calculate the first-order retarded field in the variables of Sec.~\ref{sec:perturbations/reconstruction-in-lm-modes}; that is, the $(\ell,m)$-modes $\hs{A}^{\lm}(r)$.
We follow the sourced formulation of Wardell, Kavanagh and Dolan~\cite{Wardell:2024yoi}, in which the Lorenz gauge perturbation is obtained by a gauge transformation of the metric perturbation of Aksteiner, Andersson and B\"ackdahl (AAB)~\cite{Aksteiner:2016pjt}.
Schematically,
\begin{equation} 
    \label{eq:metric-reconstruction-decomposition}
    h^{(1)}_{\mu\nu} = (M\partial_{t})^{-1}[h^{\rm AAB}_{\mu\nu} + {\cal L}_{\xi}g_{\mu\nu} + h^{\rm tr}_{\mu\nu}] + h^{\rm comp}_{\mu\nu}.
\end{equation}
The AAB perturbation $h^{\rm AAB}_{\mu\nu}$ is built from the Weyl scalars $\psi_0$ and $\psi_4$; the gauge term ${\cal L}_{\xi}g_{\mu\nu}$ takes it to the Lorenz gauge; the trace piece $h^{\rm tr}_{\mu\nu}$ restores the trace, which $h^{\rm AAB}_{\mu\nu}$ lacks; and the completion $h^{\rm comp}_{\mu\nu}$ adds the perturbations of the black hole's mass and angular momentum, together with gauge pieces, which cannot be constructed from Teukolsky scalars.

On an $m$-mode, $(M\partial_t)^{-1}$ is the factor $(-iM\omega)^{-1}$, which exists only for $m\neq0$, so the construction splits into two sectors.
For $m\neq0$ (Sec.~\ref{sec:perturbations/radiative-modes}), the first three terms of \eqn{metric-reconstruction-decomposition} are built from six radial functions, whose equations are given in Sec.~\ref{sec:perturbations/radial-equations}, and these are combined directly into $\hs{A}^{\lm}$ by \eqn{reconOp} (Sec.~\ref{sec:perturbations/Reconstruction}).
In the circular case, the completion contributes only to $m=0$; these modes are constructed separately, following Sec.~IV of Ref.~\cite{Dolan:2023enf} (Sec.~\ref{sec:perturbations/non-radiative-modes}).
Section~\ref{sec:perturbations/solution} describes the numerical solution of the radial equations, and Sec.~\ref{sec:perturbations/lm-to-m-modes} the conversion of the $(\ell,m)$-modes into the $m$-modes required at second order.

\subsection{Radiative modes (\texorpdfstring{$m \neq 0$}{m≠0})}
\label{sec:perturbations/radiative-modes}
We now construct the first three terms of \eqn{metric-reconstruction-decomposition} for $m\neq0$.

\subsubsection{The AAB, gauge and trace pieces}
\label{sec:perturbations/radiative-pieces}
The first term of \eqn{metric-reconstruction-decomposition}, $h^{\rm AAB}_{\mu\nu}$, is built from the two maximum-spin Weyl scalars together, without the need for a Hertz potential,
\begin{equation} 
    h^{\rm AAB}_{\mu\nu} = \frac{4}{3}\left( {\cal S}^{\dagger}_{4} \zeta^{4}\psi_{0} - {\cal S}^{\dagger}_{0}\zeta^4\psi_{4}\right)_{\mu\nu} + 8\pi({\cal N} T)_{\mu\nu},
    \label{eq:AAB_definition}
\end{equation}
where $\psi_{0}$ and $\zeta^4 \psi_{4}$ are the solutions of the sourced $s = \pm 2$ Teukolsky equations, ${\cal S}^{\dagger}_{0, 4}$ are the adjoints of the Teukolsky decoupling operators, $\zeta$ (Sec.~\ref{sec:perturbations/reconstruction-in-lm-modes}) is the Killing spinor coefficient, and ${\cal N}$ is the \emph{AAB corrector} \cite{Aksteiner:2016pjt}, a differential operator acting on the stress-energy tensor (Ref.~\cite{Wardell:2024yoi}, Eqs.~(13) and (18)).
By an operator identity of AAB \cite{Aksteiner:2016pjt}, $h^{\rm AAB}_{\mu\nu}$ solves the linearised Einstein equation~\eqref{eq:linEinstein} with the time derivative of the source,
\begin{equation}
    G^{(1)}_{\mu\nu}[h^{\rm AAB}] = 8\pi M \partial_{t}T_{\mu\nu}.
    \label{eq:AAB_identity}
\end{equation}
The other two terms in the bracket of \eqn{metric-reconstruction-decomposition} are pure gauge, so they do not change \eqn{AAB_identity}, and the inverse time derivative removes the factor of $M\partial_{t}$ from the whole bracket.
For a point particle, the corrector term is a distribution supported on the worldline, and \eqn{AAB_identity} holds in the distributional sense, including at the particle. 
The perturbation $h^{\rm AAB}_{\mu\nu}$ is a solution of the sourced problem everywhere. It is trace-free, and it is neither in the Lorenz gauge nor
free of distributions.

The gauge transformation to the Lorenz gauge is generated by a vector $\xi_{\mu} + \xi^{\rm tr}_{\mu}$ that obeys
\begin{equation}
    \Box(\xi_{\nu} + \xi^{\rm tr}_{\nu}) = -\nabla^{\mu} h^{\rm AAB}_{\mu\nu},
    \label{eq:gauge_transformation_aab}
\end{equation}
whose right-hand side contains distributions on the worldline through $\psi_{0}$, $\psi_{4}$ and ${\cal N}T$.
Here $\xi^{\rm tr}_{\nu}$ is the part that restores the trace (below) and $\xi_{\mu}$ the divergence-free remainder, which gives ${\cal L}_{\xi}g_{\mu\nu}$.
The latter is split into a part sourced by the Weyl scalars, the vacuum gauge vector of Ref.~\cite{Dolan:2021ijg} built from $\psi_{0}$ and $\psi_{4}$, and a part sourced by the stress-energy alone, whose two-form potential has tetrad components $\phi_{0}$ and $\phi_{2}$ that obey sourced $s = \pm1$ Teukolsky equations \cite{Wardell:2024yoi}. For each part a scalar field correction is also required to enforce the vector Lorenz gauge condition on the gauge vector; i.e., $\nabla^\alpha\xi_\alpha=0$.
Its components are therefore constructed from solutions of the Teukolsky equation with spin-weights $s = 0, \pm 1, \pm 2$ and their derivatives.

The trace piece $h^{\rm tr}_{\mu\nu} = 2 \nabla_{(\mu}\xi^{\rm tr}_{\nu)}$, with $2 \nabla^{\mu}\xi^{\rm tr}_{\mu} = M\partial_{t}h^{(1)}$, supplies the trace; the trace $h^{(1)}$ itself obeys the sourced scalar equation $\Box h^{(1)} = 16\pi T$ of \secref{perturbations/metric-perturbations}.
The vector $\xi^{\rm tr}_{\mu}$ is built from the trace and the gradient of the spin-$0$ gauge scalar sourced by $M\partial_{t}h^{(1)}$ \cite{Wardell:2024yoi}.
Together with the scalar parts of $\xi_{\mu}$ that are sourced by the stress-energy, this scalar forms a single spin-0 field, $\chi$.

In total, each $(\ell, m)$-mode with $m \neq 0$ is built from six fields: $\psi_{0}$ and $\psi_{4}$ $(s = \pm 2)$, $\phi_{0}$ and $\phi_{2}$ $(s = \pm 1)$, the trace $h$ $(s = 0)$, and the gauge scalar $\chi$ $(s=0)$, whose radial equations are given in Sec.~\ref{sec:perturbations/radial-equations}.

This differs from the construction of Dolan \emph{et al.} \cite{Dolan:2023enf} for the same configuration (though the results are consistent). In Ref.~\cite{Dolan:2023enf} the perturbation is assembled from vacuum solutions on each side of the orbit: the $s = \pm 2$ part is the difference of the ingoing and outgoing radiation gauge perturbations, each transformed to Lorenz gauge with the vector of Ref.~\cite{Dolan:2021ijg}.
In vacuum, this combination reduces the inversion relations between the Hertz potentials and the Weyl scalars to algebraic expressions, and it coincides with the AAB perturbation~\eqref{eq:AAB_definition} in the vacuum regions.
The inner and outer vacuum solutions are then glued at the worldline, and the amplitudes of the $s = 0, \pm 1$ fields are fixed by imposing continuity of the metric perturbation at the worldline and the jumps in its first radial derivative.
Instead we follow Ref.~\cite{Wardell:2024yoi} in keeping the distributional terms of the AAB construction, including the corrector ${\cal N} T$ and the distributional sources of the gauge-vector equations, so that every field obeys its own sourced equation and its amplitudes follow directly. For $m\neq0$, no continuity conditions are imposed on the metric perturbation at the worldline.

\subsubsection{Radial equations}
\label{sec:perturbations/radial-equations}

We denote the radial functions of the six fields by $\Psi_0$ and $\Psi_4$ (of $\psi_0$ and $\zeta^4\psi_4$, with $s=\pm2$), $\Phi_0$ and $\Phi_2$ (of $\phi_0$ and $\zeta^2\phi_2$, with $s=\pm1$), $h_{jm}$ (the trace, $s=0$) and $\chi_{jm}$ ($s=0$), suppressing the indices $jm$ on the first four.
The first five obey Teukolsky equations~\eqref{eq:teukolsky_radial} whose sources are supported on the orbit, and therefore take the form~\eqref{eq:inup}.
For a circular orbit, the sources of $\Psi_0$ and $\Psi_4$ contain $\delta(r-r_p)$ and its first two derivatives, and those of $\Phi_0$ and $\Phi_2$ up to its fifth derivative.
The source of the trace is $\delta(r-r_p)$ alone, so that
\begin{equation}
    [h_{j m}]_{r_p} = 0,
    \qquad
    [\partial_{r} h_{j m}]_{r_p} = -\frac{16\pi M}{u^{t}\Delta} \bigg|_{\rp} \overline{{}_{0}S^{a\omega}_{j m} \l \pi/2, 0 \r },
    \label{eq:tracejumps}
\end{equation}
where $[g]_{r_{p}}$ denotes the jump of a function $g$ across the orbit.

The sixth field, $\chi$, is an auxiliary spin-$0$ field that ensures the reconstructed metric yields the correct trace, cancels the residual violation of the Lorenz gauge condition, and restores the correct distributional behavior. In the vacuum regions it satisfies $\Box \chi = \tfrac{1}{2}M \partial_t h$, where $h$ is the trace. 
In general, its radial functions obey
\begin{multline}
    \biggl[\diff{}{r}\Bigl(\Delta\diff{}{r}\Bigr)+\frac{K^2}{\Delta}-{}_0\lambda^{a\omega}_{jm}\biggr]\chi_{jm}
    \\
    = -\frac{i M \omega}{2}\sum_{j'}\bigl[\delta_{jj'}r^{2}+a^{2}\gamma^{a\omega}_{j'jm}\bigr]h_{j'm}+{}_{\chi}T_{jm},
    \label{eq:chi_radial}
\end{multline}
whose operator is the $s=0$ operator of \eqn{teukolsky_radial}.
The bracket is the projection of $\Sigma=r^2+a^2\cos^2\theta$ onto the spheroidal harmonics, with the spin-$0$ coupling coefficients
\begin{equation}
    \gamma_{j' j m}^{a\omega} \coloneqq \oint {}_0\bar{S}_{j m}^{a\omega}(\theta,\varphi)\, \cos^2\theta \,{}_0S_{j' m}^{a\omega}(\theta,\varphi)\,\dd \Omega ,
    \label{eq:gamma_coupling}
\end{equation}
and ${}_{\chi}T_{jm}$ collects the terms built from the stress-energy, which come from the scalar parts of $\xi_\mu$ sourced by the stress-energy (Sec.~\ref{sec:perturbations/radiative-pieces}; see Eq.~(60f) of Ref.~\cite{Wardell:2024yoi}. It is supported on the orbit and gives $\chi_{jm}$ jumps at $\rp$.
Because the trace term is non-zero everywhere, not only at the orbit, the solution is not of the form~\eqref{eq:inup}; we obtain it as described in Sec.~\ref{sec:perturbations/solution}.

\subsubsection{Reconstruction}
\label{sec:perturbations/Reconstruction}
For $m\neq0$, each $(\ell,m)$-mode of the field variables, \eqn{lmmode-def}, is a linear combination of the six radial fields of \secref{perturbations/radial-equations} and their radial derivatives,
\begin{equation}
    \hs{A}^{\ell m}(r) = \sum_{F}\sum_{j}\sum_{k=0}^{4}\mathcal{C}^{(k)}_{A,F;\ell j}(r)\,\partial_r^kF_{j m}(r),
    \label{eq:reconOp}
\end{equation}
with $F\in\{\Psi_0,\Psi_4,\Phi_0,\Phi_2,h,\chi\}$ and $j$ the spheroidal multipole of $F$.
Equation~\eqref{eq:reconOp} holds for $r\neq r_p$, where the corrector ${\cal N}T$ and the parts of the fields supported on the orbit vanish; these enter only through the amplitudes of the fields.

The coefficients $\mathcal{C}^{(k)}$ follow from the operators of Refs.~\cite{Dolan:2023enf,Wardell:2024yoi}. 
They also contain the projection onto the spherical harmonics ${}_{s_A}Y_{\ell m}$.
They are rational in $r$, and their angular part is built from the coefficients ${}_sb_{j\ell}$ of \eqn{spheroidal_to_spherical} and the Clebsch--Gordan coefficients of Sec.~\ref{sec:perturbations/reconstruction-in-lm-modes}.

The sum over $k$ in Eq.~\eqref{eq:reconOp} counts radial derivatives: up to four for the spin-2 fields and up to two for the other fields.
The sum over $j$ collects the spheroidal modes that feed a given spherical mode $\ell$, through both the $\theta$-dependence of the operators and the change of basis.
We evaluate it by applying each angular operation as a matrix acting on the vector of radial fields $F_{j m}$.

\subsection{Non-radiative modes (\texorpdfstring{$m=0$}{m=0})}
\label{sec:perturbations/non-radiative-modes}
When restricting to circular orbits, the axisymmetric ($m=0$) and static ($\omega=0$) modes coincide. For these modes, the decomposition in \eqn{metric-reconstruction-decomposition} does not apply, and we construct them through a separate procedure described in Sec.~\ref{sec:perturbations/static-modes}. The $\ell=0,1$ modes also require an additional completion piece, $h^{\rm comp}_{\mu\nu}$, which depends on our choice of boundary conditions in the non-radiative sector. In Sec.~\ref{sec:perturbations/Berndtson-type-solutions}, we describe our choice of regular boundary conditions, which produces the so-called Berndtson-type solution for $h^{\rm comp}_{\mu\nu}$~\cite{Berndtson:2007gsc}, and its effect on the mass and angular momentum of the perturbed spacetime \cite{Berndtson:2007gsc}.

\subsubsection{Reconstructing static modes}
\label{sec:perturbations/static-modes}

The reconstruction described in Sec.~\ref{sec:perturbations/radiative-pieces} breaks down for static modes because inverting the time derivative in \eqn{metric-reconstruction-decomposition} produces factors of $1/\omega$ that diverge in the static limit. In particular, we cannot obtain the static Lorenz gauge perturbations using the reconstructed AAB metric perturbation as an intermediate. Therefore, we instead follow Ref.~\cite{Dolan:2023enf} and reconstruct the metric perturbation in the ingoing radiation gauge (IRG) and solve for the gauge vector that transforms it from IRG to Lorenz gauge. As in the radiative sector, the gauge contributions for the static modes are built from auxiliary fields. Spin-$0$ and spin-$2$ auxiliary fields are sourced by the Teukolsky scalars, but no spin-1 fields are required. Unlike the radiative sector, the static $\ell = 0, 1$ modes cannot be fully reconstructed from Teukolsky scalars and also require the completion piece, $h^{\rm comp}_{\mu\nu}$ in Eq.~\eqref{eq:metric-reconstruction-decomposition}, which is formed from six independent non-radiative homogeneous solutions to the field equations and gauge constraints. Furthermore, this approach does not naturally include distributional source contributions along the worldline.

Therefore, we construct the IRG perturbation, gauge contributions, and completion pieces on either side of the particle's radial position, analogous to \eqn{inup}. The jumps across $r=r_p$ (which encode the distributional content of the source along the worldline) and the relative amplitudes of each spin-0 field, spin-2 field, and the homogeneous completion solutions are then determined by the field equations, Lorenz gauge constraints, continuity of each $(\ell,m)$ mode of $h_{\mu\nu}^{(1)}$ at $r=r_p$, and boundary conditions, as was done in Sec.~IV of Ref.~\cite{Dolan:2023enf}.

The choice of boundary conditions, however, is not unique. One typically seeks the most regular solution at both boundaries that also satisfies the field equations and preserves the mass and angular momentum of the spacetime. In Lorenz gauge, no such solution exists; the metric perturbation cannot simultaneously be regular at the horizon, be asymptotically flat, and leave the black hole inside the orbit with the Kerr background mass and angular momentum, $M$ and $J=aM$ \cite{Dolan:2012jg, Berndtson:2007gsc}. Because we require a globally regular solution when using first-order perturbations for second-order calculations, we construct the so-called Berndtson-type solution \cite{Berndtson:2007gsc} by introducing additional completion terms that ensure regularity at the horizon and infinity but lead to $\mathcal{O}(\varepsilon)$ corrections to the mass and angular momentum of $g_{\mu\nu} + \varepsilon h^{(1)}_{\mu\nu}$. In~\secref{perturbations/Berndtson-type-solutions}, we summarize the construction of this Berndtson-type solution and describe how we account for the shifted mass and angular momentum of our globally-regular Lorenz gauge solution when comparing with quantities defined in terms of the original Kerr background parameters.


\subsubsection{Berndtson-type solutions (completion)}
\label{sec:perturbations/Berndtson-type-solutions}

The completion piece $h^{\rm comp}_{\mu\nu}$ is constructed from a basis of six independent homogeneous solutions, $h^{(\Lambda)}_{\mu\nu}$ with $\Lambda \in \{{\rm B, C, D, E, F, G }\}$ such that $h^{\rm comp}_{\mu\nu} = \sum_{\Lambda} h^{(\Lambda)}_{\mu\nu}$. These solutions are known analytically and given in Table III of Ref.~\cite{Dolan:2023enf}.\footnote{Table III lists seven solutions, but the (A) solution is linearly dependent on the others and can thus be omitted.} The solutions (E) and (G) carry mass and angular momentum charges and are given by,
\begin{align}
    h_{\mu\nu}^{(\mathrm{E})} ={}& M\bigl[ c^{(\mathrm{E})}_\mathrm{Bernd} + c^{(\mathrm{E})}_\mathrm{comp}\,\Theta(r - r_p) \bigr] \notag \\
    & \qquad \qquad \times \bigl( \partial_M g_{\mu\nu}\big|_a + \text{gauge terms} \bigr),
    \\
    h_{\mu\nu}^{(\mathrm{G})} ={}& M\bigl[ c^{(\mathrm{G})}_\mathrm{Bernd} + c^{(\mathrm{G})}_\mathrm{comp}\,\Theta(r - r_p) \bigr] \notag \\
    & \qquad \qquad \times \bigl( \partial_a g_{\mu\nu}\big|_M + \text{gauge terms} \bigr),
\end{align}
while (B), (C), (D), and (F) are pure gauge. The coefficients $c^{(\mathrm{E})}_\mathrm{comp} = \mathcal{E}$ and $c^{(\mathrm{G})}_\mathrm{comp} = (\mathcal{L}_z-a\mathcal{E})/M$ are fixed so that the mass and angular momentum measured outside the orbit exceed those inside it by the particle’s energy, $\mu{\cal E}$, and the angular momentum, $\mu{\cal L}_{z}$. The coefficients $c^{(\mathrm{E})}_\mathrm{Bernd}$ and $c^{(\mathrm{G})}_\mathrm{Bernd}$ shift the background mass and angular momentum and are fixed by boundary conditions. Setting $c^{(\mathrm{E})}_\mathrm{Bernd}=c^{(\mathrm{G})}_\mathrm{Bernd} = 0$ preserves the mass and angular momentum charges but produces a perturbation that is either irregular at the horizon or is no longer asymptotically flat.

Following Ref.~\cite{Dolan:2023enf}, we instead construct a Berndtson-type solution using the gauge vectors $\xi^{({\rm B,C,D,F})}_\mu$ defined in its Table III.\footnote{Here we rescale $\xi^{({\rm F})}_\mu$ by $M^{-1}$ to keep its dimensions consistent with $\xi^{({\rm D})}_\mu$ and all coefficients $c^{(\Lambda)}$ dimensionless.} We require
\begin{align}
    h_{\mu\nu}^{(\mathrm{D})} ={}& 2c^{(\mathrm{D})}_\mathrm{Bernd} \Theta(r_p-r) \nabla_{(\mu}\xi^{({\rm D})}_{\nu)},
    \\
    h_{\mu\nu}^{(\mathrm{F})} ={}& 2c^{(\mathrm{F})}_\mathrm{Bernd} \Theta(r_p-r) \nabla_{(\mu}\xi^{({\rm F})}_{\nu)},
\end{align}
to ensure regularity at infinity and impose
\begin{align} \label{eq:static-completion-condition-cE}
    c^{(\mathrm{E})}_\mathrm{Bernd} &=  c^{(\mathrm{D})}_\mathrm{Bernd},
    \\  \label{eq:static-completion-condition-cG}
    M c^{(\mathrm{G})}_\mathrm{Bernd} &= a c^{(\mathrm{D})}_\mathrm{Bernd} + \frac{r_+(r_+-r_-)^2}{2M^2}c^{(\mathrm{F})}_\mathrm{Bernd},
\end{align}
to ensure regularity at the horizon. These constraints, along with further conditions on solutions (B) and (C) described in Sec.~IV of Ref.~\cite{Dolan:2023enf}, yield a globally regular Lorenz gauge metric perturbation whose mass $M_\bullet$ and angular momentum $J_\bullet = a_\bullet M_\bullet$ differ from the Kerr background mass $M$ and angular momentum $J=aM$,
\begin{align}
    M_\bullet &= M + \varepsilon \delta M,
    &
    J_\bullet &= J + \varepsilon \delta J,
\end{align}
and $a_\bullet = a + \varepsilon \delta a$, with
\begin{align} \label{eq:deltaM}
    \delta M &= M c^{(\mathrm{E})}_\mathrm{Bernd},
    \\ \label{eq:deltaJ}
    \delta J &= M( a c^{(\mathrm{E})}_\mathrm{Bernd} + M c^{(\mathrm{G})}_\mathrm{Bernd}),
\end{align}
and $M \delta a = \delta J - a\delta M = M^2 c^{(\mathrm{G})}_\mathrm{Bernd}$. In practice, we use Eqs.~\eqref{eq:static-completion-condition-cE} and \eqref{eq:static-completion-condition-cG} to eliminate $c^{(\mathrm{E})}_\mathrm{Bernd}$ and $c^{(\mathrm{G})}_\mathrm{Bernd}$ from our linear system of equations for the static solutions. Solving this system gives the amplitudes $ c^{(\mathrm{D})}_\mathrm{Bernd}$ and $c^{(\mathrm{F})}_\mathrm{Bernd}$, from which we can calculate the corrections $\delta M$ and $\delta J$ via Eqs.~\eqref{eq:deltaM} and \eqref{eq:deltaJ}. 

To recover the metric perturbation with respect to the Kerr background parameters, which we denote by $h^{\rm Kerr}_{\mu\nu}$, we re-expand the full spacetime metric including the Berndtson-type solution (to first order), which we temporarily denote as $g_{\mu\nu} + \varepsilon h^{\rm Bernd}_{\mu\nu}$, in terms of $M$ and $J$, leading to
\begin{align} \label{eq:dM-dJ-correction}
    h^{\rm Kerr}_{\mu\nu} = h^{\rm Bernd}_{\mu\nu}
    - \delta M\,\frac{\partial g_{\mu\nu}}{\partial M}\bigg|_J
    - \delta J\,\frac{\partial g_{\mu\nu}}{\partial J}\bigg|_M
    + 2\nabla_{(\mu}\Xi_{\nu)},
\end{align}
where $\Xi_\mu$ is a gauge vector whoose form depends on whether $h^{\rm Kerr}_{\mu\nu}$ is required to be asymptotically flat or regular at the horizon.
In Kerr, horizon regularity requires terms linear in $t$, a rescaling of $t$ and a rigid rotation in $\varphi$, analogous to the $\alpha$ correction in Schwarzschild \cite{Sago:2008id}.

Here $h^{\rm Bernd}_{\mu\nu}$ is our solution, and $h^{\rm Kerr}_{\mu\nu}$ is the same perturbation for comparison with the perturbation computed in radiation-gauge and post-Newtonian calculations, in which the black hole keeps the background parameters.
For the redshift, we take $\Xi_{\mu} = 0$, so that \eqn{dM-dJ-correction} subtracts only the parameter derivatives of the E and G modes.
The resulting $h^{\rm Kerr}_{\mu\nu}$ is neither in Lorenz gauge no regular at the horizon, but it is asymptotically flat, so $\Omega$ is the orbital frequency measured at infinity and ${\tilde z}_{1}(\Omega)$ can be compared directly with other calculations (\secref{results/redshift}).
The self-force is quoted for $h^{\rm Bernd}_{\mu\nu}$, which is both asymptotically flat and regular at the horizon.

We use $h^{\rm Bernd}_{\mu\nu}$ at every spin, including $a = 0$, where its black hole acquires the angular momentum $\varepsilon\delta J = -8{\cal L}_{z}M^{4}/\rp^3$\footnote{This follows from Eqs.~(\ref{eq:static-completion-condition-cE}) and (\ref{eq:static-completion-condition-cG}). For $a = 0$, $\rplus = 2M$ and $\rminus = 0$, so $c^{(F)}_{\rm Bern} = 4c^{(F)}_{\rm Bernd}$ and $\delta J = 4M^{2} c^{(F)}_{\rm Bernd}$, with $c^{(F)}_{\rm Bernd} = -2{\cal L}_{z} M^2 / \rp^{3} \neq 0$}.
Schwarzschild calculations~\cite{Akcay:2012ea, Leather:2024mls} instead complete the $m = 0$ mode with zero black-hole spin and the same $\delta M$.
The two differ only by the spin term of \eqn{dM-dJ-correction}, made regular at the horizon by a static gauge term:
\begin{gather}
    h^{\delta J=0}_{\mu\nu} = h^{\rm Bernd}_{\mu\nu} - \delta J\,S_{\mu\nu},\label{eq:h_dJ0}\\
    S_{t\varphi} = -\frac{2\sin^{2}\theta}{r},
    \qquad
    S_{r\varphi} = -\frac{4M\sin^{2}\theta}{r(r-2M)},\nn
    \label{eq:zero_spin_berndtson}
\end{gather}
with all other components of $S_{\mu\nu}$ zero.
$S_{t\varphi}$ is $\partial_{J}g_{t\varphi}$ at $a = 0$, and $S_{r\varphi}$ is pure gauge,
$S_{\mu\nu}$ is in Lorenz gauge and asymptotically flat, and in ingoing Eddington-Finkelstein coordinates its components are $\mp 2\sin^{2}\theta/r$, so it is regular at the horizon.
The mode is static and axisymmetric, so it leaves $F^{t}$ and $F^{\varphi}$ unchanged.
Its change in ${\tilde z}^{\rm Bernd}_{1}$ is removed by the $\delta J$ term, so ${\tilde z}_{1}$ is the same for both completions.
Only the conservative self-force changes, by
\begin{equation}
    \frac{M^{2}}{\mu^{2}}\bigl(F^{r}_{\delta J = 0} - F^{r}\bigr) = \frac{16(\rho_{0} - 2)}{\rho_{0}^{11/2}(\rho_{0} - 3)^{3/2}},
    \label{eq:Fr_spin_mode}
\end{equation}
where $\rho_{0} \coloneqq \rp / M$.

\subsection{Numerical solution}
\label{sec:perturbations/solution}
In this work, we solve every radial equation on hyperboloidal slices, in the compactified radial coordinate~\cite{Leather2026}
\begin{equation}
    \sigma \coloneqq \frac{2\kappa}{r-\rminus},
    \label{eq:sigma}
\end{equation}
which runs from $\sigma=0$ at $\scri^+$ to $\sigma=1$ at $\hor^+$, following the Schwarzschild calculation of Ref.~\cite{Leather:2024mls}.
Rescaling each field removes the oscillations and the peeling powers of \eqref{eq:bcs}, so the physical solutions are regular at both boundaries and no boundary condition is imposed beyond regularity. 
Each equation is solved by multi-domain Chebyshev collocation with the particle at a subdomain boundary. The homogeneous solutions are obtained domain by domain, starting from the boundary at which each is regular, whilst the sourced $\chi$ equation is solved as a single global linear system with the jumps as interface conditions. 
Radial derivatives beyond the first are obtained from the radial equation itself rather than by spectral differentiation; this is important for the fourth derivatives required in \eqn{reconOp}.
Each angular coupling connects a spherical mode $\ell$ to $\ell \pm 2$, and \eqn{reconOp} applies several such couplings in succession, so the highest spherical modes $\ell$ draw on spheroidal modes $j > \ell$ of the radial fields.
We therefore solve for every field up to the spheroidal mode $j_{\rm max}=\ellmax+12$ and keep only the spherical modes $\ell\le\ellmax$ of the reconstructed field.
The reconstruction loses accuracy rapidly with $\ell$ in double precision, so we use arbitrary-precision arithmetic with 60 significant digits.
Using these settings at every spin, the $s=-2$ amplitudes at $\scri^+$ and $\hor^+$ agree with those of the \texttt{pybhpt} package~\cite{Nasipak_pybhpt_2026} to better than $10^{-12}$ at $m=2$ and $m=40$, for $a=0.9M$ at $\rp=5M$, $8M$ and $12M$ and for $a=0.3M$ at $\rp=8M$.

During development of this work, various codes have been used to produce data sets and perform consistency checks. Exploratory work was done using the code developed by Dolan \textit{et al.}~\cite{Dolan:2023enf} which used the continuity of the metric perturbation in spin-weighted spherical harmonic modes to fix the Teukolsky ${}_sZ^{\rm up/in}_{\ell m}$ coefficients. That code was later optimized by two of the present authors, leveraging MST methods \cite{Mano:1996vt,Mano:1996gn} and robust numerical solvers previously developed for the first calculation of the generic gravitational self-force \cite{vandeMeent:2015lxa,vandeMeent:2016pee,vandeMeent:2017bcc}. Another implementation, detailed in Refs.~\cite{Cunningham_2026, Cunningham_inprep}, used theory laid out by Wardell \textit{et al.}~\cite{Wardell:2024yoi} to source Teukolsky equations for the gauge fields, and tools from the \texttt{Black Hole Perturbation Toolkit}~\cite{BHPToolkit} to obtain the ${}_sZ^{\rm up/in}_{\ell m}$ coefficients directly using variation of parameters without reference to the metric perturbation itself. In Ref.~\cite{Cunningham_2026}, the auxiliary field ($\kappa$ in their notation) of Wardell \textit{et al.}~\cite{Wardell:2024yoi} is calculated by using a matching procedure to determine the homogeneous contribution self-consistently. The spectral, hyperboloidal solver described above is based on this code and will be described in an upcoming work \cite{Leather2026}. That matching procedure worked well for eccentric-equatorial and inclined-circular orbits, but it proved an obstacle for generic orbits, so it is replaced in Ref.~\cite{Cunningham_inprep} with a partial annihilator method, which builds on the work of Refs.~\cite{Hopper:2012ty,Durkan:2022fvm,Dolan:2023enf,Berndtson:2007gsc}.

\subsection{\texorpdfstring{$(\ell,m)$}{(ℓ,m)} to \texorpdfstring{$m$}{m}-modes}
\label{sec:perturbations/lm-to-m-modes}

Having constructed the Berndtson solution, we obtain the complete set of $(\ell, m)$ solutions for the tetrad components of $\hs{A}^{\ell m}$. The second-order calculation, however, requires the $m$-modes of the metric perturbation. One could na\"{i}vely construct the $m$-modes by summing \eqn{lmmode-def} up to a finite $\ell_{max}$; however, this leads to Gibbs ringing over the whole sphere $r = \rp$. The reason behind this is straightforward: near the particle, the $(\ell, m)$-modes and the $m$-modes of the Lorenz-gauge field have different
non-smooth behavior,
\begin{align}
    &\hs{A}^{\ell m }\big|_{|r-\rp| \rightarrow 0} \sim C_0 + C_1 |r-\rp|, \label{eq:lm-kink} \\
    &\hs{A}^{m }\big|_{|\vec{x}-\vec{x}_p|\rightarrow 0} \sim B_0 \ln|\vec{x}-\vec{x}_p|, \label{eq:m-log}
\end{align}
where $\vec{x} = (r,\theta)$ and $\vec x_{p} = (\rp, \pi/2)$. The ringing is an artifact of approximating a logarithmic divergence by a finite sum of modes, each of which has only a $C^0$ kink at $r = \rp$. Such ringing remains substantial even at high values of $\ell$, as $\ell \leq \ellmax$ resolves angular structure only down to $\sim \pi/\ellmax$, so it follows the logarithm only down to $|r - r_p| \sim r_p / \ellmax$ and rings inside that distance; see \fig{gibbs_ringing}. This poor convergence near the worldline then spoils any attempt at the delicate cancellations required at second order in the self-force calculation.

In this work, we use a puncture to remove the leading non-smooth behavior of $\hs{A}^{\ell m}$ that would otherwise produce Gibbs ringing when summed over $\ell$.
The puncture therefore serves two purposes: first, to numerically convert $(\ell, m)$-modes into accurate $m$-modes, and secondly, to extract the regular field, from which the self-force and redshift will follow.

\section{Puncture construction}
\label{sec:punctures}

In this section, we describe the construction of the first-order puncture, $h^{\calP}_{\mu\nu}$, and its $m$- and $(\ell,m)$-mode decompositions.
Building on methods developed in Ref.~\cite{Bourg:2024cgh}, we start by calculating the puncture in comoving coordinates adapted to the motion of the particle before analytically decomposing the field into $m$-modes.
We verify that the puncture has been derived correctly by checking that it satisfies the $m$-mode field equations and gauge condition to the correct order in distance from the worldline.
The $m$-modes are then numerically decomposed into $(\ell,m)$-modes so they can be used to regularize the retarded field $(\ell,m)$-modes calculated in the previous section.
This is performed using a window function that fixes the poor behavior of the puncture at the poles.
We then sum over $\ell$ to get the $m$ modes of the residual field before adding back the puncture $m$-modes (multiplied by the window function) to obtain the physical field $m$-modes.

Written in covariant form, the first-order singular field is given by~\cite{Heffernan:2012su,Pound:2014xva}
\begin{align}
    h^{\S}_{\mu\nu} ={}& \mass g^{\alpha'}_{\mu}g^{\beta'}_{\nu}\biggl\{\frac{2}{\lambda\scov}L_{\alpha'\beta'} + \frac{\lambda}{3\scov^3}\Bigl[(\rcov^2-\scov^2)L_{\alpha'\beta'}R_{u\sigma u\sigma} \nonumber \\
        & - 12\scov^{4}R_{\alpha'u\beta'u} - 12\rcov\scov^2u_{(\alpha'}R_{\beta')u\sigma u}\bigr] \nonumber \\
        & + \frac{\lambda^2}{12\scov^3}\Bigl[24\scov^4(R_{\alpha'u\beta'u;\sigma} - \rcov\dot{R}_{\alpha'u\beta'u}) \nonumber \\
        & + L_{\alpha'\beta'}\Bigl\{\rcov(\rcov^2-3\scov^2)\dot{R}_{u\sigma u\sigma}+(\scov^2-\rcov^2)R_{u\sigma u\sigma;\sigma}\Bigr\} \nonumber \\
        & - 16\scov^2(\rcov^2+\scov^2)u_{(\alpha'}\dot{R}_{\beta')u\sigma u} \nonumber \\
        & + 16\rcov\scov^2u_{(\alpha'}R_{\beta')u\sigma u;\sigma}\Bigr]\biggr\} + \order(\lambda^3), \label{eq:hS1cov}
\end{align}
where $\lambda$ counts powers of distance from the worldline, primed indices denote tensors at $x_p^\mu$, coordinates on the worldline $\gamma$, and we use the notation $A_{u\sigma} \coloneqq A_{\alpha'\beta'}u^{\alpha'}\sigma^{\beta'}$ and $A_{\alpha'u;\sigma} \coloneqq A_{\alpha'\mu';\nu'}u^{\mu'}\sigma^{\nu'}$.
In the above expression, $g^{\mu'}_{\alpha}$ is the parallel propagator connecting $x_p^\mu$ and $x^\mu$, and $\sigma(x,x_p)$ is Synge's world function, which gives half of the geodesic distance squared between $x_p^\mu$ and $x^\mu$, with $\sigma_{\mu'}\coloneqq\nabla_{\mu'}\sigma$ being its derivative; see Refs.~\cite{Synge1960,Poisson:2011nh} for more details.
For compactness, we define the tensor
\begin{equation}
    L_{\mu'\nu'} \coloneqq g_{\mu'\nu'} + 2u_{\mu'}u_{\nu'},
\end{equation}
and we additionally define the scalar quantities
\begin{align}
    \scov(x,x_p) \coloneqq{}& \sqrt{P_{\mu'\nu'}\sigma^{\mu'}\sigma^{\nu'}}, \\
    \rcov(x,x_p) \coloneqq{}& u_{\mu'}\sigma^{\mu'},
\end{align}
where $P_{\mu'\nu'}\coloneqq g_{\mu'\nu'} + u_{\mu'}u_{\nu'}$ is the projection operator.
The scalar quantities above give a notion of the proper spatial and temporal distance between $x_p$ and $x$, respectively.

For our current work, we use a second-order-in-distance puncture and neglect terms of order $\lambda^3$ and higher.
That is, we define the puncture as the terms explicitly given in Eq.~\eqref{eq:hS1cov} so that
\begin{equation}
    h^{\S}_{\mu\nu} = h^{\calP}_{\mu\nu} + \order(\lambda^3).
\end{equation}
This results in an effective source that satisfies the Einstein field equations and the Lorenz-gauge condition through order $\lambda^0$/$\lambda^1$, respectively, in four dimensions.

\subsection{Comoving coordinates}
\label{sec:punctures/co-moving-coordinates}
The puncture is a local expansion of the singular field about the particle's
worldline $\gamma$. Both its coordinate construction and subsequent decomposition into
$m$-modes are greatly simplified if it is expressed in local coordinates
$(T,X,Y,Z)$ adapted to the particle. In Ref.~\cite{Bourg:2024cgh}
such coordinates were constructed for a circular orbit in Schwarzschild
spacetime; here we generalize them to circular equatorial orbits in Kerr.

The coordinates are required to satisfy three conditions: first, the particle sits at the
spatial origin, $X=Y=Z=0$, for all $T$; second, the metric is Minkowskian on the
worldline, $g_{AB}|_\gamma=\eta_{AB}=\mathrm{diag}(-1,1,1,1)$; and third, $\partial_T$
is everywhere tangent to the helical Killing vector
$k \coloneqq \partial_t+\Omega\,\partial_\varphi$, and coincides with the four-velocity
$u^\alpha$ on $\gamma$. In those coordinates, any field that is stationary in the frame co-rotating with the particle, such as the singular field, or the particle's
stress--energy tensor, has $T$-independent components and is therefore a
function of the $X,Y,Z$ alone. Note that the $(T,X,Y,Z)$ are not normal coordinates; for instance, $\Gamma^X_{TZ}|_\gamma=-\Gamma^Z_{TX}|_\gamma=-\sqrt{M/r_p^3}$, reflecting the fact that the spatial axes co-rotate with the orbit rather than being Fermi--Walker transported.

Following Ref.~\cite{Bourg:2024cgh}, we first choose an orthonormal tetrad on
$\gamma$ whose timelike leg is the four-velocity.
Defining $w \coloneqq \Lz \,\partial_t+\En\,\partial_\varphi$, which is orthogonal to
$u^\alpha$ on $\gamma$, the tetrad reads
\begin{subequations}\label{eq:CMtetrad}
\begin{align}
    e_{0} ={}& u^t k, &
    e_{1} ={}& \frac{\sqrt{\Delta_p}}{r_p}\,\partial_r, \\
    e_{2} ={}& \frac{1}{r_p}\,\partial_\theta, &
    e_{3} ={}& \frac{w}{\sqrt{\Delta_p}},
\end{align}
\end{subequations}
where $\Delta_p\coloneqq\Delta(r_p)$. Consistently, for $a=0$, Eq.~\eqref{eq:CMtetrad} reduces to
Eqs.~(47)--(50) of Ref.~\cite{Bourg:2024cgh}, where the legs are instead labeled
$e_1,\dots,e_4$ and the azimuthal leg is $e_4=-e_3$.

\begin{figure}
\centering
\includegraphics[scale=1.2]{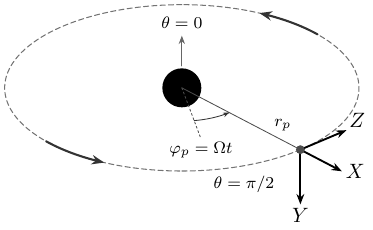}
\caption{The comoving coordinates $(X,Y,Z)$ centered on the particle. In
the black-hole frame, the particle is on the equator ($\theta=\pi/2$) at
constant orbital radius $r_p$ and azimuthal angle $\varphi_p=\Omega t$. $X$
points radially outwards, $Y$ towards increasing $\theta$, and $Z$ along the direction of
motion, so that $(X,Y,Z)$ is right-handed.}
\label{fig:comoving_axes}
\end{figure}

We then demand that the coordinate basis coincide with the tetrad on the
worldline, $\partial x^\mu/\partial X^A|_\gamma=e_A^{\ \mu}$, where
$x^\mu=(t,r,\theta,\varphi)$ are Boyer--Lindquist coordinates. This condition
only constrains the transformation on $\gamma$, and admits infinitely many
solutions. We choose
\begin{subequations}\label{eq:CMcoords}
\begin{align}
    T ={}& \En\,t - \Lz \,\varphi, \\
    X ={}& \frac{r_p\,(r-r_p)}{\sqrt{\Delta_p}}, \\
    Y ={}& {-}r_p\cos\theta, \\
    Z ={}& z_c \sin\!\left(\frac{\hat{\varphi}}{2}\right),
\end{align}
\end{subequations}
where $\phihat\coloneqq\varphi-\varphi_p(t)$ is the co-rotating azimuthal angle,
which vanishes at the particle, and
\begin{equation}\label{eq:zc}
    z_c \coloneqq 2u^t\sqrt{\Delta_p};
\end{equation}
see Fig.~\ref{fig:comoving_axes}.

As for the tetrad, Eqs.~\eqref{eq:CMcoords} and~\eqref{eq:zc} reduce to Eqs.~(57)--(61) of
Ref.~\cite{Bourg:2024cgh} except for the sign of $z_c$, which is negative
there. As a result, the spatial coordinates of Ref.~\cite{Bourg:2024cgh} are
left-handed, whereas ours are right-handed. The two are related by $Z\to-Z$.

\begin{figure*}
\centering
\includegraphics{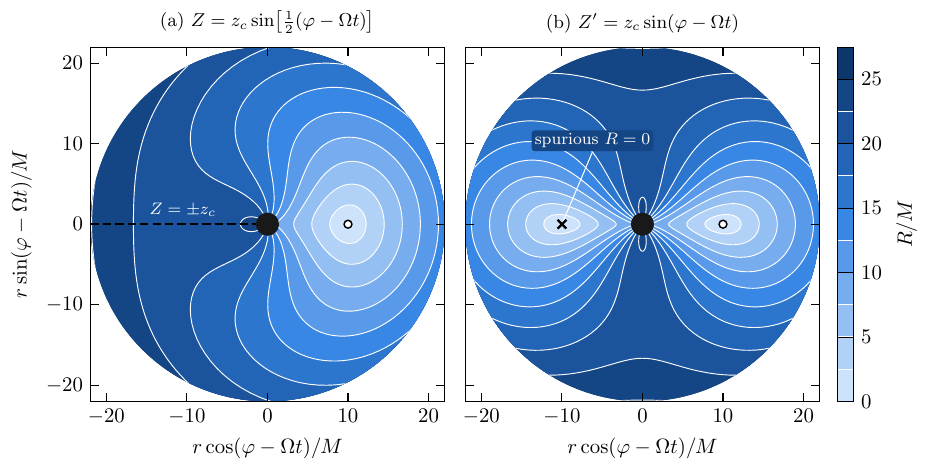}
\caption{The comoving radius $R$ of Eq.~\eqref{eq:CMradius} in the equatorial
plane ($Y=0$), viewed from above in the frame co-rotating with the particle,
for $a=0.9M$ and $r_p=10M$. A point is located by its Boyer--Lindquist radius
$r$ and its co-rotating angle $\varphi-\Omega t$. White lines are contours of
constant $R$, spaced by $\Delta R=2.5M$. The black disc is the central black hole, and the open
circle marks the particle. (a)~With $Z=z_c\sin[(\varphi-\Omega t)/2]$
[Eq.~\eqref{eq:CMcoords}], $R$ vanishes only at the particle; the dashed
half-line $\varphi-\Omega t=\pi$ is the coordinate singularity $Z=\pm z_c$.
(b)~With $Z'=z_c\sin(\varphi-\Omega t)$, $R$ also vanishes at the antipodal
point (cross), which would produce a spurious singularity in the puncture.}
\label{fig:comoving_R}
\end{figure*}

The comoving radius
\begin{equation}\label{eq:CMradius}
    R^2 \coloneqq X^2 + Y^2 + Z^2
\end{equation}
vanishes only at the particle $(r,\theta,\hat{\varphi}) = (r_p, \pi/2, 0)$; this is the purpose of the factor of $1/2$ in
Eq.~\eqref{eq:CMcoords}. Had we instead used $Z' \coloneqq z_c\sin\hat{\varphi}$, $R$ would
also vanish at the antipodal point $(r_p,\pi/2,\pi)$, introducing a
spurious singularity in the puncture there~\cite{Bourg:2024cgh}; see
Fig.~\ref{fig:comoving_R}.

Computing $\partial x^\mu/\partial X^A$ from Eq.~\eqref{eq:CMcoords}, the
coordinate basis vectors are, at every point, $\partial_T=u^t k$,
$\partial_X=e_1$, $\partial_Y=e_2/\sin\theta$ and
$\partial_Z=\sec(\hat{\varphi}/2)\,w/\sqrt{\Delta_p}$, with the constant
Boyer--Lindquist components of Eq.~\eqref{eq:CMtetrad}. The non-vanishing
components of the metric in comoving coordinates are therefore
\begin{subequations}\label{eq:CMmetric}
\begin{align}
    g_{TT} ={}& (u^t)^2\,k^\mu k_\mu, \\
    g_{TZ} ={}& \frac{u^t\sec(\hat{\varphi}/2)}{\sqrt{\Delta_p}}\,k^\mu w_\mu, \\
    g_{XX} ={}& \frac{\Delta_p}{r_p^2}\,\frac{\Sigma}{\Delta}, \\
    g_{YY} ={}& \frac{\Sigma}{r_p^2-Y^2}, \\
    g_{ZZ} ={}& \frac{\sec^2(\hat{\varphi}/2)}{\Delta_p}\,w^\mu w_\mu,
\end{align}
\end{subequations}
where, explicitly,
\begin{subequations}\label{eq:CMnorms}
\begin{align}
    k^\mu k_\mu ={}& {-}1+\Omega^2(r^2+a^2)\sin^2\theta \nonumber\\
        & + \frac{2Mr}{\Sigma}\bigl(1-a\Omega\sin^2\theta\bigr)^2, \\
    k^\mu w_\mu ={}& {-}\Lz + \Omega \En\,(r^2+a^2)\sin^2\theta \nonumber\\
        & + \frac{2Mr}{\Sigma}\bigl(1-a\Omega\sin^2\theta\bigr)
            \bigl(\Lz-a\En\sin^2\theta\bigr), \\
    w^\mu w_\mu ={}& {-}\Lz^2 + \En^2(r^2+a^2)\sin^2\theta \nonumber\\
        & + \frac{2Mr}{\Sigma}\bigl(\Lz-a\En\sin^2\theta\bigr)^2.
\end{align}
\end{subequations}
In particular, the metric is independent of $T$, reduces to $\eta_{AB}$ on
$\gamma$, and depends on $Z$ only through the factor $\sec(\hat{\varphi}/2)$,
which diverges on the half-plane $\hat{\varphi}=\pi$ antipodal to the particle,
where $Z=\pm z_c$.

Far from the particle, the comoving coordinates break down.
Since $g_{TT}=(u^t)^2 k_\mu k^\mu$, $\partial_T$ is timelike only inside the light cylinder $r\sin\theta \lesssim 1/\Omega$ and outside an inner light surface close to the horizon.
Moreover, since $\partial_X$ and $\partial_Y$ are spacelike and orthogonal to $\partial_Z$, the hypersurfaces of constant $T$ are spacelike if and only if $g_{ZZ}>0$, i.e.\ only for $r\sin\theta\gtrsim \Lz/\En$.
Both conditions therefore hold only in a shell around the orbit.
For example, for $a=0.9M$ and $r_p=10M$, it extends from $r\approx2.6M$ to $r\approx31.5M$ in the equatorial plane.

\subsection{Coordinate expressions}
\label{sec:punctures/coordinate-expressions}

To write the puncture in comoving coordinates, we expand covariant quantities in terms of coordinate differences.
For more details on computing such expansions, see e.g.\ Refs.~\cite{Ottewill:2008uu,Heffernan:2012su,Heffernan:2022zgc,Upton:2023tcv}.
The coordinate difference is naturally given in comoving coordinates by
\begin{equation}
    \Delta x^{\mu'} = (0,X,Y,Z),
\end{equation}
so that
\begin{align}
    g^{\mu'}_{\alpha} ={}& \delta_{\alpha'}^{\mu'} + \sum_{n=1}^{\infty}\lambda^ng^{(n)\mu'}_{\alpha'}{}_{\beta'_1\ldots\beta'_n}\Delta x^{\beta'_1}\cdots\Delta x^{\beta'_n}, \label{eq:pp_expansion} \\
    \sigma_{\mu'} ={}& {-}\lambda\Delta x_{\mu'} + \sum_{n=2}^\infty \lambda^n\sigma^{(n)}_{\mu'\alpha'_1\ldots\alpha'_n}\Delta x^{\alpha'_1}\cdots\Delta x^{\alpha'_n}. \label{eq:sig_expansion}
\end{align}
The coefficients are functions of the metric, Christoffel symbols, Riemann tensor, and their derivatives, all evaluated on the worldline.
The form of these quantities is dramatically simplified through the adoption of the comoving coordinates, as many components are identically zero.
As mentioned previously, an additional benefit of the comoving coordinates is that it significantly reduces the complexity of the structure of the puncture.
To see how, we note that
\begin{align}
    \rcov ={}& \sigma_{\mu'}u^{\mu'} \nonumber \\
        ={}& {-}\lambda\Delta x_{\mu'}u^{\mu'} + \sum_{n=2}^\infty \lambda^n u^{\mu'}\sigma^{(n)}_{\mu'\alpha'_1\ldots\alpha'_n}\Delta x^{\alpha'_1}\cdots\Delta x^{\alpha'_n}.
\end{align}
By construction, the leading term is identically zero in comoving coordinates.
Inspecting Eq.~\eqref{eq:hS1cov}, we see that this immediately simplifies the coordinate form of the puncture, as it has eliminated the leading-order-in-distance term in $\rcov$.
Finally, we have that
\begin{align}
    \scov ={}& \sqrt{P_{\mu'\nu'}\Delta x^{\mu'}\Delta x^{\nu'}} + \order(\lambda^2) \nonumber \\
        ={}& R + \order(\lambda^2), \label{eq:scovExp}
\end{align}
where $P_{\mu'\nu'}=\diag(0,1,1,1)$ in comoving coordinates and $R$ is defined in Eq.~\eqref{eq:CMradius}.

After substituting in the coordinate expansions and truncating at the appropriate order in $\lambda$, the puncture takes the form
\begin{equation}
    h^{\calP}_{\mu\nu} = \sum_{ijkn} c_{\mu\nu}^{ijkn} X^i Y^j Z^k R^n, \label{eq:hP1sumZ}
\end{equation}
where $n$ is an odd integer, $c_{\mu\nu}^{ijkn}$ is a constant function of $M$, $a$ and $r_p$, and the bounds of the sum are such that, due to the order of puncture being used, $-1\leq i+j+k+n\leq 2$ (with some coefficients identically zero depending on the component).
To further simplify the expression, we rewrite powers of $Z$ in terms of $R$, $X$ and $Y$ such that $k$ only takes the values $0$ or $1$.
As an example, at leading order, the puncture then takes the simple form
\begin{equation}
    h^{\calP}_{\mu\nu} \sim \frac{2}{R}\diag(1,1,1,1).
\end{equation}

The $(\ell,m)$-mode reconstruction from Sec.~\ref{sec:perturbations/reconstruction-in-lm-modes} utilises the Newman--Penrose formalism.
We therefore project the puncture field onto the Kinnersley tetrad by contracting with the relevant tetrad legs, given in comoving coordinates in Eq.~\eqref{eq:NPCM}.
The resulting expressions can then be rescaled using Eq.~\eqref{eq:ddkwlegs}\footnote{This is done after the $m$-mode decomposition is performed, as there is no $\phihat$ dependence in the rescalings.}.
Following this, we perform the same procedure as before: truncating at order $\lambda^2$ and reducing powers of $Z$.

To prepare for decomposition into $m$ modes and to make the integration analytically tractable, we expand
\begin{equation}
    Z = \frac{Z'}{2} + \lambda^2\frac{Z'^3}{16z_c^2} + \order(\lambda^4),
\end{equation}
where we recall $Z'=z_c\sin(\phihat)$, and repeatedly apply
\begin{equation}\label{eq:ZpReduce}
    Z'^n = \begin{cases}\begin{aligned}[b]
        \MoveEqLeft[1] 2^n z_{c}^{-n}(R^2-X^2-Y^2)^{n/2} \\
            & \times (z_c^2-R^2+X^2+Y^2)^{n/2},
        \end{aligned} & n \text{ even}, \\
        \begin{aligned}[b]
        \MoveEqLeft[1] 2^{n-1} z_{c}^{1-n}Z'(R^2-X^2-Y^2)^{(n-1)/2} \\
            & \times (z_c^2-R^2+X^2+Y^2)^{(n-1)/2},
        \end{aligned} & n \text{ odd}.
    \end{cases}
\end{equation}
Note that this substitution is only done to explicit powers of $Z$ that appear in $h^{\calP}_{ab}$, we do \emph{not} expand the $Z$ that appear in $R$.
With the substitutions, re-expansions and replacements completed, we have our final expressions for the tetrad legs of the puncture field which are ready to be decomposed into $m$ modes.
Similarly to Eq.~\eqref{eq:hP1sumZ}, we write the puncture as
\begin{equation}
    h^{\calP}_{ab} = \sum_{ijkn} c_{ab}^{ijkn} X^i Y^j (Z')^k R^n, \label{eq:hP1sum}
\end{equation}
where, again, $k\in\{0,1\}$, $n$ is an odd integer, $-1\leq i+j+k+n\leq 2$ and $c_{ab}^{ijkn}$ is a function of $M$, $a$ and $r_p$.

\subsection{Analytical decomposition into \texorpdfstring{$m$}{m}-modes}
\label{sec:punctures/m-modes}

Because the orbit is circular, the puncture depends on $t$ and $\varphi$ only through $\phihat = \varphi-\varphi_p(t)$.
We decompose it as in \eqn{mmode-def},
\begin{equation}
    h^{\calP}_{ab} = \sum_{m=-\infty}^{\infty} h^{\calP,m}_{ab}(r,\theta)\,e^{im\phihat},
\end{equation}
with
\begin{align}
    h^{\calP,m}_{ab} ={}& \frac{1}{2\pi}\int_{-\pi}^{\pi}h^{\calP}_{ab}\, e^{-im\phihat}d\phihat \nonumber\\
        ={}& \sum_{ijkn}c_{ab}^{ijkn}X^i Y^j\,\mathcal{J}_{k,n}(m), \label{eq:hP1mint}
\end{align}
where
\begin{equation}
    \mathcal{J}_{k,n}(m) \coloneqq \frac{1}{2\pi}\int_{-\pi}^{\pi}(Z')^k R^n e^{-im\phihat}d\phihat. \label{eq:Jdef}
\end{equation}
The coefficients and $X^iY^j$ come out of the integral because, at fixed $(r,\theta)$, $X$ and $Y$, and hence $\varrho\coloneqq\sqrt{X^2+Y^2}$, do not depend on $\phihat$.
All of the $\phihat$-dependence is in
\begin{equation}
    R^2 = \varrho^2 + z_c^2\sin^2(\phihat/2),
\end{equation}
As we have used Eq.~\eqref{eq:ZpReduce} to reduce powers of $Z'$, the $m$-modes of the puncture involve only $k=0$ and $k=1$.
We first give $\mathcal{J}_{0,n}$ in closed form and reduce $k\ge1$ to it, and then discuss the behavior of the $m$-modes near the particle and their convergence in $m$.

The integral with $k=0$ was previously derived in Eq.~(115) of Ref.~\cite{Bourg:2024cgh} using Eq.~(3.664.3) of Ref.~\cite{Grad14}:
\begin{equation}
    \mathcal{J}_{0,n}(m) = \frac{(-1)^m(\varrho\calR)^{n/2}}{(1+n/2)_{m}}\,P^m_{n/2}(z), \label{eq:J0}
\end{equation}
where
\begin{equation}
    \calR^2\coloneqq\varrho^2+z_c^2, \qquad z\coloneqq\frac{\varrho^2+z_c^2/2}{\varrho\calR}. \label{eq:R2exp}
\end{equation}
Here $(a)_k=\Gamma(a+k)/\Gamma(a)$ is the Pochhammer symbol and $P^\mu_\nu(x)$ is the Legendre function of the first kind on $x>1$~\cite{DLMF} (\texttt{LegendreP[$\nu$,$\mu$,3,$x$]} in \texttt{Mathematica}).
Since $z-1=(\calR-\varrho)^2/(2\varrho\calR)>0$, the argument always lies in that range.

Equation~\eqref{eq:J0} holds for all integer $m$.
When $n\le-2$ is even, however, two things happen once $|m|\ge-n/2$:
\begin{itemize}
    \item the Pochhammer symbol $(1+n/2)_{|m|}=(1+\tfrac{n}{2})(2+\tfrac{n}{2})\cdots(|m|+\tfrac{n}{2})$ contains a zero factor;
    \item $P^{|m|}_{n/2}=P^{|m|}_{-n/2-1}$ vanishes, because its order exceeds its degree.
\end{itemize}
The right-hand side of Eq.~\eqref{eq:J0} is then $0/0$, although the integral is finite.
Its value can be recovered as the limit $\nu\to n/2$ of $P^{|m|}_\nu/(1+\nu)_{|m|}$.
By l'H\^opital's rule, however, this requires the derivative of $P^{|m|}_\nu$ with respect to its degree, which is slow to evaluate.
A simpler route uses a symmetry of the integral~\eqref{eq:Jdef}: with $k=0$ it is even in $m$ (send $\phihat\to-\phihat$), so $\mathcal{J}_{0,n}(|m|)=\mathcal{J}_{0,n}(-|m|)$.
Equation~\eqref{eq:J0} can therefore be evaluated at $-|m|$ instead, where neither $(1+n/2)_{-|m|}$ nor $P^{-|m|}_{n/2}$ vanishes.
In practice, we use $+|m|$ for $n\ge-1$ and $-|m|$ for $n\le-2$.

Integrals with $k\ge1$ reduce to the case $k=0$.
Writing $Z'=z_c(e^{i\phihat}-e^{-i\phihat})/(2i)$ shifts the mode number:
\begin{equation}
    \mathcal{J}_{k,n}(m) = \frac{iz_c}{2}\bigl[\mathcal{J}_{k-1,n}(m+1)-\mathcal{J}_{k-1,n}(m-1)\bigr]. \label{eq:Jk}
\end{equation}
Derivatives of the puncture, whose $m$-modes we also need, contain higher powers of $Z'$.
For these, Eq.~\eqref{eq:ZpReduce}, $Z'^2=4(R^2-\varrho^2)(z_c^2+\varrho^2-R^2)/z_c^2$, lowers $k$ by two at the cost of shifting $n$:
\begin{align}
    \mathcal{J}_{k,n}(m) ={}& \frac{4}{z_c^2}\Bigl[(z_c^2+2\varrho^2)\,\mathcal{J}_{k-2,n+2}(m) - \mathcal{J}_{k-2,n+4}(m) \nonumber\\
        & - \varrho^2\mathcal{R}^2\,\mathcal{J}_{k-2,n}(m)\Bigr]. \label{eq:Jk2}
\end{align}
For even $k$, we apply Eq.~\eqref{eq:Jk2} repeatedly.
For odd $k$, we also apply Eq.~\eqref{eq:Jk} once.

Equations~\eqref{eq:hP1mint}, \eqref{eq:J0}, \eqref{eq:Jk} and~\eqref{eq:Jk2} thus give the $m$-modes of the puncture and its derivatives exactly.
For all ten tetrad components of the puncture at one $(r,\theta,m)$, this takes on the order of milliseconds in \texttt{Mathematica} at 30-digit precision.

The closed form~\eqref{eq:J0} also shows how the $m$-modes behave near the particle.
They are singular only at the particle, $\varrho=0$, where $z\simeq z_c/(2\varrho)\to\infty$.
For $n=-1$, the large-$z$ behavior of $P^m_{-1/2}(z)$~(Sec.~14.8(iii) of Ref.~\cite{DLMF}) gives
\begin{equation}
    \mathcal{J}_{0,-1}(m) = -\frac{2}{\pi z_c}\ln\varrho + \mathcal{O}(\varrho^0). \label{eq:Jlog}
\end{equation}
Note that the coefficient of the logarithm does not depend on $m$. This is most easily understood from the integral representation Eq.~\eqref{eq:Jdef}, where the peak of the integral occurs where $|\phihat| \approx 0$.

This term dominates the $m$-modes of the puncture, because $c_{ab}^{000,-1}R^{-1}$ is its only term of order $\lambda^{-1}$.
Every other term $X^iY^j(Z')^kR^n$ has $p\coloneqq i+j+k+n\ge0$.
Since $|X|,|Y|\le R$ and $|Z'|\le2R$, such a term is bounded by $2^kR^p$, so its $m$-modes remain bounded as $\varrho\to0$.
Hence
\begin{equation}
    h^{\calP,m}_{ab} = -\frac{2}{\pi z_c}\,c_{ab}^{000,-1}\ln\varrho + C_{ab}^m + \mathcal{O}(\varrho\ln\varrho). \label{eq:logdiv}
\end{equation}
The constant $C_{ab}^m$ depends on $m$ but not on the direction of approach.

Away from the particle, $\varrho>0$, the $m$-modes are finite and decay exponentially in $|m|$, at a rate set by the distance to the orbit.
This follows from a standard property of Fourier coefficients:
If a $2\pi$-periodic function is analytic in the strip $|\mathrm{Im}\,\phihat|<\eta$ of the complex $\phihat$-plane, its Fourier coefficients fall off as $e^{-\eta|m|}$, up to powers of $|m|$.
Here $R^n$ can fail to be analytic only where $R$ vanishes.
Since
\begin{equation}
    R^2=\varrho^2+\tfrac{1}{2}z_c^2\,(1-\cos\phihat),
\end{equation}
this happens at $\cos\phihat=1+2\varrho^2/z_c^2>1$, that is, at the complex points $\phihat=\pm i\eta$ with
\begin{equation}
    \cosh\eta = 1+\frac{2\varrho^2}{z_c^2}. \label{eq:eta}
\end{equation}
These points are the particle's singularity continued to complex $\phihat$, and they approach the real axis as $\varrho\to0$.
The $m$-modes therefore decay as $e^{-\eta |m|}$, with
\begin{equation}
    e^{-\eta}=\frac{\calR-\varrho}{\calR+\varrho} = \sqrt{\frac{z-1}{z+1}}. \label{eq:alpha}
\end{equation}
By Eq.~\eqref{eq:Jk}, the same $\eta$ governs $\mathcal{J}_{k,n}$, and hence the puncture.
For $\varrho\ll z_c$, Eq.~\eqref{eq:eta} gives $\eta\simeq2\varrho/z_c$.
Reaching a relative accuracy $\epsilon$ therefore takes about $\ln(1/\epsilon)/\eta\simeq(z_c/2\varrho)\ln(1/\epsilon)$ modes, a number that grows as the field point approaches the orbit.

Finally, to check that the puncture is correct, we have verified that it satisfies the field equations and the gauge condition to the appropriate order in distance from the particle.
For a puncture through order $\lambda^2$, where terms of order $\lambda^3$ and higher are omitted, the $m$-modes should satisfy
\begin{equation}
    E^m_{ab}[\barh^{\calP}] = \text{smooth terms} + \mathcal{O}(\lambda^2),
\end{equation}
and the gauge condition to act as
\begin{equation}
    Z^m_{a}[\barh^{\calP}] = \text{smooth terms} + \mathcal{O}(\lambda^3),
\end{equation}
where $\lambda$ counts powers of $\varrho$ and the omitted terms may include $\ln\lambda$.
The omitted terms are functions of $X$ and $Y$ that are not smooth at $\varrho=0$; for example, they depend on the direction of approach or contain $\ln\varrho$.

The smooth terms appear because each $m$-mode is an integral over the full range
of $\hat\varphi$ at fixed $(r,\theta)$, and hence also over regions far from the
particle. Being smooth at $\varrho=0$, they do not affect the regularity of the
residual field. At the particle, however, their sum over $m$ must vanish since
$E_{ab}[\bar h^{\calP}]=\mathcal{O}(\lambda)$ and
$Z_{a}[\bar h^{\calP}]=\mathcal{O}(\lambda^2)$ on $\gamma$. Note the $m$-modes are one order more regular than the corresponding 4D quantities, because the integration over $\hat\varphi$ gains one order, as in Eq.~\eqref{eq:logdiv}.
Explicitly, near $\gamma$ and for all components,
\begin{align}
    E^m_{ab}[\bar h^{\calP}] \sim{}&
        \frac{e_{ab}^{(0,0)} + m\,e_{ab}^{(0,1)}
              + m^2\,e_{ab}^{(0,2)}}{(4m^2-1)(4m^2-9)}
        + \mathcal{O}(\lambda), \\
    Z^m_{a}[\bar h^{\calP}] \sim{}&
        \frac{f_{a}^{(0,0)} + m\,f_{a}^{(0,1)}}{(4m^2-1)(4m^2-9)}
        \nonumber \\
    & + \lambda\,\frac{f_{a}^{(1,0)} + m\,f_{a}^{(1,1)}
              + m^2\,f_{a}^{(1,2)}}{(4m^2-1)(4m^2-9)} + \mathcal{O}(\lambda^2),
\end{align}
where the coefficients $e^{(i,j)}_{ab}$ and $f^{(i,j)}_{a}$ depend only on $M$, $r_p$ and $a$.
The displayed terms vanish when summed over $m\in\mathbb{Z}$, whatever the values of the coefficients: $\sum_{m\in\mathbb{Z}} m^{k}/[(4m^2-1)(4m^2-9)]=0$ for $k=0,1,2$.
Note that even though the next order expressions do not have any directional discontinuities or divergences, they feature terms of the form $m^3/[(4m^2-1)(4m^2-9)]$ whose sum over $m$ is not absolutely convergent: the terms are odd in $m$, so the symmetric sum over $m\in\mathbb{Z}$ vanishes, but the sum over $m\geq0$ diverges logarithmically.
On the particle, for sufficiently large $m$, we see that the effective source $E^m_{ab}[h^{\calP}]\sim m^{-2}$, meaning that the residual field's $m$-modes fall off as $\sim m^{-4}$.
Therefore, the error of the sum truncated at $m$ falls off as $m^{-3}$, matching expected behavior for the order of the puncture we use~\cite{Barack:2007we,Dolan:2010mt}.

\subsection{Numerical decomposition into \texorpdfstring{$(\ell,m)$}{(ℓ,m)}-modes}
\label{sec:punctures/lm-modes}

\begin{figure}
    \centering
    \includegraphics[width=\columnwidth]{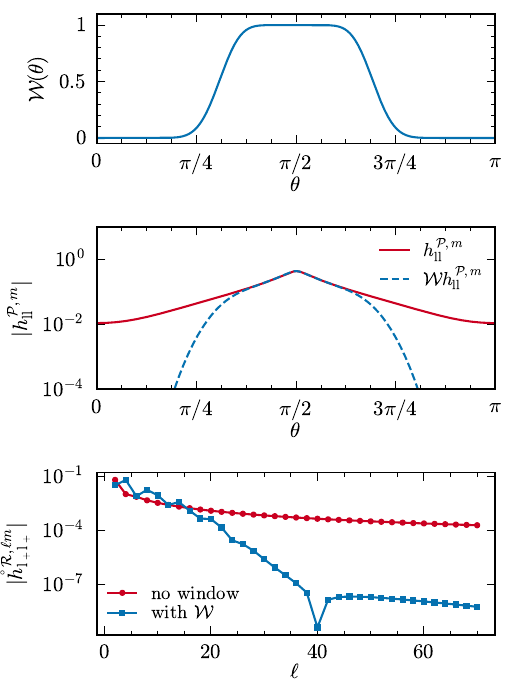}
    \caption{The polar window and its effect on the $(\ell,m)$-modes of the residual field, for $a=0.6M$, $\rp=8M$, $m=2$. 
    Top: the window function, \eqn{window_function}. 
    Middle: the $m=2$ mode of the $\ltetrad\ltetrad$ component of the puncture along $r=\rp+0.5M$, without (solid) and with (dashed) the window; the unwindowed puncture tends to a constant at the poles. 
    Bottom: the residual $(\ell,m)$-modes of the component $\hs{\ltetrad_+\ltetrad_+}$ on the same row, with the puncture projected without and with the window. 
    Without it, they decay only slowly; with it, they fall below $10^{-7}$ by $\ell\simeq40$. 
    Only $\ell-m$ even is shown, since the odd modes of this component vanish.}
    \label{fig:window_puncture}
\end{figure}

In order to appropriately regularize the retarded field $(\ell, m)$-modes, we next need to project the $m$-mode punctures down to spherical harmonic $(\ell, m)$-modes. The $m$-modes of the puncture are the azimuthal average of a local expansion about the worldline, which knows nothing about the poles; a regular field of spin-weight $s$ and azimuthal number $m$ vanishes at $\theta = 0$ and $\pi$ like $\sin^{|m\pm s|}\theta$, but the puncture $m$-modes tend to non-zero constants there. 
See \fig{window_puncture}, middle panel.
Their harmonic coefficients therefore fall off slowly in $\ell$ at every radius, not only near the particle, and a sum over $\ell \leq \ellmax$ of them rings over the whole sphere.
Subtracting those coefficients from the retarded field then leaves a residual that is biased by the omitted tail everywhere, including at the worldline as shown by the bottom panel in \fig{window_puncture}.

We therefore follow previous self-force calculations, e.g.\ Refs.~\cite{Vega:2009qb,Warburton:2013lea,Wardell:2015ada,Miller:2016hjv,Upton:2025bja}, and multiply the puncture by a window in the polar direction, ${\cal W}(\theta)$, equal to one at the equator and vanishing at the poles, and subtract ${\cal W} h^{\calP}$ in place of $h^{\calP}$.
Since ${\cal W} = 1$ on the worldline, the residual field still agrees with the regular field there, and since the windowed puncture vanishes at the poles like a regular field, its harmonic coefficients converge.
No window is applied in $r$: the puncture is only ever used inside a worldtube about the worldline, and outside of it the retarded modes are used.

We take our window function to be
\begin{equation}
    {\cal W}(\theta) = I_{\sin^{2}\theta}(10, 5),
    \label{eq:window_function}
\end{equation}
where $I_{x}$ is the \emph{regularised incomplete beta function}, defined as~\cite{DLMF}
\begin{equation}
    I_{x}(p, q) \coloneqq \frac{{\rm B}(x; p, q)}{{\rm B}(a, b)} = \frac{\Gamma(p + q)}{\Gamma(p)\Gamma(q)}\int^{x}_{0} t^{p - 1}(1 - t)^{q - 1}dt,
    \label{eq:incomplete_beta_function}
\end{equation}
where ${\rm B}(x; p, q) \rho_{0} \coloneqq \int^{x}_{0} t^{p - 1} (1 - t)^{q -1} dt$ is the \emph{incomplete} beta function\footnote{The incomplete beta function may also be written in terms of the hypergeometric function as ${\rm B}(x; p, q) = \frac{x^{p}}{p} {}_{2}F_{1}(p, 1 - q; p + 1; x)$.} and ${\rm B}(p, q) = \int^{1}_{0} t^{p - 1}(1 - t)^{q - 1}dt$ is simply the beta function.
Our window is therefore a polynomial of degree $14$ in $\sin^2\theta$, equal to one at the equator with $1-\calW=\order(\cos^{10}\theta)$ and vanishing like $\sin^{20}\theta$ at the poles, so a field multiplied by it has the pole behavior of a regular field whenever $|m \pm s|\leq 20$.
For larger $|m \pm s|$ it vanishes more slowly than a regular field, but still like $\sin^{20}\theta$, so its harmonic coefficients fall off as a high power of $\ell$.
The partial sums over $\ell$ of the residual at the worldline converge monotonically.

At each $r$, the puncture components are projected to spherical modes such that
\begin{equation}
    \hs{A}^{\calP,\lm}(r) = 2\pi\int_0^\pi \calW(\theta)\,\hs{A}^{\calP,m}(r,\theta)\,\overline{\sY{s_A}_{\lm}(\theta,0)}\sin\theta\,d\theta,
    \label{eq:puncture_lm}
\end{equation}
for every $\ell$ up to the retained $\ellmax$ and for $m$ up to the largest computed.
The same is done for the first and second derivatives of the puncture with respect to $r$ and $\theta$, which are needed for the derivatives of the residual: the $r$ derivatives are taken on the closed-form $m$-modes before the projection, and the $\theta$ derivatives act on $\calW\hs{A}^{\calP,m}$ by the product rule. 
Away from $r=\rp$, the integrand is smooth, and a Gauss-Legendre rule converges exponentially. On the radial slice $r=\rp$ it contains $\ln|\theta-\pi/2|$, on which Gauss--Legendre quadrature converges only as the inverse square of the number of nodes; that error is flat in $\ell$, so it survives the subtraction as a floor that looks like a residual that fails to decay. 
We use there a graded rule, Gauss panels that close geometrically on $\theta=\pi/2$, whose error falls as $4^{-n}$ in the number $n$ of grading levels.
The radial-derivative modes of the puncture on that row do not converge, since the puncture's $m$-mode is logarithmic there; they are not projected, and the residual's radial derivatives on the worldline are obtained as described in \secref{punctures/m-mode-retarded-and-residual}.

\subsection{\texorpdfstring{$m$}{m}-mode retarded and residual fields}
\label{sec:punctures/m-mode-retarded-and-residual}

Now, with the $(\ell,m)$-modes of the windowed puncture in hand, the subtraction is done before the sum. 
For each $m$ we form the residual $(\ell,m)$-modes
\begin{equation}
    \hs{A}^{\res,\lm}(r) \coloneqq \hs{A}^{\lm}(r)-\hs{A}^{\calP,\lm}(r),
    \label{eq:residual_lm}
\end{equation}
with the windowed puncture modes \eqref{eq:puncture_lm}, sum them over $\ell$, and, where the retarded field itself is needed, add back the windowed puncture in closed form,
\begin{equation}
    \hs{A}^{m}(r,\theta) = \sum_{\ell=\ell_{\min}}^{\ellmax}\hs{A}^{\res,\lm}(r)\,\sY{s_A}_{\lm}(\theta,0)+\calW(\theta)\,\hs{A}^{\calP,m}(r,\theta).
    \label{eq:resummed}
\end{equation}
The singular part is then supplied analytically and only the smooth residual is approximated by a finite sum of harmonics; derivatives in $\theta$ act on the harmonics and on $\calW\hs{A}^{\calP,m}$, and derivatives in $r$ on the coefficients.

We follow three rules in applying \eqn{resummed}. 
The sum runs to the retained $\ellmax$ and never into the modes computed only as padding (\secref{perturbations/solution}). 
The lower limit is $\ell_{\min}=\max(|m|,|s_A|)$, so that the $\ell=1$, $m=1$ modes of the spin-0 and spin-$\pm1$ components are kept. 
And the static sector stops at the $\ell$ to which its radial functions were generated, even when the radiative modes go further.

At the worldline $r=\rp$ itself, the $(\ell,m)$ projections of the puncture are finite but have a kink, like the retarded modes in \eqn{lm-kink}: the $m$-mode puncture diverges only logarithmically at the particle, \eqn{logdiv}, which is integrable over $\theta$, but its radial derivative changes sign across $r=\rp$.
The projected radial derivatives therefore jump at $r=\rp$.

We therefore never take radial derivatives at the node $r=\rp$.
The value on the worldline is computed directly. The radial derivatives there come from one polynomial of degree five fitted to the residual at the four nearest nodes on each side of $\rp$, so the two sides give the same worldline values by construction.

\begin{figure*}
    \centering
    \includegraphics[width=\textwidth]{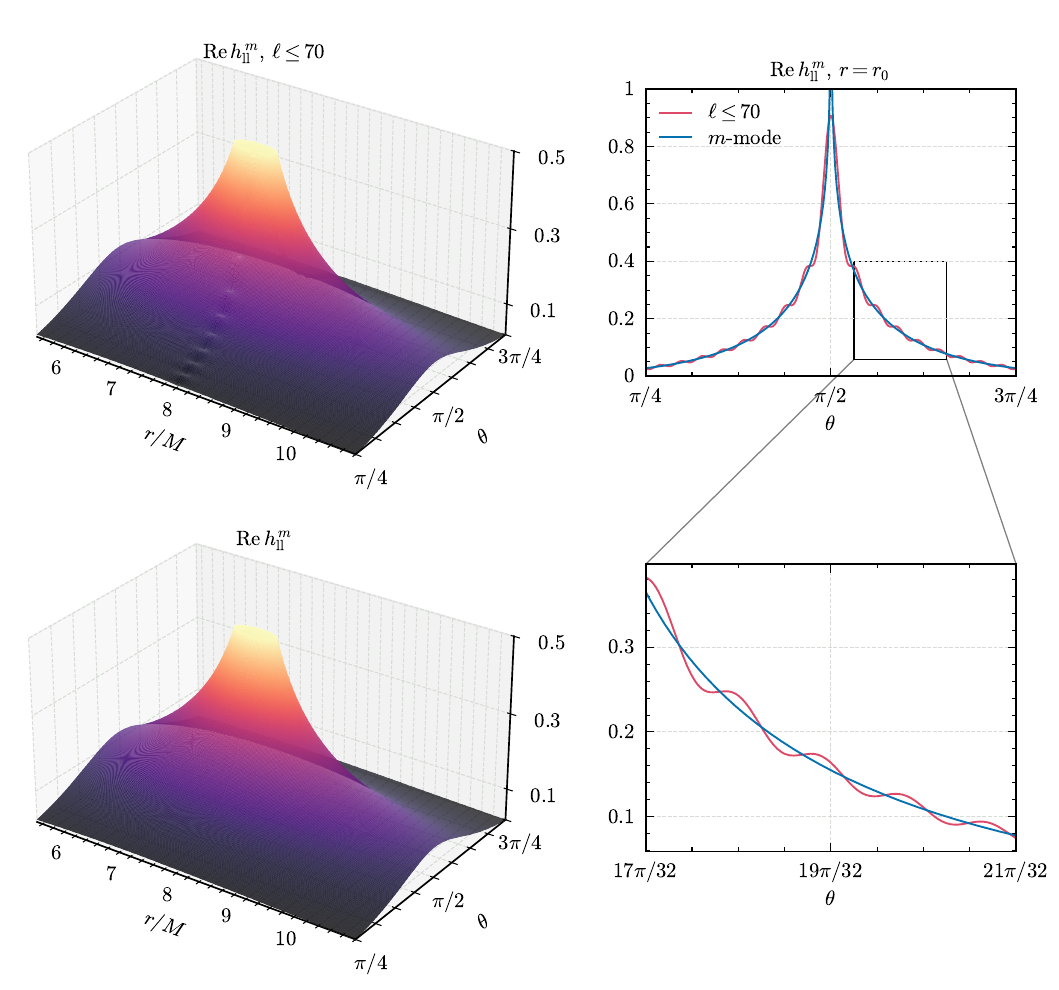}
    \caption{The $m=2$ mode of the retarded field, ${\rm Re}\,h^{m}_{\ltetrad\ltetrad}$, near the worldline for $a/M=0.6$, $r_0/M=8$.
    Top left: the $(\ell,m)$-modes summed directly up to $\ell=70$. 
    Each $(\ell,m)$-mode is finite but has a kink at $r=r_0$, so the truncated sum converges only algebraically near the worldline and shows Gibbs oscillations along $r=r_0$.
    Bottom left: the same $m$-mode with the windowed puncture $\calW h^{\calP}$ subtracted mode by mode in $\ell$ and added back in closed form; the oscillations are absent.
    The $m$-mode diverges logarithmically at the particle, so both surfaces are cut at ${\rm Re}\,h^{m}_{\ltetrad\ltetrad}=0.5$.
    Top right: a one-dimensional cut along $r=\rp$, comparing the direct sum (red), which oscillates, with the $m$-mode computed with the windowed puncture (blue), which does not. 
    Bottom right: an enlargement of the boxed region, where the $\ell\le70$ sum oscillates about the $m$-mode with a wavelength of order $\pi/\ellmax$.}
    \label{fig:gibbs_ringing}
\end{figure*}

\begin{figure*}
    \centering
    \includegraphics[width=\textwidth]{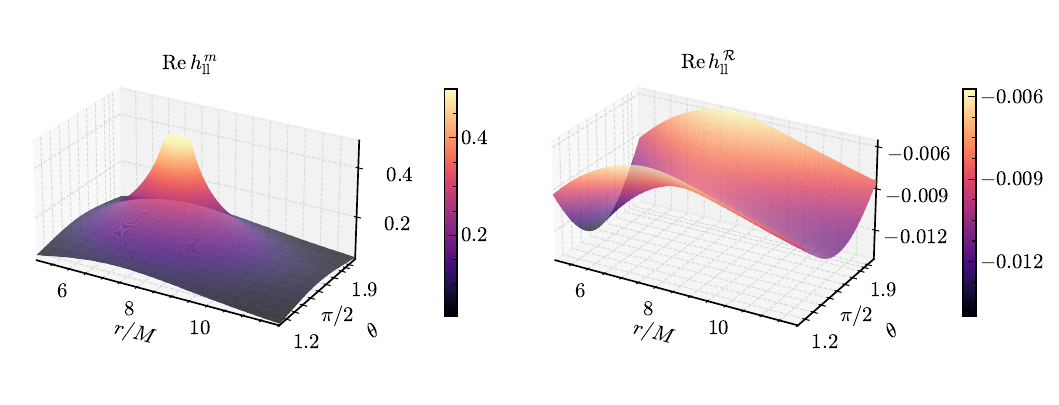}
    \caption{Surface plots of the real part of the $\ltetrad\ltetrad$ tetrad component of the $m = 2$ mode of the metric perturbation calculated with our reconstruction scheme for a particle at $\rp / M = 8.0$, with a background spin of $a/M = 0.6$.
    The retarded field, $h^{\rm ret}_{\ltetrad\ltetrad}$,  exhibits the expected logarithmic divergence towards the worldline, whereas the residual field, $h^{\cal R}_{\ltetrad\ltetrad}$, smoothly varies across the same domain.}
    \label{fig:ret_res_3d}
\end{figure*}

\section{Bondi--Sachs transformation}
\label{sec:Bondi--Sachs}

\subsection{Asymptotic flatness at second order}
\label{sec:Bondi--Sachs/asymptotic-flatness}

The first-order field of Sec.~\ref{sec:perturbations} is in the Lorenz gauge, which is well suited to first-order calculations: its singular structure at the particle is well understood, the punctures of Sec.~\ref{sec:punctures} are constructed in this gauge, and the field is regular everywhere else, falling off as $1/r$ at $\scri^{+}$.
At second order, however, calculations in most gauges, including the Lorenz gauge, run into problems near $\scri^{+}$~\cite{Pound:2015wva, Cunningham_2026, spiers2023second, Spiers:2026yqx}: the second-order Lorenz-gauge perturbation is not asymptotically flat.
Its static ($m=0$) modes do not decay at $\scri^{+}$: they generically tend to a constant, and the $\ell=2$ mode diverges logarithmically (Appendix~B of Ref.~\cite{Leather:2026zhl}).
Its radiative ($m\neq0$) modes contain spurious $(\ln r)/r$ terms, which prevent the waveform from being read off directly at $\scri^{+}$.
These features are gauge artifacts, but they also affect the field equations: for example, the retarded Green's-function integral for the second-order field can fail to converge~\cite{Pound:2015wva, spiers2023second, Leather:2026zhl}.
Such integrals can be made convergent by introducing a far-zone puncture that cancels the divergence~\cite{Pound:2015wva, Cunningham_2026, Bucciotti:2024zyp}.
Alternatively, we can avoid gauge divergences near $\scri^{+}$ by putting the first-order perturbation in an asymptotically flat gauge~\cite{spiers2023second, Spiers:2026yqx, Leather:2026zhl}, which is the approach we take.

We use the definition of second-order asymptotic flatness of Ref.~\cite{spiers2023second}.
For the source of the second-order Teukolsky equation~\cite{Campanelli:1998jv, spiers2023second, Green:2019nam} to be asymptotically flat, it suffices that the Kinnersley tetrad components of the first-order perturbation fall off, as $r\to\infty$ at fixed retarded time, as
\begin{align}
    h^{(1)}_{ab}=\TeukO(r^{-1}), \ h^{(1)}_{\ltetrad\ntetrad}=& \ \TeukO(r^{-2}), \ h^{(1)}_{\ltetrad\mtetrad}=\TeukO(r^{-2}),\notag \\ h^{(1)}_{\ltetrad\mb}=\TeukO(r^{-2}),& \
    h^{(1)}_{\mtetrad\mb}=\TeukO(r^{-2}),\notag \\ h^{(1)}_{\ltetrad\ltetrad}=\ &\TeukO(r^{-3}).\label{eq:asymptotically-flat-conditions}
\end{align}
The Lorenz-gauge perturbation falls off as $1/r$, but it does not satisfy all of these conditions.
Equations~\eqref{eq:asymptotically-flat-conditions} are satisfied by a class of gauges, including the Bondi--Sachs gauge~\cite{bondi1962gravitational, sachs1962BMS-fav, Spiers:2026yqx}, the CCK outgoing radiation gauge~\cite{cohen1975space, chrzanowski1975vector, kegeles1979constructive}, and the Newman--Unti (Bondi-like) gauge~\cite{newman1962behavior, barnich2013note, Hollands:2024iqp}.
To express which conditions the Lorenz gauge violates, we expand its $m$-modes at large $r$ in the outgoing Kerr coordinates of Sec.~\ref{sec:perturbations/background},
\begin{align}
    \label{eq:h1 large r expansion}
    h^{(1)}_{ab} ={}& \sum_m e^{-i\omega u+im\varphi_-}\Bigg[\frac{h^{(1,1)}_{ab,m}(\theta)}{r} 
    + \frac{h^{(1,2)}_{ab,m}(\theta)}{r^2} \Bigg] \nonumber \\
        & + \TeukO(1/r^3).
\end{align}
The second superscript counts powers of $1/r$, and $h^{(1,i)}_{ab,\lm}$ denotes the $(\ell,m)$-modes of $h^{(1,i)}_{ab,m}(\theta)$, as in \eqn{lmmode-def}.

We remove the violations of \cref{eq:asymptotically-flat-conditions} by transforming the first-order perturbation to the Bondi--Sachs gauge at large radius, with the scheme of Ref.~\cite{Spiers:2026yqx}, while remaining in the Lorenz gauge near the particle, where the punctures are used (Sec.~\ref{sec:Bondi--Sachs/transition}).
Since we truncate the transformation at a finite order in $1/r$, the Bondi--Sachs conditions hold only to that order, and we call the result the (near) Bondi--Sachs gauge; it satisfies \eqn{asymptotically-flat-conditions}, which is all we need.

Working in the (near) Bondi--Sachs gauge at first order is sufficient to obtain the waveform in an asymptotically flat gauge when solving the second-order Teukolsky equation.
Under a gauge transformation, the second-order Weyl scalar $\psi^{(2)}_{4}$ changes only through the first-order gauge vector $\xi^c_{(1)}$, via $\mathcal{L}_{\xi_{(1)}}\psi^{(1)}_4$~\cite{Campanelli:1998jv, spiers2023second}; it is invariant under second-order gauge transformations because $\psi_4$ vanishes on the Kerr background.
The second-order waveform is given by the $1/r$ part of $\psi^{(2)}_{4}$ at $\scri^{+}$, so it depends on the gauge only through the first-order gauge vector.
If we leave the first-order field in the Lorenz gauge, the resulting $\psi^{(2)}_{4}$ is not asymptotically flat, and the waveform cannot be read off directly at $\scri^{+}$.
If instead we transform the first-order metric perturbation to an asymptotically flat gauge, we can calculate $\psi^{(2)}_{4}$ and extract the waveform without a second-order gauge transformation.

\subsection{Gauge vector to the (near) Bondi--Sachs gauge}
\label{sec:Bondi--Sachs/gauge-vector}

Reference~\cite{Spiers:2026yqx} gives the gauge vector that takes a generic first-order perturbation with $h^{(1)}_{ab}=\mathcal{O}(\frac1r)$ to the (near) Bondi--Sachs gauge, in Kerr coordinates (their Eqs.~(C11)--(C13)).
Throughout this section, we denote this gauge vector to the (near) Bondi--Sachs gauge by $\xi$; it is distinct from the vector $\xi$ used in the Lorenz-gauge reconstruction of Sec.~\ref{sec:perturbations/radiative-pieces}.

Three simplifications reduce it to a few terms here.
\begin{enumerate}
    \item The Lorenz gauge already satisfies most of the conditions in \cref{eq:asymptotically-flat-conditions}, which simplifies the transformation.
    In the $m\neq0$ sector, the only non-vanishing coefficients that violate \cref{eq:asymptotically-flat-conditions} are $h_{\ltetrad\ltetrad,\lm}^{(1,2)}$ and $h_{\ltetrad\ntetrad,\lm}^{(1,1)}$, which satisfy $h_{\ltetrad\ntetrad,\lm}^{(1,1)}= \frac{ i\omega}{2}\,h_{\ltetrad\ltetrad,\lm}^{(1,2)}$.
    The four further coefficients that enter Eqs.~(C11)--(C13) at the orders we keep (item 3), $h_{\ltetrad\ltetrad,\lm}^{(1,1)}$, $h_{\ltetrad\mtetrad,\lm}^{(1,1)}$, $h_{\ltetrad\mb,\lm}^{(1,1)}$, and $h_{\mtetrad\mb,\lm}^{(1,1)}$, vanish in the Lorenz gauge for $m\neq 0$.
    For $m=0$, the Lorenz gauge satisfies \cref{eq:asymptotically-flat-conditions} except for the $\ell=0$ mode, where $h_{\ltetrad\ltetrad,00}^{(1,1)}$, $h_{\ltetrad\ltetrad,00}^{(1,2)}$, and $h_{\mtetrad\mb,00}^{(1,1)}$ are non-zero. 
    \item We do not fix the Bondi--Metzner--Sachs frame: we set the integration constants $\xi_\circ^u=\xi_\circ^{\hat \theta}=\xi_\circ^{\hat \phi}=0$ of Ref.~\cite{Spiers:2026yqx}.
    \item We use the minimal gauge transformation that satisfies~\cref{eq:asymptotically-flat-conditions}: we keep $\xi_\ltetrad$ up to and including $\frac{1}{r}$ terms, and $\xi_\ntetrad$, $\xi_\mtetrad$, and $\xi_\mb$ up to and including $r^0$ terms.
\end{enumerate}
To write the simplified Eqs.~(C11)--(C13) in modes, we decompose the gauge vector in spin-weighted spherical harmonics, as for the metric perturbation in \eqn{lmmode-def},
\begin{align}
    \xi_{a}=\sum_{\ell=|s|}^\infty \sum_{m=-\ell}^{\ell}\xi^{\lm}_{a}(r) \ {}_sY_{\lm}(\theta,\varphi_-)\,e^{-i\omega u},
\end{align}
where $s=0$ for $a\in\{\ltetrad,\ntetrad\}$, $s=1$ for $a=\mtetrad$ and $s=-1$ for $a=\mb$, and in $m$-modes,
\begin{align}
    \xi_{a}= \sum_{m=-\infty}^{\infty}\xi^{\m}_{a}(r,\theta)\ e^{-i\omega u+im\varphi_-}.
\end{align}
We use the same phase as in \eqn{h1 large r expansion}, so that the coefficients of $\xi$ follow directly from those of $h^{(1)}$; this is also the convention in which the expressions of Sec.~\ref{sec:Bondi--Sachs/mmode-gauge-transformation} hold.
The $m$-modes of \eqn{mmode-def} differ from these by the radial phase $e^{i(\omega r_*-m\rsharp)}$.

In the $m\neq0$ sector, the spherical-harmonic modes of Eqs.~(C11)--(C13) reduce to\footnote{The $m\neq0$ gauge transformation eliminates both $h_{\ltetrad\ltetrad,\lm}^{(1,2)}$ and $h_{\ltetrad\ntetrad,\lm}^{(1,1)}$ due to $h_{\ltetrad\ntetrad,\lm}^{(1,1)}= \frac{i\omega}{2}\,h_{\ltetrad\ltetrad,\lm}^{(1,2)}$ in the Lorenz gauge. }
\begin{align}
    \xi^{\lm}_{\ltetrad}&= \frac{h_{\ltetrad\ltetrad,\lm}^{(1,2)}}{2r},\\
    \xi^{\lm}_{\ntetrad}&= 0,\\
    \xi^{\lm}_{\mtetrad}&=\xi^{\lm}_{\mb}= 0.
\end{align}
The $m$-mode gauge vector for $m\neq 0$ is the sum over $\ell$,
\begin{align}
    \xi^{\m}_{\ltetrad}&=\sum_{\ell=|m|}^\infty \frac{h_{\ltetrad\ltetrad,\lm}^{(1,2)}}{2r}\, {}_{0}Y_{\lm}(\theta,0),\label{eq:xi_l-mneq0}\\
    \xi^{\m}_{\ntetrad}&=0,\label{eq:xi_n-mneq0}\\
    \xi^{\m}_{\mtetrad}&=\xi^{\m}_{\mb}=0.\label{eq:xi_m-mneq0}
\end{align}
In the $m=0$ sector, Eqs.~(C11)--(C13) give
\begin{align}
    \xi^{00}_{\ltetrad}&=-\frac{h_{\ltetrad\ltetrad,00}^{(1,1)}}{2}\ln \frac{r}{M}+\frac{h_{\ltetrad\ltetrad,00}^{(1,2)}}{2r}, \\
    \xi^{00}_{\ntetrad}&=-\frac{h_{\ltetrad\ltetrad,00}^{(1,1)}}{4}\ln \frac{r}{M}+\frac{h_{\mtetrad\mb,00}^{(1,1)}}{2},\\
    \xi^{00}_{\mtetrad}&=\xi^{00}_{\mb}=0,
\end{align}
and all modes with $\ell\neq0$ vanish, so the $m=0$ gauge vector is
\begin{align}
    \xi^{0}_{\ltetrad}&=\sqrt{\frac{1}{4\pi}}\Biggl(-\frac{h_{\ltetrad\ltetrad,00}^{(1,1)}}{2}\ln \frac{r}{M}+\frac{h_{\ltetrad\ltetrad,00}^{(1,2)}}{2r}\Biggr),\label{eq:xi_l-m=0} \\
    \xi^{0}_{\ntetrad}&=\sqrt{\frac{1}{4\pi}}\Biggl(-\frac{ h_{\ltetrad\ltetrad,00}^{(1,1)}}{4}\ln \frac{r}{M}+\frac{h_{\mtetrad\mb,00}^{(1,1)}}{2}\Biggr),\label{eq:xi_n-m=0}\\
    \xi^{0}_{\mtetrad}&=\xi^{0}_{\mb}=0.\label{eq:xi_m-m=0}
\end{align}
The coefficient $h_{\ltetrad\ltetrad,\lm}^{(1,2)}$ for $m\neq0$ is obtained analytically in Sec.~\ref{sec:Bondi--Sachs/large-r-expansion}. The coefficients $h_{\ltetrad\ltetrad,00}^{(1,1)}$, $h_{\ltetrad\ltetrad,00}^{(1,2)}$, and $h_{\mtetrad\mb,00}^{(1,1)}$ for $m=0$ are obtained using a fit of numerical data in Sec.~\ref{sec:Bondi--Sachs/large-r-expansion-fit}.

\subsection{Large-\texorpdfstring{$r$}{r} expansion of the Lorenz gauge metric perturbation for \texorpdfstring{$m\neq 0$}{m≠0}}
\label{sec:Bondi--Sachs/large-r-expansion}

A large-$r$ expansion of the $m\neq0$ Lorenz-gauge metric perturbation is straightforward to obtain from the reconstruction of Sec.~\ref{sec:perturbations/radiative-modes}.
The metric perturbation is built from six fields; all but $\chi$ obey Teukolsky equations with sources on the orbit and therefore take the form~\eqref{eq:inup} (Sec.~\ref{sec:perturbations/radial-equations}), so outside the orbit each is a multiple of its up solution.
These extend the asymptotic form~\eqref{eq:bcs} to the series~\cite{Pound:2021qin}
\begin{equation}
    {}_{s}{R}^{\rm up}_{j m}(r) = \frac{1}{r^{1+2s}}e^{i m\Omega r_*}\sum_{n=0}^{n_{\rm trunc}} \frac{{}_{s}{c}^{(n)}_{j m}}{r^n},
\end{equation}
where $n_{\rm trunc}$ is the truncation order, chosen either to reach the required accuracy or to stop before the series starts to diverge, and ${}_{s}c^{(0)}_{jm}=1$ by the normalization of \eqn{bcs}.
The metric perturbation inherits this behavior.
The exception is $\chi$, whose source, the trace, extends to infinity (\eqn{chi_radial}); it does not enter at the order we need, so we do not discuss its expansion further.

The expansion coefficients follow from recursion relations derived from the Teukolsky radial equation.
Some of these were given in Refs.~\cite{Hughes:1999bq,Warburton:2010eq,Dolan:2023enf}; Ref.~\cite{Cunningham_2026} gives the relations for any spin weight, together with those needed at higher orders in the expansion.
Substituting these series into the reconstruction gives, at the orders we need, series for the metric perturbation in terms of the coefficients ${}_{s}c^{(n)}_{jm}$ and the amplitudes ${}_{s}{Z}_{jm}^{\rm up}$ of \eqn{inup}.

For $m\neq0$, the gauge vector of Sec.~\ref{sec:Bondi--Sachs/gauge-vector} needs only the first two coefficients of $h_{\ltetrad\ltetrad}$; in spheroidal modes, its expansion is
\begin{equation}
    h_{\ltetrad\ltetrad,j m}^{(1)}(r) =  e^{i m\Omega r_*}
    \biggl[\frac{h^{(1,1)}_{\ltetrad\ltetrad, j m}}{r}+\frac{h^{(1,2)}_{\ltetrad\ltetrad, j m}}{r^2}+\mathcal{O}(r^{-3})\biggr],
\end{equation}
with the coefficients $h^{(1,i)}_{\ltetrad\ltetrad,j m}$ given by
\begin{align}
    h^{(1,1)}_{\ltetrad\ltetrad,j m} ={}& 0, \\
    h^{(1,2)}_{\ltetrad\ltetrad,j m} ={}& \frac{i Z^{h }_{j m }}{m \Omega},
\end{align}
where $Z^{h}_{jm}$ is the amplitude ${}_{0}Z^{\rm up}_{jm}$ in \eqn{inup} of the trace $h_{jm}$ of Sec.~\ref{sec:perturbations/radial-equations}.

Since only the up solutions contribute outside the orbit, we drop the label ``up'' from the amplitudes and expansion coefficients, and label the amplitude of the trace by $h$ rather than $0$, since two of the fields have spin weight $0$.

The factor $e^{im\Omega r_*}$ combines with the $e^{im(\varphi-\Omega t)}$ of \eqn{mmode-def} into $e^{-i\omega u+im\varphi_-}e^{im\rsharp}$; since $\rsharp=\mathcal{O}(1/r)$ and $h^{(1,1)}_{\ltetrad\ltetrad,jm}=0$, $h^{(1,1)}_{\ltetrad\ltetrad,jm}$ and $h^{(1,2)}_{\ltetrad\ltetrad,jm}$ are the spheroidal modes of the coefficients in \eqn{h1 large r expansion}.
At this order, $h_{\ltetrad\ltetrad}$ depends only on the trace, a spin-0 field, so its spherical-harmonic $(\ell, m)$ coefficients follow from the spheroidal $(j, m)$ coefficients by the spin-0 change of basis~\eqref{eq:spheroidal_to_spherical},
\begin{equation}
    h^{(1,i)}_{\ltetrad\ltetrad,\ell m} = \sum_{j=j_{\rm min}}^\infty{}_0b^{(m, a \omega)}_{j \ell}h^{(1,i)}_{\ltetrad\ltetrad,j m}, \quad i\in\{1,2\},
\end{equation}
where $j_{\rm min} = |m|$.
In other components, and at higher orders, several fields contribute, with different spin weights or coupled across $j$, which complicates the projection.
The amplitude $Z^{h}_{j m}$ is exactly four times that of the $s=0$ field of a scalar charge equal to the particle's mass, the toy model of early self-force calculations~\cite{Warburton:2010eq}.
Both Dolan's~\cite{Dolan:2023enf} and Cunningham's~\cite{Cunningham_2026, Cunningham_inprep} codes therefore compute it with the \texttt{Teukolsky} package~\cite{BHPToolkit:Teukolsky} of the \texttt{Black Hole Perturbation Toolkit}~\cite{BHPToolkit}.

\subsection{Large-\texorpdfstring{$r$}{r} expansion fit for the \texorpdfstring{$m=0$}{m=0} Lorenz gauge metric perturbation }
\label{sec:Bondi--Sachs/large-r-expansion-fit}

For $m=0$, the gauge vector of Sec.~\ref{sec:Bondi--Sachs/gauge-vector} needs three coefficients of the $\ell=0$ mode, $h_{\ltetrad \ltetrad,00}^{(1,1)}$, $h_{\ltetrad \ltetrad,00}^{(1,2)}$, and $h_{\mtetrad\mb,00}^{(1,1)}$, and we obtain them by a least-squares fit. Early testing showed that a numerical fit at large-$r$ was sufficiently accurate, so we forgo an analytical expansion of the complicated Lorenz-gauge reconstruction of the static modes (Sec.~\ref{sec:perturbations/non-radiative-modes})~\cite{Dolan:2021ijg, Dolan:2023enf, Wardell:2024yoi}. In future work, we will attempt to extend a WKD~\cite{Wardell:2024yoi} type reconstruction procedure to the static case, at which stage we shall revisit the expansion.

We fit the radial data of the $(\ell,m)$-modes $h^{(1)}_{\ltetrad \ltetrad,00}$ and $h_{\mtetrad\mb,00}^{(1)}$, constructed from the variables of Sec.~\ref{sec:perturbations/reconstruction-in-lm-modes}, on a range of large radii,
 e.g., from $r=450M$ to $r_{\rm max}=500M$, with the powers $r^0$, $r^{-1}$, $r^{-2}$, and $r^{-3}$.
The three coefficients are stable under variation of the fit range and the number of free parameters in the fit.

\subsection{General \texorpdfstring{$m$}{m}-mode gauge transformation formula}
\label{sec:Bondi--Sachs/mmode-gauge-transformation}

To apply the gauge vectors of \cref{eq:xi_l-mneq0,eq:xi_n-mneq0,eq:xi_m-mneq0,eq:xi_l-m=0,eq:xi_n-m=0,eq:xi_m-m=0}, we need the $m$-modes of $\mathcal{L}_{\xi} g_{\mu\nu}$, the change of the first-order perturbation under the gauge transformation.
For a general gauge vector, the Kinnersley tetrad components of a first-order gauge transformation in the outgoing Kerr coordinates of Sec.~\ref{sec:perturbations/background} are
\begin{align}
(\mathcal{L}_{\xi} g)^\m_{\ltetrad\ltetrad} &= 2\,\partial_r \xi^\m_\ltetrad ,
\\[6pt]
(\mathcal{L}_{\xi} g)^\m_{\ltetrad\ntetrad} &= \partial_r \xi^\m_\ntetrad - \frac{\Delta}{2\Sigma}\,\partial_r \xi^\m_\ltetrad
 + \frac{1}{\Sigma^2}  \Big[\, M(a^2\cos^2\theta - r^2) \notag \\ &\quad+ a^2 r\sin^2\theta - i\,\Sigma K \,\Big]\xi^\m_\ltetrad
 \notag \\ &\quad+ \frac{\sqrt{2}\,a^2 \sin\theta\cos\theta}{\Sigma^2}\Big(\bar\zeta\,\xi^\m_\mtetrad + \zeta\,\xi^\m_\mb\Big), \displaybreak[0]
\\[6pt]
(\mathcal{L}_{\xi} g)^\m_{\ltetrad\mtetrad} &= \partial_r \xi^\m_\mtetrad + \frac{\partial_\theta \xi^\m_\ltetrad}{\sqrt{2}\,\bar\zeta} - \frac{\xi^\m_\mtetrad}{\bar\zeta}
 \notag \\ &\quad - \frac{1}{\sqrt{2}}\Bigg[\frac{m\csc\theta - a\omega\sin\theta}{\bar\zeta}  - \frac{2 i a\sin\theta}{\bar\zeta^{\,2}}\Bigg]\xi^\m_\ltetrad , \displaybreak[0]
\\[6pt]
(\mathcal{L}_{\xi} g)^\m_{\ntetrad\ntetrad} &= -\frac{\Delta}{\Sigma}\,\partial_r \xi^\m_\ntetrad
 - \frac{2}{\Sigma^2}\Big[\, M(a^2\cos^2\theta - r^2) \notag \\ &\quad + a^2 r\sin^2\theta + i\,\Sigma K \,\Big]\xi^\m_\ntetrad , \displaybreak[0]
\\[6pt]
(\mathcal{L}_{\xi} g)^\m_{\ntetrad\mtetrad} &= -\frac{\Delta}{2\Sigma}\,\partial_r \xi^\m_\mtetrad + \frac{\partial_\theta \xi^\m_\ntetrad}{\sqrt{2}\,\bar\zeta}
 + \Biggl[-\frac{iK}{\Sigma}  \notag \\ \quad &+ \frac{\Delta\,(r + 3 i a\cos\theta)}{2\Sigma^2}\Biggr]\xi^\m_\mtetrad
  - \frac{\zeta}{\sqrt{2}\,\Sigma^2}\Big[\,\Sigma\,(m\csc\theta  \notag \\ &\quad- a\omega\sin\theta) + 2 i a r\sin\theta\,\Big]\xi^\m_\ntetrad , \displaybreak[0]
\\[6pt]
(\mathcal{L}_{\xi} g)^\m_{\mtetrad\mtetrad} &= \frac{\sqrt{2}}{\bar\zeta}\,\partial_\theta \xi^\m_\mtetrad
 - \sqrt{2}\Bigg[\frac{m\csc\theta - a\omega\sin\theta}{\bar\zeta}  \notag \\ &\quad + \frac{r\cot\theta + i a\csc\theta}{\bar\zeta^{\,2}}\Bigg]\xi^\m_\mtetrad , \displaybreak[0]
\\[6pt]
(\mathcal{L}_{\xi} g)^\m_{\mtetrad\mb} &= \frac{r\Delta}{\Sigma^2}\,\xi^\m_\ltetrad - \frac{2r}{\Sigma}\,\xi^\m_\ntetrad
 + \frac{\partial_\theta \xi^\m_\mtetrad}{\sqrt{2}\,\zeta} + \frac{\partial_\theta \xi^\m_\mb}{\sqrt{2}\,\bar\zeta}
 \notag \\ \quad + \frac{1}{\sqrt{2}}&\left[\frac{m\csc\theta - a\omega\sin\theta}{\zeta} + \frac{r\cot\theta - i a\csc\theta}{\zeta^{2}}\right]\xi^\m_\mtetrad \notag \\ \quad
 - \frac{1}{\sqrt{2}}&\left[\frac{m\csc\theta - a\omega\sin\theta}{\bar\zeta} - \frac{r\cot\theta + i a\csc\theta}{\bar\zeta^{\,2}}\right]\xi^\m_\mb ,
\end{align}
where $\zeta \coloneqq r - i a\cos\theta$ and $K \coloneqq \omega\,(r^2+a^2) - a m$.
The three components with $\mb$ in place of $\mtetrad$ follow from the $\ltetrad\mtetrad$, $\ntetrad\mtetrad$ and $\mtetrad\mtetrad$ components by conjugating the coefficients, reversing the sign of the mode number $m$ (and hence of $\omega$), and replacing $\xi^\m_\mtetrad$ by $\xi^\m_\mb$.

\subsection{Transition function}
\label{sec:Bondi--Sachs/transition}

We keep the first-order perturbation in the Lorenz gauge near the particle and apply the (near) Bondi--Sachs transformation only at larger radii, as the punctures used inside the worldtube are in the Lorenz gauge. 
We switch the transformation on smoothly over a finite radial region, with a transition function.
This differs from the Schwarzschild calculation of Ref.~\cite{Leather:2026zhl}, in which four of the present authors switched between the Lorenz and (near) Bondi--Sachs gauges at a single radius.
A sharp switch makes the source and the field variable jump at that radius, and integrating the field equation across it requires the jump in the field variable.
At second order, this jump is quadratic in the gauge vector and contains products of the gauge vector and the first-order metric perturbation, so deriving and implementing the jump conditions is involved and error-prone.
A smooth transition avoids the jumps altogether.
Since the gauge vector enters the second-order Teukolsky source with up to five radial derivatives, the transition function must have a high degree of smoothness.

For a given transition function ${\cal T}(r)$, we define the transition gauge vector $\xi_{\cal T}^a={\cal T}(r)\xi^a$.
The transformed first-order perturbation is then
\begin{equation}
    h + {\cal L}_{ \xi_{\cal T}}g,
    \quad
    ({\cal L}_{\xi_{\cal T}}g)_{\alpha\beta} = {\cal T}(r)({\cal L}_{\xi}g)_{\alpha\beta} + 2\xi_{(\alpha}\partial_{\beta)}{\cal T}(r).    \label{eq:transition_gauge_transformation}
\end{equation}
At second order, $\xi$ is likewise replaced by ${\cal T}\xi$ everywhere in the gauge transformation.
This is an exact gauge transformation at every $r$, so the transformed field satisfies the field equations, with the source transformed consistently.

For the transition function we take the $C^{\infty}$ step
\begin{equation}
    {\cal T}(r) = \frac{1}{1 + e^{\frac1x - \frac{1}{(1-x)}}},
    \quad
    x = \frac{\ln(r/r_{s})}{\ln(r_{e}/r_{s})},
    \label{eq:transition_function}
\end{equation}
for $r_{s}<r<r_{e}$, with ${\cal T} = 0$ for $r\leq r_{s}$ and ${\cal T} = 1$ for $r \geq r_{e}$.
For the $\rp = 8M$ data shown here, $r_{s}/M = 20$ and $r_{e}/M = 1000$.
The starting point $r_s$ must lie outside the worldtube, which ends at $r \simeq 10.9M$ for $a/M = 0.6$ and $\rp = 8M$, so that the field is in the Lorenz gauge wherever the puncture is used; for wider orbits, $r_s$ moves out with the orbit.
Since all derivatives of ${\cal T}$ vanish at both ends, the transformation is smooth across $r_s$ and $r_e$, and the field is exactly in the Lorenz gauge for $r \leq r_{s}$ and exactly in the (near) Bondi--Sachs gauge for $r \geq r_{e}$.

At the same width, the first four derivatives of ${\cal T}$ are smaller than those of a $C^{8}$ polynomial smoothstep, but its fifth and sixth derivatives are larger.

\begin{figure*}
    \centering
    \includegraphics[width=\textwidth]{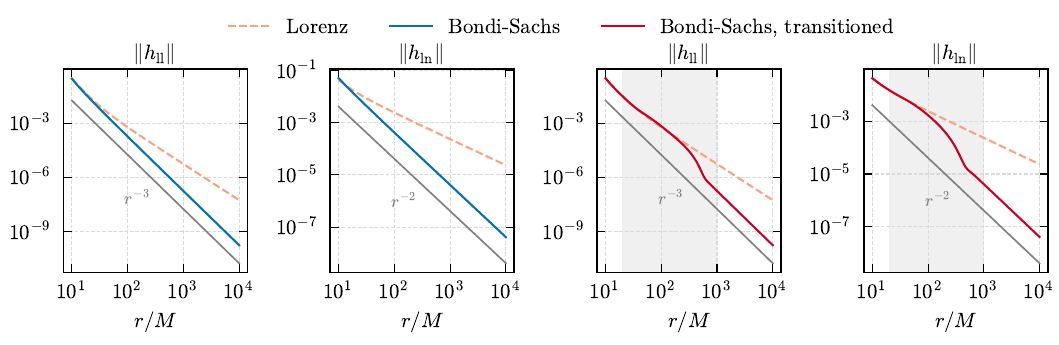}
    \caption{Angular $L^2$ norms of two Kinnersley components of the first-order $m=2$ mode outside the worldtube, $a=0.6M$, $\rp=8M$, in Lorenz gauge (dashed) and after the (near) Bondi--Sachs transformation.
    Left pair: the full transformation $h+\mathcal{L}_\xi g$; $h_{\ltetrad\ltetrad}$ falls off as $r^{-3}$ instead of $r^{-2}$ and $h_{\ltetrad\ntetrad}$ as $r^{-2}$ instead of $r^{-1}$.
    Right pair: the transformation switched on by the transition function \eqref{eq:transition_function}, $h+\mathcal{L}_{\xi_{\cal T}}g$ (shaded: $20M<r<1000M$); the field is in the Lorenz gauge below the window and in the (near) Bondi--Sachs gauge above it. Grey lines: $r^{-3}$ and $r^{-2}$ for reference.}
    \label{fig:bondi_h1_falloff}
\end{figure*}

The argument is linear in $\ln r$ because the derivative terms in \eqn{transition_gauge_transformation} multiply $\xi^a$, so a step that is steep at small $r$ produces a large transient.
Figure~\ref{fig:bondi_h1_falloff} shows the first-order field across the transition.

\section{Results}
\label{sec:results}

In this section we present our construction of the metric perturbation for circular equatorial orbits. 
We first show the retarded and residual $m$-modes of the metric perturbation for $a/M = 0.6$, $\rp/M = 8$, and then the self-force and Detweiler redshift, computed from the residual field, for a range of spins and radii.
The orbit $a/M=0.6$, $\rp/M=8$ is computed to $m\le40$ for the figures of the field and of the convergence in $m$. 
Five spins, $a/M\in\{-0.6,0,0.2,0.6,0.9\}$, each at ten radii starting from $\rp/M = 9, 7, 6, 5$, and $3$ respectively (between $1.13$ and $1.31$ times the ISCO radius), and extending to $\rp/M = 50$, are computed to $m\le64$.
These give the self-force and redshift tables and their dependence on $\rp$.
A further scan of thirteen spins, $-0.9\le a/M\le0.9$, at $\rp/M=8$, computed in the same way, gives the dependence on spin.
Every run keeps the spherical modes $\ell \leq 70$ of the metric perturbation, with the six radial fields solved for the spheroidal modes $j \leq j_{\rm max} = 82$.
The puncture is windowed with \eqn{window_function} before it is projected onto $(\ell,m)$ modes, and the radial derivatives at the particle are taken from a single polynomial fitted through the nodes on both sides of $\rp$. The $m = 0$ mode is completed so that the perturbation is asymptotically flat and regular at the horizon. 
As described in \secref{perturbations/Berndtson-type-solutions}, this shifts the primary black hole's parameters to $M_\bullet=M+\varepsilon\,\delta M$ and $J_\bullet=J+\varepsilon\,\delta J$, at every spin including $a = 0$.

\subsection{Metric perturbation}
\label{sec:results/metric-perturbations}
\begin{figure*}
    \centering
    \includegraphics[width=\textwidth]{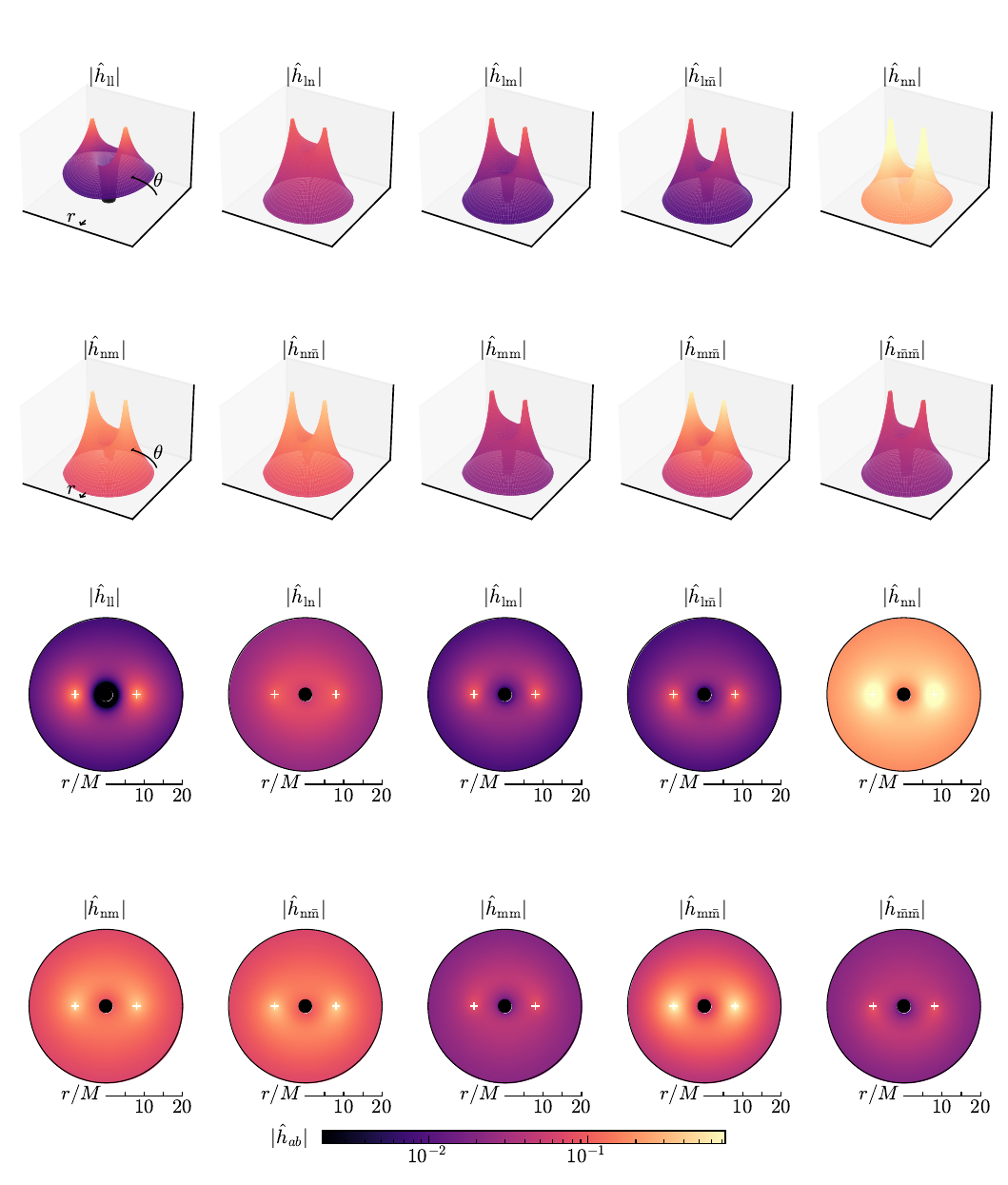}
    \caption{The ten Kinnersley components of the $m=2$ mode of the retarded field in Lorenz gauge, $a/M = 0.6$, $\rp/M = 8$, from the horizon to $r/M = 20$, with the horizon and axis factors divided out; the colour shows $|\hat h_{ab}|$ of \eqn{h_plotted} on a logarithmic scale.}
    \label{fig:tetrad_m2_polar_regular}
\end{figure*}

Figure~\ref{fig:tetrad_m2_polar_regular} shows the ten Kinnersley components of the $m=2$ mode over the exterior, from the horizon to $r=20M$, in Lorenz gauge, rescaled as described below.
Outside the worldtube, each component is the sum of its $(\ell, m)$-modes, whereas inside the worldtube it is the sum of the residual modes plus the windowed puncture in closed form, see \eqn{resummed}.
Before plotting, we divide out two factors that are properties of the tetrad rather than of the field.
A component of spin weight $s$ behaves on the axis as $\sin^{|m+s|}(\theta/2)$ at $\theta=0$ and as $\cos^{|m-s|}(\theta/2)$ at $\theta=\pi$; and, since $\ltetrad^{\mu}\sim1/\Delta$ at the horizon, a component with more $\ltetrad$ legs than $\ntetrad$ legs diverges there, while one with more $\ntetrad$ legs vanishes.
We therefore plot
\begin{equation}
    \hat h_{ab} \coloneqq \frac{(\Delta/2\Sigma)^{b}\,h^{m}_{ab}}{\sin^{|m+s|}(\theta/2)\,\cos^{|m-s|}(\theta/2)},
    \label{eq:h_plotted}
\end{equation}
where $b$ and $s$ are the boost and spin weights of the component (Appendix~\ref{sec:newman_penrose_conventions}), so that $b$ is the number of $\ltetrad$ legs minus the number of $\ntetrad$ legs, and $(\Delta/2\Sigma)^{b}h^{m}_{ab}$ is the component on the Hartle--Hawking tetrad.
The rescaled components are finite and smooth over the whole domain shown, from the horizon to $r=20M$ and across the axis, apart from the logarithmic divergence at the particle, and the components themselves vary by two orders of magnitude over the domain.
   
The first and second derivatives of the $m$-modes with respect to $r$ and $\theta$, which the second-order source requires, come from the same $(\ell, m)$-modes, the $\theta$ derivatives acting on the harmonics and on the windowed puncture and the $r$ derivatives on the radial coefficients, which are obtained from the radial equations rather than by numerical differentiation (\secref{perturbations/solution}).

The residual field in the worldtube is shown in \fig{ret_res_3d}.
The logarithmic divergence of the retarded field is absent, and the residual is $C^2$ through the particle, since the puncture is accurate through $\lambda^2$.
Its value and radial derivative at $r=\rp$ can be extracted independently from the $(\ell,m)$-modes on either side of the worldline.
The radial derivatives of the retarded modes and of the projected puncture are both discontinuous at $r=\rp$, and these discontinuities cancel in the residual only if the jump conditions of Sec.~\ref{sec:perturbations/radiative-modes} and the puncture are both correct; the agreement of the two sides is therefore a check of the construction at the particle, while the boundary conditions are checked by the comparison with \texttt{SpECTRE} below.
Extracted by polynomial fits from either side, the values of $h^{{\cal R}}_{\mu\nu}u^\mu u^\nu$ at $r=\rp$ agree to $\sim10^{-10}$, and so do the radial derivatives, which enter the self-force, for each $m$-mode before the two sides are fitted together.

\begin{figure}[h]
    \centering
    \includegraphics[width=\columnwidth]{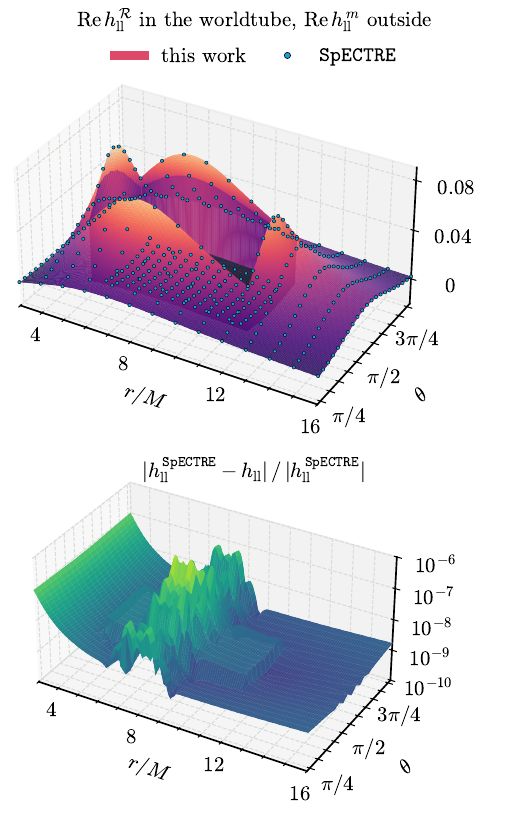}
    \caption{Comparison of the $m = 2$ mode of $h_{\ltetrad\ltetrad}$ with an independent \texttt{SpECTRE} calculation~\cite{SpectreGR}, for $a/M = 0.6$, $\rp/M = 8$, in the Kinnersley tetrad and $t$-slicing.
    Top: ${\rm Re} \, h^{{\cal R}, m}_{\ltetrad\ltetrad}$ inside \texttt{SpECTRE}'s worldtube $(5M < r < 12M,\, 11\pi/30 \leq \theta \leq 19\pi/30)$ and the retarded ${\rm Re} \, h^{m}_{\ltetrad\ltetrad}$ outside it.
    The surface is this work, and the points are \texttt{SpECTRE}'s values, every $1M$ in $r$ outside the worldtube, every $0.5M$ inside, and every $\pi/30$ in $\theta$.
    The steps are the worldtube walls, where the plotted field changes from retarded to residual.
    Bottom: the point-wise relative difference at every \texttt{SpECTRE} grid point.}
    \label{fig:spectre_comparison}
\end{figure}

As an independent check of the field, we compare the $m = 2$ mode for $a/M = 0.6$, $\rp/M = 8$ with the Lorenz-gauge $m$-mode solution computed with the spectral elliptic solver of \texttt{SpECTRE}~\cite{SpectreGR}.
That calculation shares neither the method nor the field variables with ours: it solves the coupled $(r, \theta)$ equations for each $m$-mode directly, for the trace-reversed perturbation in ingoing Kerr coordinates, rather than reconstructing the field from $(\ell, m)$-modes of Teukolsky solutions.
Inside its worldtube, $5 < r/M < 12$ and $11\pi/30 \leq \theta \leq 19\pi/30$, it solves for the residual field $h^{m} - h^{\calP,m}$, with the same puncture but no window, and outside it for the retarded field.
We convert its output to Kinnersley components in $t$-slicing and compare on its grid for $r \leq 16M$, where its time function differs from $t$ by a known function of $r$ only.
Figure~\ref{fig:spectre_comparison} shows the comparison for $h_{\ltetrad\ltetrad}$.
Outside the worldtube the retarded fields agree to a relative difference between $4\times10^{-10}$ and $1\times10^{-7}$, with a median of $1.4\times10^{-9}$.
The difference grows towards the smallest radius compared, $r/M = 3$.
Inside the worldtube, the residual fields agree to $5.4\times10^{-7}$, with the largest difference next to the particle on the equator.
Our residual, \eqn{residual_lm}, summed over $\ell$, is taken with respect to ${\cal W}h^{\calP}$, so we subtract $(1 - {\cal W})h^{\calP}$ in closed form before comparing.
The agreement inside the worldtube therefore tests the $\ell$-space subtraction of \secref{punctures/m-mode-retarded-and-residual} next to the particle, where the plain sum over $\ell$ converges slowly (and diverges at the particle).

\subsection{Self-force}
\label{sec:results/self-force}
The self-force \eqn{GSF} is linear in the residual field and its first derivatives on the worldline, so it is a sum of $m$-modes, $F^{\alpha} = \sum_{m}F^{\alpha}_{m}$, each read from the residual $m$-mode and its derivatives at the particle, and each $m$-mode is in turn a sum over $\ell$.

We compute the $m$-modes up to $m = 64$ and the $(\ell, m)$-modes up to $\ell = 70$, and complete both sums with fitted tails, taking the uncertainty of each fit as an estimate of its error.
Because the truncation contaminates the higher $m$-modes, the $m$ tail is fitted only over $11 \leq m \leq 23$ and replaces every computed mode above $m_{\rm max} = 23$, as described below; the modes with $m > 23$ do not enter the sums.
The residual field is $C^{2}$ at the particle, so the $m$-modes of its first derivatives fall off as $m^{-4}$.
\begin{figure*}
    \centering
    \includegraphics[width=\textwidth]{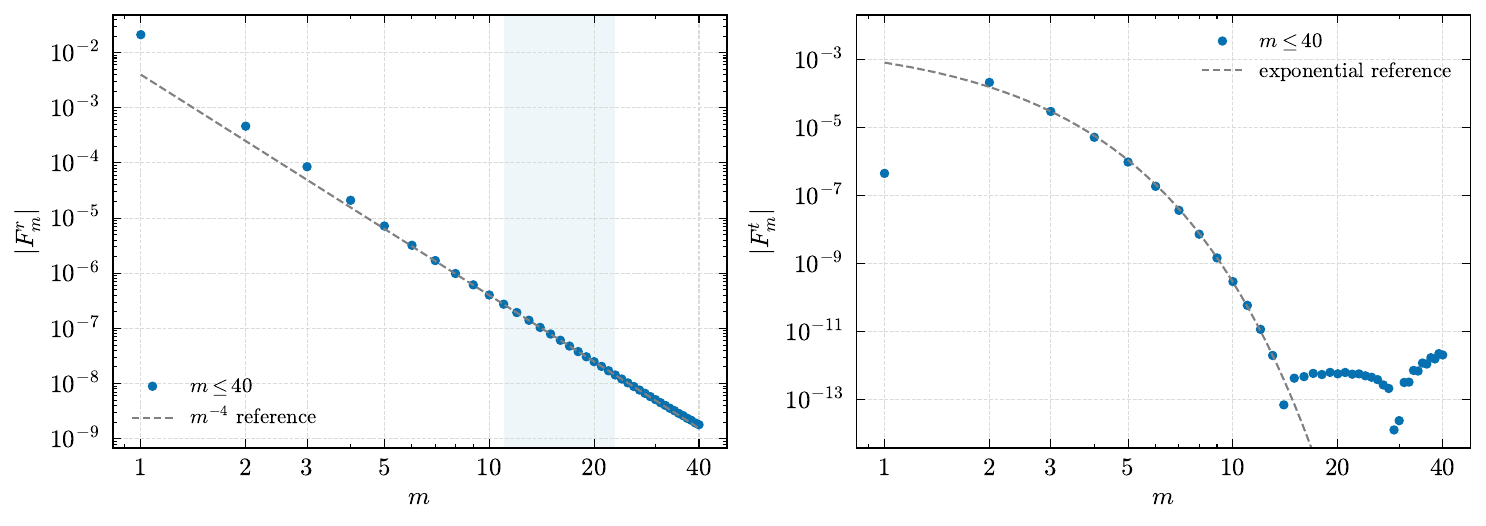}
    \caption{$m$-modes of the self-force for $a=0.6M$, $\rp=8M$. Left: $|F^r_m|$, with the $m^{-4}$ reference (dashed) fitted over the shaded window. Right: $|F^{t}_{m}|$, which reaches the numerical floor at $m\simeq14$.}
    \label{fig:gsf_m}
\end{figure*}
\fig{gsf_m} shows this for the radial component of the self-force, $F^{r}_{m}$, at $a/M = 0.6$, $\rp/M = 8$.
We see for the majority of the modes a fall-off with the expected power law, but at larger $m$ the points bend upwards to a smaller exponent.
This is not a property of the components of the self-force but of the $\ell$-truncation.
The modes with $\ell > 70$ make up $\sim 1\%$ of $F^{r}_{m}$ at $m = 20$ and $\sim 15\%$ for $35 < m < 40$.
A tail fitted to the highest computed $m$-modes would therefore fit this truncation artifact.
We instead fit the per-$m$ contributions $c_m$ only over a window of modes unaffected by the truncation, $m_0\le m\le m_{\rm max}$ with $m_{\rm max}=\lfloor\ellmax/3\rfloor=23$ and $m_0=\lfloor m_{\rm max}/2 \rfloor=11$, with the exponent held fixed,
\begin{equation}
    c_m = \sum_{k=0}^{K-1}A_k\,m^{-(4+k)},\qquad K=2,
    \label{eq:mtail_fit}
\end{equation}
by least squares in relative weight, and we replace every computed mode above the window by the fit. 
The sum is then
\begin{equation}
    \sum_{m=0}^{\infty}c_m = 
    \sum_{m=0}^{m_{\rm max}}c_m+\sum_{k=0}^{K-1}A_k\,\zeta(4+k,\,m_{\rm max}+1),
    \label{eq:mtail_sum}
\end{equation}
with $\zeta(s,a)$ the Hurwitz zeta function. 
We take the error of the tail to be the largest change in it when $K$ is varied over $\{1,2,3\}$ and the window is shifted to $m_0\pm4$.
For $F^r$, we fit the $m$-modes of each first derivative of the residual field, and of the field itself, separately.
The fitted tails enter $F^r$ through the linear map of \eqn{GSF}, and their errors are added in absolute value, each weighted by the modulus of its coefficient in that map.
At $a/M=0.6$, $\rp/M=8$, the $m$ tail of $F^r$ is $-9\times10^{-10}$ with an error of $6\times10^{-9}$. 
The dissipative components behave differently. 
They are unaffected by the puncture, which is time-symmetric, and their $m$-modes fall off exponentially, reaching the numerical floor of the construction, $\sim 10^{-12}$, by $m \simeq 14$ (\fig{gsf_m}, right). 
We sum them to the onset of the floor and take the sum of the floor modes as the error.
The second tail is in $\ell$. 
The $(\ell,m)$-modes of the first derivatives at the particle fall off as a power of $\ell$ with a parity alternation, so each $m$-mode is treated as two sequences, even and odd $\ell$. 
On each we fit
\begin{equation}
    c_\ell = \sum_{k=0}^{K-1}A_k\,\ell^{-(q+k)}
    \label{eq:ltail_fit}
\end{equation}
over $\ell_a\le\ell\le70$ with $\ell_a=\max(44,2m+4)$, by least squares in relative weight, and sum the fit beyond $\ell=70$ in closed form, again through the Hurwitz zeta function. 
The exponent $q$ is held fixed, and an ensemble of fits is used, the tail being the median of its members and the error their half-range. 
The local exponents of the pieces that dominate the tail of $F^r$, fitted on the top six modes of each parity are all between $\sim 4-5$ over the fifty orbits, although individual pieces among them decay more slowly, with local exponents as low as $2.2$.
For $F^r$, the ensemble is therefore $q \in \{3,4,5\}$ with $K \in \{2, 3\}$, which brackets the dominant fall-off.
We also check during the fitting procedure that we also capture the fall-off of the slower pieces.
The fit is done on the ``in'' and ``up'' sides separately and the two tails averaged, with the larger half-range kept. 
The window $\ell_a$ leaves at least ten modes of each parity for $m<m_{\rm max} = 23$, and above that the modes are replaced by the $m$ tail in any case. 
We test the rule on every orbit by fitting to $\ell \le 58$ and predicting the computed modes $59\le \ell \le70$; for $h^{\rm R}_{uu}$, the prediction error is within $0.57$ of the ensemble half-range at every orbit, with a median of $0.10$.
At $a/M = 0.6$, $\rp/M = 8$, the $\ell$-tail of $F^{r}$ is $\sim -7\times10^{-9}$ with an error of $\sim 5\times10^{-9}$, comparable to the $m$ tail.
The quoted error of every quantity is the sum of the $m$-tail error, the $\ell$-tail error, and the difference between the one-sided extrapolations at the worldline, which is below $10^{-13}$ once both sides are fitted together.

The dissipative components must satisfy a balance law,
\begin{equation}
    \frac{F_t}{u^t} = \Bigl\langle\diff{E}{t}\Bigr\rangle_{\scri^+}+\Bigl\langle\diff{E}{t}\Bigr\rangle_{\hor^+},
    \label{eq:balance}
\end{equation}
with the fluxes computed from the asymptotic amplitudes of $\psi_{4}$.
The balance law holds with a relative difference of $1.1\times10^{-8}$ at $a/M = 0.6$, $\rp/M = 8$, within $4 \times 10^{-11}$ at every $a = 0$ orbit, and within $\sim 10^{-7}$ at every orbit computed.

\begin{figure}[h]
    \centering
    \includegraphics[width=\columnwidth]{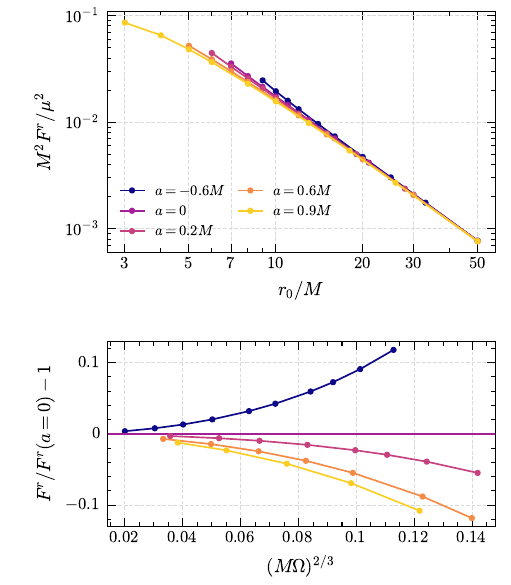}
    \caption{Top: the conservative self-force $F^r$ against $\rp$ for five spins, from the $m\le64$ runs with the $m$ and $\ell$ tails. 
    Bottom: the spin dependence at fixed orbital frequency, $F^r/F^r(a=0)-1$ against $y=(M\Omega)^{2/3}$, where $F^r(a=0)$ is the $a=0$ value at the same frequency, interpolated through the $a=0$ orbits; as in \fig{redshift_spins} the panel covers $y\le1/7$.}
    \label{fig:Fr_r0}
\end{figure}

\begin{table}
    \centering
    \begin{tabular}{lc}
    \hline\hline
    $M^2F^r/\mu^2$ & $2.3880872(11)\times10^{-2}$ \\
    $M^2F^t/\mu^2$ & $-2.50305114(13)\times10^{-4}$ \\
    $M^2F^\varphi/\mu^2$ & $-7.21387304(38)\times10^{-5}$ \\
    energy balance, \eqn{balance} & $1.1\times10^{-8}$ \\
    $h^{\rm R}_{uu}$ & $-0.29754101(11)$ \\
    $\Delta U$ & $-0.15511065(7)$ \\
    $\tilde{z}_{1}$ & $0.099806901(45)$ \\
    \hline\hline
    \end{tabular}
    \caption{Self-force and redshift for $a=0.6M$, $\rp=8M$, from the $m\le64$ run with the $m$ and $\ell$ tails. Parenthesized digits are the error in the last digits. $h^{\rm R}_{uu}$ and the redshift quantities are the coefficients of $\varepsilon$.}
    \label{tbl:results_a06}
\end{table}
Table~\ref{tbl:results_a06} gives the results for $a/M = 0.6$, $\rp/M = 8$ and Table~\ref{tbl:results_spins} for the fifty orbits of the five-spin set.
Figure~\ref{fig:Fr_r0} shows $F^r$ against $\rp$ for the five spins and the effect of spin at fixed orbital frequency: prograde spin lowers $F^r$, by $10.5$-$11.3\%$ at $y\simeq0.12$-$0.14$ for $a=0.6M$ and $0.9M$, and retrograde spin raises it by $12\%$ at $y\simeq0.11$ for $a=-0.6M$. 
The relative error of $F^r$ is $2$-$3\times10^{-7}$ for $\rp\gtrsim10M$ at every spin and grows towards the ISCO, to $2\times10^{-5}$ at $a/M=0.9$, $\rp/M=3$, where the $\ell$ tail is largest. 
The relative error of $F^t$ is below $4\times10^{-9}$ at $a=0$ (below $10^{-10}$ for $\rp \leq 11M$) and between $10^{-9}$ and $2\times10^{-6}$ at $a\neq0$.

At $a/M = 0$, several comparisons are available.
After converting to the $\delta J = 0$ completion with \eqn{Fr_spin_mode}, our $F^{r}$ agrees with an independent Schwarzschild Lorenz-gauge calculation \cite{Akcay:2010dx, Leather:2024mls} to an absolute difference between $4\times10^{-11}$ and $2.3\times10^{-9}$ (a relative difference of $5-7\times10^{-8}$) for all ten radii, within $0.3$ of our error.
The values of Barack and Sago~\cite{Barack:2007tm} use the Lorenz-gauge monopole of Detweiler and Poisson, in which $h_{tt}$ tends to a constant at infinity; they differ from ours by a factor $\sim 1.48$ at $\rp/M=8$ and $\sim 1.03$ at $\rp/M=50$. 
The two completions differ by a known gauge transformation, and once it is applied, the values agree to within $\sim 0.6$ of their errors at the eight radii in common, $\rp/M \in \{7, 8, 9, 10, 11, 13, 30, 50\}$, and $F^t$ agrees with theirs to within $1.1$ of their errors at the same eight radii.
To our knowledge, no Lorenz-gauge self-force has been published for $a/M\neq0$; Refs.~\cite{Dolan:2021ijg,Dolan:2023enf} give the method and the completion, but no self-force values.
For $a \neq 0$, our values are therefore tested by the balance law, by the redshift comparisons below, and by the $a=0$ checks, which use the same code with no special treatment of $a=0$.
At $a = 0$, the native completion differs from the Schwarzschild convention only by the black-hole spin mode of \eqn{Fr_spin_mode}, which we remove exactly before comparing.

\begin{table}
    \centering
    \setlength{\tabcolsep}{3.5pt}\small
    \begin{tabular}{rlll}
    \hline\hline
    $a/M$ & $\Delta U$ & ${\tilde z}_{1}$ & $10^{2}M^2F^r/\mu^2$ \\
    \hline
    $-0.90$ & $-0.232717723(85)$ & $0.137669760(50)$ & $3.6747510(18)$ \\
    $-0.75$ & $-0.221237420(83)$ & $0.132227541(49)$ & $3.4540999(17)$ \\
    $-0.60$ & $-0.210806209(81)$ & $0.127232241(49)$ & $3.2608418(16)$ \\
    $-0.45$ & $-0.201323211(79)$ & $0.122649261(48)$ & $3.0915710(15)$ \\
    $-0.30$ & $-0.192699148(77)$ & $0.118447246(47)$ & $2.9435410(14)$ \\
    $-0.15$ & $-0.184854810(76)$ & $0.114597734(47)$ & $2.8143774(13)$ \\
    $+0.00$ & $-0.177719742(74)$ & $0.111074839(46)$ & $2.7020117(13)$ \\
    $+0.15$ & $-0.171231122(73)$ & $0.107854968(46)$ & $2.6046450(12)$ \\
    $+0.30$ & $-0.165332782(72)$ & $0.104916573(46)$ & $2.5207215(12)$ \\
    $+0.45$ & $-0.159974374(71)$ & $0.102239921(45)$ & $2.4489084(11)$ \\
    $+0.60$ & $-0.155110646(70)$ & $0.099806901(45)$ & $2.3880872(11)$ \\
    $+0.75$ & $-0.150700814(68)$ & $0.097600845(44)$ & $2.3373757(11)$ \\
    $+0.90$ & $-0.146708033(67)$ & $0.095606380(44)$ & $2.2963101(10)$ \\
    \hline\hline
    \end{tabular}
    \caption{Redshift and conservative self-force at $\rp=8M$ against spin, from the $m\le64$, $\ell\le70$ runs with the $m$ and $\ell$ tails; parenthesized digits are the error in the last digits. $\Delta U$ and $\tilde{z}_{1}$ are the coefficients of $\varepsilon$. For $a<-0.65M$ the orbit lies inside the retrograde ISCO; the circular orbit and its field are still well defined.}
    \label{tbl:redshift_spins}
\end{table}

\begin{figure}
    \centering
    \includegraphics[width=\columnwidth]{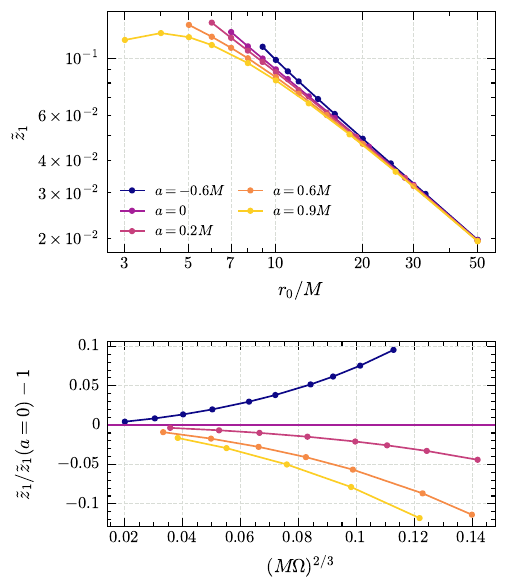}
    \caption{Top: the redshift ${\tilde z}_{1}$ against $\rp$ for five spins, from the $m\le64$ runs with the $m$ and $\ell$ tails. 
    Bottom: the spin dependence at fixed orbital frequency, ${\tilde z}_{1}/{\tilde z}_{1}(a=0)-1$ against $y=(M\Omega)^{2/3}$, where ${\tilde z}_1(a=0)$ is the $a=0$ value at the same frequency, interpolated through the $a=0$ orbits; the panel covers their range, $y\le1/7$, so the innermost $a=0.9M$ orbits are not shown.}
    \label{fig:redshift_spins}
\end{figure}

\begin{figure}
    \centering
    \includegraphics[width=\columnwidth]{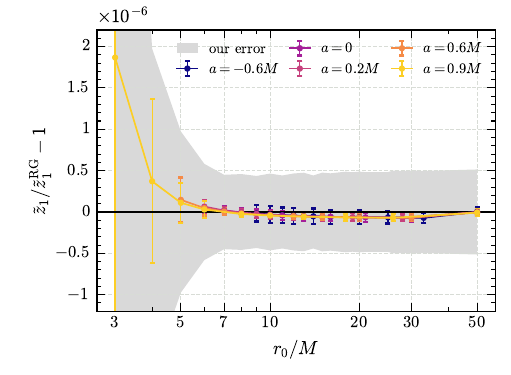}
    \caption{Relative difference between our redshift $\tilde{z}_{1}$ and the radiation-gauge value ${\tilde z}_{1}^{\rm RG}$ of Nasipak~\cite{Nasipak_pybhpt_2026,Nasipak:2025tby}, against $\rp$ for the five spins of the $m\le64$ runs. 
    The error bars are those of the radiation-gauge values; the grey band is our largest relative error at each radius.}
    \label{fig:redshift_vs_rg}
\end{figure}

\subsection{Redshift}
\label{sec:results/redshift}
For an equatorial circular orbit, the redshift $\tilde z \coloneqq 1/u^t$, expressed as a function of the orbital frequency $\Omega$, is invariant under gauge transformations that respect the helical symmetry of the system \cite{Detweiler:2008ft, Sago:2008id}. Expanding in the mass ratio at fixed frequency, $\tilde z(\Omega;\varepsilon) = \tilde z_0(\Omega) + \varepsilon\,\tilde z_1(\Omega) + O(\varepsilon^2)$, the background redshift $\tilde z_0$ follows from inverting \eqn{orbit}, while the first-order correction is determined by the regular field at the particle,
\begin{gather}
    \tilde z^{\rm Bernd}_{1} = -\half\tilde z_0\,h^{\rm R}_{uu},\qquad h^{\rm R}_{uu}\coloneqq h^{{\rm R}1}_{\mu\nu}u^\mu u^\nu,\nn\\
    \Delta U^{\rm Bernd}\coloneqq-\frac{\tilde z^{\rm Bernd}_1}{\tilde z_0^2}=\frac{h^{\rm R}_{uu}}{2\tilde z_0}.
    \label{eq:redshift}
\end{gather}
Here $h^{\rm R}_{uu}$ is evaluated at $r=r_p$, and $\Delta U$ is the corresponding first-order correction to $u^t$. 
The superscript indicates that these refer to the black hole of our Berndtson-type completion; below, we convert them to ${\tilde z}_{1}$ and $\Delta U$ for a black hole with the background parameters.
Because $\tilde z_1$ is gauge invariant, it provides a standard benchmark for comparing self-force calculations performed in different gauges. It also encodes the conservative first-order dynamics. Through the first law of binary mechanics \cite{Blanchet:2012at}, $\tilde z_1$ determines the conservative corrections to the binding energy and angular momentum, which enter the inspiral at the same post-adiabatic order as the second-order dissipative effects \cite{Hinderer:2008dm, Nasipak:2025tby}.

The redshift correction requires only the value of the residual field at the particle, summed over $\ell$ and $m$. The retarded $(\ell,m)$-modes of $h_{uu}$ at the particle decay only slowly in $\ell$, as the logarithmic divergence of the retarded $m$-mode requires, whereas the residual modes fall by six orders of magnitude by $\ell=70$. The two sums are completed as for the self-force. 
The $m$-modes of $h^{\rm R}_{uu}$ fall off as $m^{-4}$ (the local exponent over the window $11 \leq m \leq 23$ is $\sim 4$ at $a/M = 0.6$, $\rp/M = 8$) and the fit \eqref{eq:mtail_fit} is applied to them directly over the same window; at $a=0.6M$, $\rp/M=8$ the $m$ tail is $2.3\times10^{-7}$ with an error of $6\times10^{-8}$. 
The $\ell$ tail is fitted with \eqn{ltail_fit}, with the ensemble $q\in\{4,5\}$ and $K\in\{2,3\}$: the large-$\ell$ exponent of the pieces of $h^{\rm R}_{uu}$, fitted on the top six modes of each parity for the $m$ that dominate the tail, is at least $4.16$ at every one of the fifty orbits, with a weighted median of $5.2$, and at least $4.57$ at the thirteen orbits of the spin scan, so a leading $\ell^{-3}$ term is excluded. 
At $a=0.6M$, $\rp/M=8$ the $\ell$ tail is $-5\times10^{-8}$ with an error of $5\times10^{-8}$, and over the fifty orbits it is a median $42\%$ of the error of the redshift.

Comparisons with other calculations must be made at fixed $\Omega$ in a spacetime with the same black-hole parameters. The radiation-gauge results of Refs.~\cite{Shah:2012gu,vandeMeent:2015lxa} and the post-Newtonian series refer to a primary with the background parameters $(M,J)$, whereas our completion gives the primary inside the orbit the parameters $M_\bullet$ and $J_\bullet$, so the coefficient we compute is $\tilde z_1^{\rm Bernd}(\Omega;M_\bullet,J_\bullet)$, as described in Sec.~\ref{sec:perturbations/Berndtson-type-solutions}. We correct for these Berndtson mass and angular momentum shifts using \eqn{dM-dJ-correction}, which for $\tilde z_1$ reduces to 
\begin{equation}
    \tilde z_1(\Omega;M,J) = \tilde z_1^{\rm Bernd}(\Omega;M_\bullet,J_\bullet)-\pdiff{\tilde z_0}{M}\,\delta M-\pdiff{\tilde z_0}{J}\,\delta J,
    \label{eq:DeltaUBH}
\end{equation}
where the derivatives of $\tilde z_0(\Omega;M,J)$ are performed at fixed $\Omega$.
This $\tilde{z}_{1}$, together with $\Delta U := -\tilde{z}_{1}/\tilde{z}^{2}_{0}$, is what we quote in the tables and figures.

Figure~\ref{fig:redshift_spins} shows $\tilde{z}_{1}$ against $\rp$ for the five spins. Its relative error is between $4\times10^{-7}$ and $2\times10^{-6}$ for orbits more than $1.5$ ISCO radii out, with a median of $5\times10^{-7}$ over all fifty, rising to $9\times10^{-6}$ at $a=0.9M$, $\rp=3M$. The lower panel shows the effect of the spin at fixed orbital frequency, against $y=(M\Omega)^{2/3}$: prograde spin lowers the redshift correction by $12\%$ at $y\simeq0.12$--$0.14$ for $a=0.6M$ and $0.9M$, and retrograde spin raises it, by $10\%$ at $y=0.11$ for $a=-0.6M$.

We test these values against three independent calculations. 
At $a=0$, they agree with the Lorenz-gauge results of Ref.~\cite{Dolan:2014pja} at the seven radii in common, $\rp/M\in\{7,8,9,10,16,30,50\}$, to within $4.2\times10^{-9}$ in $\Delta U$, which is $0.13$ of our error or less at each. At $\rp=8M$ they agree with the corrected radiation-gauge data of Ref.~\cite{vandeMeent:2015lxa} at $a=0.9M$ to $4.2\times10^{-9}$, and at $a=0.6M$ with the reference obtained by interpolating in $a$ the difference between those data and the 8.5PN series of Ref.~\cite{Kavanagh:2016idg}, to $5.0\times10^{-8}$. Interpolating our own values in $a$ to $-0.5M$, $0.5M$ and $0.7M$, the differences from Ref.~\cite{vandeMeent:2015lxa} are $2.8\times10^{-9}$, $6.1\times10^{-9}$ and $3.9\times10^{-8}$. 
The five radiation-gauge comparisons lie within $0.73$ of our error, and three of them within $0.1$. Against the 8.5PN series itself the relative difference along $a=0.6M$ falls from $2\times10^{-2}$ at $\rp=5M$ to $10^{-6}$ at $\rp=15M$ and then stays at $1$--$2\times10^{-7}$, where our own error dominates; at $a=0.9M$ the series is eight times our value at $\rp=3M$ and within $5\times10^{-4}$ at $\rp=8M$.

A fourth comparison covers every orbit. 
We computed $z_1$ for all fifty orbits of the five-spin set in radiation gauge with \texttt{pybhpt}~\cite{Nasipak_pybhpt_2026,Nasipak:2025tby}, with errors a median ten times smaller than ours. 
Figure~\ref{fig:redshift_vs_rg} shows the relative difference. 
Our $\tilde{z}_{1}$ agrees at every orbit within $0.17$ of the combined error, the median being $0.10$; the relative difference has a median of $-3.7\times10^{-8}$ and is at most $7.1\times10^{-8}$ in magnitude for $\rp\ge6M$.
Every comparison in this section and in \secref{results/self-force} that is measured against our own error lies within $0.73$ of it.
For Gaussian errors, the median ratio would be $\sim 0.67$, compared with $\sim 0.10$ here, so our quoted errors are conservative.
This is expected, since the dominant contributions are the tail errors, which are half-ranges of fit ensembles and so bound the truncation error rather than estimate a standard deviation.

Table~\ref{tbl:redshift_spins} and Fig.~\ref{fig:redshift_spins} give the dependence on spin at $\rp=8M$. The retrograde orbits with $a<-0.65M$ lie inside the ISCO; the circular orbit and its field are still well defined. 
The error is $0.7$-$0.9\times10^{-7}$ in $\Delta U$ at every spin.

\begin{table*}
    \centering
    \begin{tabular}{ccccc}
    \hline\hline
    $\rp/M$ & $\tilde{z}_{1}$ & $M^2F^r/\mu^2$ & $M^2F^t/\mu^2$ & $(F_t/u^t)/\langle\dd E/\dd t\rangle-1$ \\
    \hline
    \multicolumn{5}{l}{$a = -0.6M$} \\
    $9$ & $1.10604309(43)\times10^{-1}$ & $2.46996449(83)\times10^{-2}$ & $-2.15903328(41)\times10^{-4}$ & $-7.3\times10^{-9}$ \\
    $10$ & $9.8360767(40)\times10^{-2}$ & $1.95210436(50)\times10^{-2}$ & $-1.13471243(11)\times10^{-4}$ & $-6.2\times10^{-9}$ \\
    $11$ & $8.8827352(37)\times10^{-2}$ & $1.58965538(37)\times10^{-2}$ & $-6.46234756(20)\times10^{-5}$ & $-6.2\times10^{-9}$ \\
    $12$ & $8.1124174(35)\times10^{-2}$ & $1.32379195(29)\times10^{-2}$ & $-3.914043423(98)\times10^{-5}$ & $-7.5\times10^{-9}$ \\
    $14$ & $6.9328197(31)\times10^{-2}$ & $9.6422881(20)\times10^{-3}$ & $-1.64749551(17)\times10^{-5}$ & $-5.7\times10^{-9}$ \\
    $16$ & $6.0643943(27)\times10^{-2}$ & $7.3642614(14)\times10^{-3}$ & $-7.9301919(12)\times10^{-6}$ & $-5.6\times10^{-9}$ \\
    $20$ & $4.8607118(23)\times10^{-2}$ & $4.72134984(92)\times10^{-3}$ & $-2.40203162(54)\times10^{-6}$ & $-5.8\times10^{-9}$ \\
    $25$ & $3.9009167(19)\times10^{-2}$ & $3.03786121(59)\times10^{-3}$ & $-7.4542765(20)\times10^{-7}$ & $-4.2\times10^{-9}$ \\
    $33$ & $2.9679526(15)\times10^{-2}$ & $1.75792894(34)\times10^{-3}$ & $-1.77760222(56)\times10^{-7}$ & $-3.7\times10^{-9}$ \\
    $50$ & $1.9701670(10)\times10^{-2}$ & $7.7487457(15)\times10^{-4}$ & $-2.13950646(77)\times10^{-8}$ & $-3.4\times10^{-9}$ \\
    \hline
    \multicolumn{5}{l}{$a = 0M$} \\
    $7$ & $1.26198589(53)\times10^{-1}$ & $3.5399779(26)\times10^{-2}$ & $-7.41112784891(21)\times10^{-4}$ & $-3.9\times10^{-12}$ \\
    $8$ & $1.11074839(46)\times10^{-1}$ & $2.7020117(13)\times10^{-2}$ & $-3.30739750955(11)\times10^{-4}$ & $-1.9\times10^{-12}$ \\
    $9$ & $9.9573737(42)\times10^{-2}$ & $2.14096695(72)\times10^{-2}$ & $-1.668101230638(18)\times10^{-4}$ & $-1.6\times10^{-12}$ \\
    $10$ & $9.0385589(39)\times10^{-2}$ & $1.74327562(47)\times10^{-2}$ & $-9.19075772023(32)\times10^{-5}$ & $-6.3\times10^{-13}$ \\
    $11$ & $8.2817924(37)\times10^{-2}$ & $1.44953158(36)\times10^{-2}$ & $-5.416230027386(64)\times10^{-5}$ & $+1.5\times10^{-12}$ \\
    $13$ & $7.1010236(32)\times10^{-2}$ & $1.05070901(24)\times10^{-2}$ & $-2.18392491042(54)\times10^{-5}$ & $-8.9\times10^{-12}$ \\
    $16$ & $5.8544731(27)\times10^{-2}$ & $7.0493501(15)\times10^{-3}$ & $-7.23715385880(68)\times10^{-6}$ & $+6.3\times10^{-13}$ \\
    $21$ & $4.5307950(22)\times10^{-2}$ & $4.17476348(85)\times10^{-3}$ & $-1.75174515410(62)\times10^{-6}$ & $-9.0\times10^{-12}$ \\
    $30$ & $3.2200480(16)\times10^{-2}$ & $2.09121601(41)\times10^{-3}$ & $-2.80818783141(88)\times10^{-7}$ & $+9.2\times10^{-12}$ \\
    $50$ & $1.9593977(10)\times10^{-2}$ & $7.7022666(15)\times10^{-4}$ & $-2.1084590913(71)\times10^{-8}$ & $-4.0\times10^{-11}$ \\
    \hline
    \multicolumn{5}{l}{$a = 0.2M$} \\
    $6$ & $1.37241625(80)\times10^{-1}$ & $4.4641890(57)\times10^{-2}$ & $-1.671354571(66)\times10^{-3}$ & $-1.7\times10^{-9}$ \\
    $7$ & $1.19801439(52)\times10^{-1}$ & $3.3177720(24)\times10^{-2}$ & $-6.52587165(13)\times10^{-4}$ & $-1.5\times10^{-9}$ \\
    $8$ & $1.06845238(46)\times10^{-1}$ & $2.5752471(12)\times10^{-2}$ & $-2.999845741(24)\times10^{-4}$ & $-1.4\times10^{-9}$ \\
    $9$ & $9.6580954(42)\times10^{-2}$ & $2.06242544(70)\times10^{-2}$ & $-1.5427356904(19)\times10^{-4}$ & $-1.2\times10^{-9}$ \\
    $10$ & $8.8165661(39)\times10^{-2}$ & $1.69157926(47)\times10^{-2}$ & $-8.617017706(65)\times10^{-5}$ & $-1.2\times10^{-9}$ \\
    $12$ & $7.5105712(34)\times10^{-2}$ & $1.20014863(29)\times10^{-2}$ & $-3.212246355(52)\times10^{-5}$ & $-1.1\times10^{-9}$ \\
    $15$ & $6.1436926(29)\times10^{-2}$ & $7.8708030(17)\times10^{-3}$ & $-9.85663015(24)\times10^{-6}$ & $-9.2\times10^{-10}$ \\
    $19$ & $4.9410772(24)\times10^{-2}$ & $5.0181767(11)\times10^{-3}$ & $-2.880959482(86)\times10^{-6}$ & $-8.2\times10^{-10}$ \\
    $28$ & $3.4264598(17)\times10^{-2}$ & $2.38022625(48)\times10^{-3}$ & $-3.94741329(14)\times10^{-7}$ & $-6.0\times10^{-10}$ \\
    $50$ & $1.9559639(10)\times10^{-2}$ & $7.6882008(15)\times10^{-4}$ & $-2.098512272(78)\times10^{-8}$ & $-4.1\times10^{-10}$ \\
    \hline
    \multicolumn{5}{l}{$a = 0.6M$} \\
    $5$ & $1.3449225(13)\times10^{-1}$ & $5.220687(13)\times10^{-2}$ & $-3.4878808(11)\times10^{-3}$ & $-1.9\times10^{-8}$ \\
    $6$ & $1.20961324(57)\times10^{-1}$ & $3.8944289(44)\times10^{-2}$ & $-1.21799936(16)\times10^{-3}$ & $-1.6\times10^{-8}$ \\
    $7$ & $1.09547089(49)\times10^{-1}$ & $3.0059985(20)\times10^{-2}$ & $-5.17052935(12)\times10^{-4}$ & $-1.3\times10^{-8}$ \\
    $8$ & $9.9806901(45)\times10^{-2}$ & $2.3880872(11)\times10^{-2}$ & $-2.50305114(13)\times10^{-4}$ & $-1.1\times10^{-8}$ \\
    $10$ & $8.4302362(38)\times10^{-2}$ & $1.61037560(46)\times10^{-2}$ & $-7.6317158(11)\times10^{-5}$ & $-9.0\times10^{-9}$ \\
    $12$ & $7.2702052(34)\times10^{-2}$ & $1.15859911(29)\times10^{-2}$ & $-2.93925335(60)\times10^{-5}$ & $-8.0\times10^{-9}$ \\
    $15$ & $6.0078300(29)\times10^{-2}$ & $7.6858462(18)\times10^{-3}$ & $-9.2707350(24)\times10^{-6}$ & $-6.9\times10^{-9}$ \\
    $20$ & $4.6444750(23)\times10^{-2}$ & $4.48315166(98)\times10^{-3}$ & $-2.12738939(65)\times10^{-6}$ & $-5.9\times10^{-9}$ \\
    $30$ & $3.1837908(16)\times10^{-2}$ & $2.06683018(42)\times10^{-3}$ & $-2.72303193(94)\times10^{-7}$ & $-4.4\times10^{-9}$ \\
    $50$ & $1.9493279(10)\times10^{-2}$ & $7.6621854(15)\times10^{-4}$ & $-2.07918085(78)\times10^{-8}$ & $-3.5\times10^{-9}$ \\
    \hline
    \multicolumn{5}{l}{$a = 0.9M$} \\
    $3$ & $1.175332(11)\times10^{-1}$ & $8.59621(20)\times10^{-2}$ & $-4.2852569(65)\times10^{-2}$ & $-1.0\times10^{-7}$ \\
    $4$ & $1.2500215(25)\times10^{-1}$ & $6.540437(34)\times10^{-2}$ & $-8.5675032(50)\times10^{-3}$ & $-5.8\times10^{-8}$ \\
    $5$ & $1.20479151(88)\times10^{-1}$ & $4.8263013(97)\times10^{-2}$ & $-2.58239908(44)\times10^{-3}$ & $-4.1\times10^{-8}$ \\
    $6$ & $1.12340894(53)\times10^{-1}$ & $3.6635495(37)\times10^{-2}$ & $-9.86683464(69)\times10^{-4}$ & $-3.5\times10^{-8}$ \\
    $8$ & $9.5606380(44)\times10^{-2}$ & $2.2963101(10)\times10^{-2}$ & $-2.20898033(74)\times10^{-4}$ & $-2.7\times10^{-8}$ \\
    $10$ & $8.1863204(38)\times10^{-2}$ & $1.56675792(46)\times10^{-2}$ & $-7.0084080(33)\times10^{-5}$ & $-2.2\times10^{-8}$ \\
    $13$ & $6.6676446(32)\times10^{-2}$ & $9.8292223(25)\times10^{-3}$ & $-1.8359389(11)\times10^{-5}$ & $-1.7\times10^{-8}$ \\
    $18$ & $5.0514195(25)\times10^{-2}$ & $5.4088504(12)\times10^{-3}$ & $-3.5240459(25)\times10^{-6}$ & $-1.3\times10^{-8}$ \\
    $26$ & $3.6194307(18)\times10^{-2}$ & $2.70406076(57)\times10^{-3}$ & $-5.5122447(43)\times10^{-7}$ & $-1.0\times10^{-8}$ \\
    $50$ & $1.9445519(10)\times10^{-2}$ & $7.6445114(15)\times10^{-4}$ & $-2.0651484(17)\times10^{-8}$ & $-6.2\times10^{-9}$ \\
    \hline\hline
    \end{tabular}
    \caption{First-order redshift and self-force for five spins and ten radii each, from the $m\le64$, $\ell\le70$ runs with the $m$ and $\ell$ tails. Parenthesized digits are the error in the last digits. The self-force components are Boyer-Lindquist contravariant, in the asymptotically flat, horizon-regular completion of the $m=0$ mode at every spin, including $a = 0$ (the $\delta J = 0$ values follow from \eqn{Fr_spin_mode}); $\tilde{z}_{1}$ is the coefficient of $\varepsilon$ and is defined by \eqn{DeltaUBH}. 
    $u_\alpha F^\alpha=0$ fixes $F^\varphi=(\mathcal{E}/\mathcal{L}_z)F^t$, which our values satisfy to $2\times10^{-15}$.
    The last column is the relative residual of the energy balance \eqref{eq:balance}, $(F_t/u^t)/\langle\dd E/\dd t\rangle-1$, with $\langle\dd E/\dd t\rangle$ the sum of the horizon and infinity fluxes, computed independently from $\psifour$.}
    \label{tbl:results_spins}
\end{table*}

\section{Conclusion}
\label{sec:conclusion}

In this work, we construct the $m$-modes of the first-order Lorenz-gauge metric perturbation sourced by a point mass on a circular, equatorial orbit in Kerr spacetime. As an example, Fig.~\ref{fig:tetrad_m2_polar_regular} presents the ten Kinnersley components of the $m = 2$ mode of the retarded field in Lorenz gauge for the orbit $a/M = 0.6$, $\rp/M = 8$, from
the horizon to $r/M = 20$. Our perturbation data are specifically constructed to meet the requirements of second-order self-force calculations: the data are highly accurate, and thus robust to the delicate cancellations that occur when constructing the source of the second-order metric perturbation; complete, including all radiative, static, and gauge contributions; comprehensive, spanning the whole spacetime exterior from the horizon to infinity; in a globally regular combination of the Lorenz and (near) Bondi--Sachs gauges; and expressed in coordinates free of unphysical singularities. First-order data enter the second-order calculation in three ways: (i) through the second-order source, constructed from quadratic combinations of the first-order perturbation; (ii) through the first-order regular field and self-force on the worldline, which appear in the second-order punctures; and (iii) through the parametric derivative of the first-order perturbation, which enters the second-order field equations as a source through the multiscale expansion. This work provides the inputs for the first two; the third will be addressed in separate work.

To meet these requirements, we calculate the first-order perturbation using the sourced Lorenz gauge $(\ell,m)$-mode reconstruction procedure of Refs.~\cite{Dolan:2023enf, Wardell:2024yoi}. In the radiative sector $(\omega \neq 0)$, we follow Ref.~\cite{Wardell:2024yoi} and construct the modes from distributional sources, as summarized in Sec.~\ref{sec:perturbations/radiative-modes}. In the static sector $(\omega=0)$, where this method breaks down, we instead follow Ref.~\cite{Dolan:2023enf} and impose continuity conditions along the worldline to recover distributional source contributions. We complete the static modes with a Berndtson-type solution \cite{Berndtson:2007gsc} that is regular at both the horizon and infinity but with shifted primary mass and angular momentum $(M_\bullet, J_\bullet) = (M + \varepsilon \delta M, J + \varepsilon \delta J)$. We produce global solutions by solving for the fields in compactified hyperboloidal coordinates using a new multi-domain pseudospectral Chebyshev collocation routine, based on that of Ref.~\cite{Leather:2024mls}. The $s=-2$ Teukolsky amplitudes produced by our solver agree with those of the independent solver \texttt{pybhpt}~\cite{Nasipak_pybhpt_2026} to better than $10^{-12}$.

The disadvantage of constructing the metric perturbation from $(\ell,m)$-modes is that $\ell$-mode sum converges slowly and introduces Gibbs ringing near the worldline due to the divergence of the metric perturbation at $r=r_p$, as shown in Fig.~\ref{fig:gibbs_ringing}. To address this, we introduce a new $(\ell,m)$-to-$m$-mode puncture scheme that removes the singular structure at the level of the $(\ell,m)$-modes. Generalizing Ref.~\cite{Bourg:2024cgh} to circular orbits in Kerr, we construct a puncture $h^{\calP}_{\mu\nu}$ in comoving coordinates through order $\lambda^2$ in distance from the worldline, whose $m$-modes can be expressed in closed form. We then numerically project the windowed $m$-mode puncture onto spin-weighted spherical harmonics and subtract it mode by mode, forming the residual modes $h^{\calR,\ell m}_{\mu\nu} = h^{\ell m}_{\mu\nu} - h^{\calP,\ell m}_{\mu\nu}$. These converge rapidly in $\ell$, falling below $10^{-7}$ by $\ell \simeq 40$ (Fig.~\ref{fig:window_puncture}), and the Gibbs ringing is removed (Fig.~\ref{fig:gibbs_ringing}). The resulting residual field is $C^2$ at the particle, so the $m$-modes of its first derivatives decay as $m^{-4}$. We then recover the full retarded field $h^{(1)}_{\mu\nu}$ by adding back the windowed puncture in closed form. This procedure not only provides highly-accurate metric perturbation data, but also delivers them as $m$-modes, which are better suited to second-order calculations by significantly reducing the mode-coupling sums in the quadratic part of the second-order source.

While our Lorenz gauge perturbation is asymptotically flat at first order, it does not decay rapidly enough in the far field to produce an asymptotically flat second-order perturbation. We therefore transform to a (near) Bondi--Sachs gauge at large $r$, following Ref.~\cite{Spiers:2026yqx}, so that the asymptotic-flatness conditions of Ref.~\cite{Spiers:2023cip} are satisfied. A smooth transition function over a buffer region, $20M < r < 1000M$, joins the Lorenz-gauge interior to the (near) Bondi--Sachs exterior. This avoids the gauge discontinuity and the associated jump conditions of the single-radius approach of Ref.~\cite{Leather:2026zhl}. The transformation improves the fall-off of $h_{\ltetrad\ltetrad}$ and $h_{\ltetrad\ntetrad}$ by one order in $1/r$ for the $m\neq0$ modes, as demonstrated in Fig.~\ref{fig:bondi_h1_falloff}. For the $\m=0$ mode, the transformation improves the fall-off of $h_{\ltetrad\ltetrad}$ by two orders in $1/r$ and $h_{\mtetrad\mb}$ by one order in $1/r$.

From the metric perturbation, we calculate the redshift correction $\tilde{z}_1$ and first-order self-force $F^\mu$. We present numerical values for thirteen different spins at fixed radius $r_p = 8M$ in Table \ref{tbl:redshift_spins} and for five different spins across ten radii in Table \ref{tbl:results_spins}. The estimated errors are largest for high spins and small radii, owing to the slower convergence of the $\ell$-mode sum. The relative error in $F^r$ is $\sim2\times10^{-7}$ for $r_p \gtrsim 10M$, rising to $\sim2\times10^{-5}$ at $(a, r_p) = (0.9M,3M)$, while the median relative error in $\tilde{z}_1$ is $\sim5\times10^{-7}$. The redshift values agree with independent calculations \cite{Nasipak:2025tby,Dolan:2014pja,vandeMeent:2015lxa} within the numerical error of our results. Likewise, for $a=0$, our self-force values agree with an independent Lorenz-gauge Schwarzschild calculation~\cite{Leather:2024mls,Akcay:2012ea} to relative differences $\lesssim 7\times10^{-9}$, and with the results of Ref.~\cite{Barack:2007tm} within their quoted errors once the two completions are related by a gauge transformation. We highlight that these are the first calculations of both the full self-force and redshift from Lorenz gauge perturbations in Kerr spacetime.

Future work will look to incorporate this first-order data in second-order calculations, ultimately producing the second-order fluxes that are required for the phase-accurate evolution of small-mass-ratio binaries. As mentioned above, this will require, in addition to the first-order data presented here, computing parametric derivatives of the metric perturbation. One way to achieve this is to densely tile the parameter space and take numerical derivatives with respect to the orbital parameters (i.e., $\Omega$). Alternatively, one can reconstruct the parametric derivatives of the metric perturbation directly from the field equations. We leave both avenues for future work. 

Beyond second-order calculations, these first-order data can also be applied to the study of boson clouds and accretion discs around Kerr black holes \cite{Dyson:2025dlj,Li:2025ffh,Li:2026foy,Dyson:2026ddd}, which require regular perturbation data across the entire spacetime exterior. Similarly, the data can be used to study non-linear effects in modeling scalar-charged secondary black holes~\cite{maselli2020detecting,Spiers:2023cva, Barsanti:2026ulr}. They can also serve as useful input for the study of quasinormal modes and black hole spectroscopy in a perturbed Kerr spacetime \cite{Berti:2025hly,Bucciotti:2024zyp}. Furthermore, the $m$-mode punctures presented in this work can be applied to other methods for computing the first-order metric perturbation for circular orbits, including the elliptic and spectral solvers developed in Refs.~\cite{Osburn:2022bby,Osburn:2026uct,Vu:2026ypc} and the GHZ metric reconstruction schemes of Refs.~\cite{Green:2019nam,Hollands:2024iqp,Toomani:2021jlo}. Our results also provide a useful benchmark for other methods of computing the metric perturbation, including those listed above, as well as time-domain Lorenz-gauge solvers \cite{Barack:2007we,Dolan:2010mt,Dolan:2011dx,Dolan:2012jg,Thornburg:2016msc,Thornburg:2026eeb} and other approaches \cite{Loutrel:2020wbw,Ripley:2020xby,Casals:2024ynr,Li:2026rkf}.

Astrophysical EMRIs are expected to be eccentric and precessing, which will necessitate extending these calculations beyond circular, equatorial orbits. The work done in Ref.~\cite{Wardell:2024yoi} has laid the early groundwork for this extension by showing how the methods derived in Ref.~\cite{Dolan:2023enf} can be applied to compute non-vacuum, retarded metric perturbations in Lorenz gauge. First implementations of this sourced calculation were presented in Ref.~\cite{Cunningham_2026} for circular, equatorial-eccentric, and inclined-circular orbits, dealing only with radiative ($\omega\neq 0$) contributions. The generic case is presented in Ref.~\cite{Cunningham_inprep}, including an \textit{ad hoc} method of fitting the static sector, based on Ref.~\cite{Dolan:2023enf}. The comoving coordinates used by our puncture scheme, however, are specialized to circular orbits, and a straightforward generalization to more generic orbits has not yet been identified.
Extending our punctures to more generic orbits will therefore require a different choice of coordinates.
This is under way~\cite{Upton2026Capra}, with the $m$-modes computed by the method of Ref.~\cite{Thornburg:2016msc}.
First, a shift of the azimuthal coordinate, introduced in Refs.~\cite{Mino:2001mq,Haas:2006ne}, removes the $\Delta r\,\Delta\varphi$ and $\Delta\theta\,\Delta\varphi$ cross-terms from $\scov_1$, the leading-order part of \eqn{scovExp}.
Second, the $Q$--$R$ scheme of Ref.~\cite{Wardell:2011gb}, in which powers of $\Delta\varphi$ are rewritten in terms of $Q=\sin(\Delta\varphi/2)$ and $R=\sin\Delta\varphi$, puts $\scov_1$ in the form of \eqn{R2exp}, which can be integrated analytically.
However, it is likely that the decomposition into radial ($n$) and polar ($k$) frequency modes will have to be done numerically.
This is because the arguments of the Legendre functions in \eqn{J0} depend on the radial and polar coordinates in a complicated way.

For circular orbits, the accuracy of the present results can be improved in two ways.
Computing more $\ell$-modes in the reconstruction would reduce the $\ell$-tail error, which dominates at high spin and close to the ISCO (Sec.~\ref{sec:results/self-force}).
Alternatively, extending the $m$-mode punctures to higher order in $\lambda$ would make the residual field smoother and so accelerate the convergence in both $\ell$ and $m$.


\begin{acknowledgments}
We thank Adam Pound for extensive discussions and guidance throughout this project. We thank Leor Barack for discussions of the $m$-mode formulation, gauge issues and metric completion. 

BL gratefully acknowledges funding from the European Union’s Horizon Europe research and innovation programme under the Marie Sklodowska-Curie grant agreement No. 101209791.
SDU and ZN acknowledge the support of a Royal Society University Research Fellowship and the ERC Consolidator/UKRI Frontier Research Grant GWModels (selected by the ERC and funded by UKRI [grant number EP/Y008251/1]).
PB acknowledges support from the Dutch Research Council (NWO) with file number OCENW.M.21.119.
KC acknowledges support from the Irish
Research Council under grant number GOIPG/2021/725, and from STFC grant number ST/B001170/1. 
SRD acknowledges financial support from the Science and Technology Facilities Council (STFC) under Grant No. ST/X000621/1. 
AS gratefully acknowledges funding from the European Union's Horizon Europe research and innovation programme under the Marie Sklodowska-Curie grant agreement No. 101199153. 
The Center of Gravity is a Center of Excellence funded by the Danish National Research Foundation under grant No.\ 184. 
CD is funded by the European Union (ERC Starting Grant, Unleash-TDEs, project number 101163093). 
CD and  MvdM \ acknowledge support by VILLUM Foundation (grant no.\ VIL37766 and no.\ VIL53101)
and the DNRF Chair program (grant no.\ DNRF162) by the Danish National Research
Foundation. 
MvdM is  supported by the European Union’s Horizon ERC Synergy Grant ``Making Sense of the Unexpected in the Gravitational-Wave Sky'' (GWSky--101167314).
This work is supported by ERC grant EMRIWaveforms (DOI: \href{https://doi.org/10.3030/101200625}{10.3030/101200625}).
Views and opinions expressed are those of the authors only and do not necessarily reflect those of the European Union, the European Research Council, or any of the granting agencies. Neither the European Union nor the granting authority can be held responsible for them. 

This work makes use of the \texttt{Black Hole Perturbation Toolkit}~\cite{BHPToolkit} (\texttt{Teukolsky}~\cite{BHPToolkit:Teukolsky}, \texttt{SpinWeightedSpheroidalHarmonics}~\cite{BHPToolkit:SpinWeightedSpheroidalHarmonics}), Wolfram Mathematica~\cite{Mathematica} and xAct~\cite{xAct,Martin-Garcia:2008ysv,xCoba}.

\textit{AI usage:}
The formulation and conceptualization of the problem, derivations, calculations, and initial drafts of this manuscript were prepared entirely by the authors without the assistance of AI tools. Claude (Opus 5.5) was used in the late stages of the project to improve figure presentation, to assist with drafting and refining the final manuscript, and with checking equations. All AI products were independently verified by the authors, who take full responsibility for the accuracy and content of this paper.
\end{acknowledgments}

\appendix
\section{Newman--Penrose conventions}
\label{sec:newman_penrose_conventions}

We use a null tetrad $e^\mu_{a} = (\ltetrad^\mu,\ntetrad^\mu,\mtetrad^\mu,\barm^\mu)$, normalised by
\begin{equation}
    \ltetrad^{\mu}\ntetrad_{\mu} = -1,
    \qquad
    \mtetrad^{\mu}\barm_{\mu} = 1,
    \label{eq:null_tetrad}
\end{equation}
with all other inner products vanishing, so that
\begin{equation}
    g_{\mu\nu} = -2\ltetrad_{(\mu}\ntetrad_{\nu)} + 2\mtetrad_{(\mu}\barm_{\nu)},
\end{equation}
where round brackets denote symmetrisation with unit weight, $A_{(\mu}B_{\nu)}\coloneqq\tfrac12(A_\mu B_\nu+A_\nu B_\mu)$.
The dual coframe is $e^{a}_{\mu} = (-\ntetrad_{\mu}, -\ltetrad_{\mu}, \barm_{\mu}, \mtetrad_{\mu})$, so that $e^a_\mu e^\mu_b=\delta^a_b$.
Tetrad components carry Latin indices, $T_{ab} = T_{\mu\nu}e^{\mu}_{a}e^{\nu}_{b}$ and $T^a=e^a_\mu T^\mu$, and are labelled by the legs, as in $T_{\ltetrad\ltetrad}=T_{\mu\nu}\ltetrad^\mu \ltetrad^\nu$.

Under a boost $(\ltetrad^\mu,\ntetrad^\mu)\to(\gamma\,\ltetrad^\mu,\gamma^{-1}\ntetrad^\mu)$ and a spin $\mtetrad^\mu\to e^{i\vartheta}\mtetrad^\mu$, with $\gamma>0$ and $\vartheta$ real, a quantity $\eta$ of boost weight $b$ and spin weight $s$ transforms as $\eta\to\gamma^{b}e^{is\vartheta}\eta$.
The legs $\ltetrad$, $\ntetrad$, $\mtetrad$ and $\barm$ have $\{b,s\}=\{1,0\}$, $\{-1,0\}$, $\{0,1\}$ and $\{0,-1\}$.
Accordingly, the boost weight of a tetrad component is the number of its $\ltetrad$ indices minus the number of its $\ntetrad$ indices, and its spin weight is the number of its $\mtetrad$ indices minus the number of its $\barm$ indices; for example, $T_{\ltetrad\mtetrad}$ has $\{b,s\}=\{1,1\}$.
This gives the spin weights $s_A$ of the reconstruction, Sec.~\ref{sec:perturbations/reconstruction-in-lm-modes}.

In Boyer--Lindquist coordinates $(t,r,\theta,\varphi)$, the Kinnersley tetrad~\cite{Kinnersley:1969zza} is
\begin{subequations}
\begin{align}
    \ltetrad^{\mu} ={}& \frac{1}{\Delta}\left(r^2+a^2,\ \Delta,\ 0,\ a\right), \\
    \ntetrad^{\mu} ={}& \frac{1}{2\Sigma}\left(r^2+a^2,\ -\Delta,\ 0,\ a\right), \\
    \mtetrad^{\mu} ={}& \frac{i}{\sqrt{2}\,\bar\zeta}\left(a\sin\theta,\ 0,\ -i,\ \csc\theta\right), \\
    \barm^{\mu} ={}& {-}\frac{i}{\sqrt{2}\,\zeta}\left(a\sin\theta,\ 0,\ i,\ \csc\theta\right),
\end{align}
\end{subequations}
where $\zeta=r-ia\cos\theta$, so that $\zeta\bar\zeta=\Sigma$.
The covariant components are
\begin{subequations}
\begin{align}
    \ltetrad_{\mu} ={}& \left(-1,\ \frac{\Sigma}{\Delta},\ 0,\ a\sin^{2}\theta\right), \\
    \ntetrad_{\mu} ={}& \frac{1}{2\Sigma}\left(-\Delta,\ -\Sigma,\ 0,\ a\Delta\sin^{2}\theta\right), \\
    \mtetrad_{\mu} ={}& \frac{i}{\sqrt{2}\,\bar\zeta}\left(-a\sin\theta,\ 0,\ -i\Sigma,\ (r^2 + a^2)\sin\theta\right),
\end{align}
\end{subequations}
and $\barm_\mu=\overline{\mtetrad_\mu}$.

We define the spin coefficients,
\begin{subequations}\label{eq:spin_coefficients}
\begin{align}
    \kappa &= -\mtetrad^\mu \ltetrad^\nu\nabla_\nu \ltetrad_\mu, &
    \sigma &= -\mtetrad^\mu \mtetrad^\nu\nabla_\nu \ltetrad_\mu,\\
    \rho &= -\mtetrad^\mu \barm^\nu\nabla_\nu \ltetrad_\mu, &
    \tau &= -\mtetrad^\mu \ntetrad^\nu\nabla_\nu \ltetrad_\mu,\\
    \beta &= \mathrlap{\tfrac12\bigl(\barm^\mu \mtetrad^\nu\nabla_\nu \mtetrad_\mu - \ntetrad^\mu \mtetrad^\nu\nabla_\nu \ltetrad_\mu\bigr),}\\
    \epsilon &= \mathrlap{\tfrac12\bigl(\barm^\mu \ltetrad^\nu\nabla_\nu \mtetrad_\mu - \ntetrad^\mu \ltetrad^\nu\nabla_\nu \ltetrad_\mu\bigr).}
\end{align}
\end{subequations}
The primed coefficients follow from the interchange $l\leftrightarrow \ntetrad$, $\mtetrad\leftrightarrow\barm$; for example, $\rho'=-\barm^\mu \mtetrad^\nu\nabla_\nu \ntetrad_\mu$.

For the Kinnersley tetrad, $\kappa=\kappa'=\sigma=\sigma'=\epsilon=0$, and the other spin coefficients are
\begin{subequations}
\begin{alignat}{2}
    \rho &= -\frac{1}{\zeta}, &\qquad \rho' &= -\frac{\rho^2\bar\rho\Delta}{2},\\
    \tau &= -\frac{ia\rho\bar\rho\sin\theta}{\sqrt2}, & \tau' &= -\frac{ia\rho^2\sin\theta}{\sqrt2},\\
    \beta &= -\frac{\bar\rho\cot\theta}{2\sqrt2}, & \beta' &= \bar\beta+\tau',\\
    \epsilon' &= \rho'-\frac{\rho\bar\rho(r-M)}{2}. &&
\end{alignat}
\end{subequations}
The only non-vanishing Weyl scalar of the background is
\begin{equation}
    \psi_2 \coloneqq C_{\alpha\beta\gamma\delta}\,\ltetrad^\alpha \mtetrad^\beta\barm^\gamma \ntetrad^\delta = M\rho^3 .
\end{equation}

On the unit sphere, the spin-weighted angular derivatives of a quantity $\eta$ of spin weight $s$ are $\eth\eta = -(\partial_\theta + i\csc\theta\,\partial_\varphi - s\cot\theta)\eta$ and $\ethb\eta = -(\partial_\theta - i\csc\theta\,\partial_\varphi + s\cot\theta)\eta$.
On an $m$-mode, $\eta\propto e^{im\varphi}$, they become
\begin{subequations}
\begin{align}
    \eth\eta ={}& {-}\bigl(\partial_\theta-m\csc\theta-s\cot\theta\bigr)\eta, \\
    \ethb\eta ={}& {-}\bigl(\partial_\theta+m\csc\theta+s\cot\theta\bigr)\eta,
\end{align}
\end{subequations}
and, with the spin-weighted spherical harmonics of Ref.~\cite{Goldberg:1966uu},
\begin{subequations}
\begin{align}
    \eth\,\sY{s}_{\lm}&=\sqrt{(\ell-s)(\ell+s+1)}\,\sY{s+1}_{\lm},\\
    \ethb\,\sY{s}_{\lm}&=-\sqrt{(\ell+s)(\ell-s+1)}\,\sY{s-1}_{\lm}.
\end{align}
\end{subequations}

The unnormalised legs of the reconstruction are
\begin{subequations}
\begin{align}
    \ltetrad^\mu_\pm&=\Bigl(\pm\frac{r^2+a^2}{\Delta},\ 1,\ 0,\ \pm\frac{a}{\Delta}\Bigr),\\
    \mtetrad^\mu_\pm&=\bigl(\pm ia\sin\theta,\ 0,\ 1,\ \pm i\csc\theta\bigr).
\end{align}
\end{subequations}
They are related to the Kinnersley legs by the inverse of Eq.~\eqref{eq:ddkwlegs},
\begin{subequations}
\begin{align}
    \ltetrad^\mu &= \ltetrad^\mu_+, & \ntetrad^\mu &= -\frac{\Delta}{2\Sigma}\,\ltetrad^\mu_-,\\
    \mtetrad^\mu &= \frac{\mtetrad^\mu_+}{\sqrt{2}\,\bar\zeta}, & \barm^\mu &= \frac{\mtetrad^\mu_-}{\sqrt{2}\,\zeta}.
\end{align}
\end{subequations}

On the worldline, $r=\rp$ and $\theta=\pi/2$, the four-velocity $u^\mu=u^t(1,0,0,\Omega)$ has the tetrad components $u^a=e^a_\mu u^\mu$,
\begin{subequations}\label{eq:ucomponents}
\begin{align}
    u^{\ltetrad} &= -\ntetrad_\mu u^\mu = \frac{u^t\Delta_p(1-a\Omega)}{2\rp^2}, \\
    u^{\ntetrad} &= -\ltetrad_\mu u^\mu = u^t(1-a\Omega),\\
    u^{\mtetrad} &= \barm_\mu u^\mu = -\frac{iu^t[\Omega(\rp^2+a^2)-a]}{\sqrt2\,\rp},
\end{align}
\end{subequations}
and $u^{\barm}=\overline{u^{\mtetrad}}$.

In the comoving coordinates $(T,X,Y,Z)$ of Eq.~\eqref{eq:CMcoords}, the components of the Kinnersley legs, denoted by the subscript ``cm'', are
\begin{widetext}
\begin{subequations}\label{eq:NPCM}
\begin{align}
    \ltetrad^{\mu}_{\rm cm} ={}& \left(\frac{\En(r^2+a^2)-a\Lz}{\Delta},\ \frac{\rp}{\sqrt{\Delta_p}},\ 0,\ -\frac{[\Omega(r^2+a^2)-a]\,z_c\cos(\phihat/2)}{2\Delta}\right), \label{eq:NPlCM} \\
    \ntetrad^{\mu}_{\rm cm} ={}& \left(\frac{\En(r^2+a^2)-a\Lz}{2\Sigma},\ -\frac{\rp\,\Delta}{2\Sigma\sqrt{\Delta_p}},\ 0,\ -\frac{[\Omega(r^2+a^2)-a]\,z_c\cos(\phihat/2)}{4\Sigma}\right), \label{eq:NPnCM} \\
    \mtetrad^{\mu}_{\rm cm} ={}& \frac{i}{\sqrt{2}\,\bar\zeta\sin\theta}\left(a\En\sin^2\theta - \Lz,\ 0,\ -i \rp\sin^2\theta,\ \frac{1}{2}(1-a\Omega\sin^2\theta)\,z_c\cos(\phihat/2)\right), \label{eq:NPmCM}
\end{align}
\end{subequations}
\end{widetext}
and $\barm^{\mu}_{\rm cm}=\overline{\mtetrad^{\mu}_{\rm cm}}$.
Here $r$, $\theta$ and $\phihat=\varphi-\Omega t$, and hence $\Delta$, $\Sigma$, $\zeta$ and $\bar\zeta$, are viewed as functions of the comoving coordinates by inverting Eq.~\eqref{eq:CMcoords},
\begin{equation}
    r = \rp+\frac{\sqrt{\Delta_p}}{\rp}X,\qquad \cos\theta=-\frac{Y}{\rp},\qquad \frac{\phihat}{2} = \arcsin\frac{Z}{z_c}.
\end{equation}

\bibliography{ref}

@unpublished{Leather2026,
    author = "Benjamin Leather",
    title = "{Lorenz-gauge metric reconstruction in Kerr spacetime with hyperboloidal slicing and spectral methods}",
    note = {(in preparation)}
}

@article{Aksteiner:2016pjt,
    author = {Aksteiner, Steffen and Andersson, Lars and B{\"a}ckdahl, Thomas},
    title = "{New identities for linearized gravity on the Kerr spacetime}",
    eprint = "1601.06084",
    archivePrefix = "arXiv",
    primaryClass = "gr-qc",
    doi = "10.1103/PhysRevD.99.044043",
    journal = "Phys. Rev. D",
    volume = "99",
    number = "4",
    pages = "044043",
    year = "2019"
}

@article{Dyson:2025dlj,
    author = "Dyson, Conor and Spieksma, Thomas F. M. and Brito, Richard and van de Meent, Maarten and Dolan, Sam",
    title = "{Environmental Effects in Extreme-Mass-Ratio Inspirals: Perturbations to the Environment in {Kerr} Spacetimes}",
    eprint = "2501.09806",
    archivePrefix = "arXiv",
    primaryClass = "gr-qc",
    doi = "10.1103/PhysRevLett.134.211403",
    journal = "Phys. Rev. Lett.",
    volume = "134",
    number = "21",
    pages = "211403",
    year = "2025"
}

@article{Mano:1996vt,
    author = "Mano, Shuhei and Suzuki, Hisao and Takasugi, Eiichi",
    title = "{Analytic solutions of the Teukolsky equation and their low frequency expansions}",
    eprint = "gr-qc/9603020",
    archivePrefix = "arXiv",
    reportNumber = "OU-HET-238",
    doi = "10.1143/PTP.95.1079",
    journal = "Prog. Theor. Phys.",
    volume = "95",
    pages = "1079--1096",
    year = "1996"
}

@article{Mano:1996gn,
    author = "Mano, Shuhei and Takasugi, Eiichi",
    title = "{Analytic solutions of the Teukolsky equation and their properties}",
    eprint = "gr-qc/9611014",
    archivePrefix = "arXiv",
    reportNumber = "OU-HET-254",
    doi = "10.1143/PTP.97.213",
    journal = "Prog. Theor. Phys.",
    volume = "97",
    pages = "213--232",
    year = "1997"
}

@article{vandeMeent:2017bcc,
    author = "van de Meent, Maarten",
    title = "{Gravitational self-force on generic bound geodesics in Kerr spacetime}",
    eprint = "1711.09607",
    archivePrefix = "arXiv",
    primaryClass = "gr-qc",
    doi = "10.1103/PhysRevD.97.104033",
    journal = "Phys. Rev. D",
    volume = "97",
    number = "10",
    pages = "104033",
    year = "2018"
}

@misc{Dyson:2026ddd,
    author = "Dyson, Conor and D'Orazio, Daniel J.",
    title = "{Spiral Density Waves and Torque Balance in the Kerr Geometry}",
    eprint = "2601.19123",
    archivePrefix = "arXiv",
    primaryClass = "gr-qc",
    month = "1",
    year = "2026"
}

@article{Li:2025ffh,
    author = "Li, Dongjun and Weller, Colin and Bourg, Patrick and LaHaye, Michael and Yunes, Nicol{\'a}s and Yang, Huan",
    title = "{Extreme mass-ratio inspiral within an ultralight scalar cloud: Scalar radiation}",
    eprint = "2507.02045",
    archivePrefix = "arXiv",
    primaryClass = "gr-qc",
    doi = "10.1103/7l9s-g21j",
    journal = "Phys. Rev. D",
    volume = "112",
    number = "8",
    pages = "084057",
    year = "2025"
}

@article{Thornburg:2016msc,
    author = "Thornburg, Jonathan and Wardell, Barry",
    title = "{Scalar self-force for highly eccentric equatorial orbits in {Kerr} spacetime}",
    eprint = "1610.09319",
    archivePrefix = "arXiv",
    primaryClass = "gr-qc",
    doi = "10.1103/PhysRevD.95.084043",
    journal = "Phys. Rev. D",
    volume = "95",
    number = "8",
    pages = "084043",
    year = "2017"
}

@article{Warburton:2013lea,
    author = "Warburton, Niels and Wardell, Barry",
    title = "{Applying the effective-source approach to frequency-domain self-force calculations}",
    eprint = "1311.3104",
    archivePrefix = "arXiv",
    primaryClass = "gr-qc",
    doi = "10.1103/PhysRevD.89.044046",
    journal = "Phys. Rev. D",
    volume = "89",
    number = "4",
    pages = "044046",
    year = "2014"
}

@article{Wardell:2015ada,
    author = "Wardell, Barry and Warburton, Niels",
    title = "{Applying the effective-source approach to frequency-domain self-force calculations: Lorenz-gauge gravitational perturbations}",
    eprint = "1505.07841",
    archivePrefix = "arXiv",
    primaryClass = "gr-qc",
    doi = "10.1103/PhysRevD.92.084019",
    journal = "Phys. Rev. D",
    volume = "92",
    number = "8",
    pages = "084019",
    year = "2015"
}

@article{Bourg:2024cgh,
    author = "Bourg, Patrick and Pound, Adam and Upton, Samuel D. and Panosso Macedo, Rodrigo",
    title = "{Simple, efficient method of calculating the Detweiler-Whiting singular field to very high order}",
    eprint = "2404.10082",
    archivePrefix = "arXiv",
    primaryClass = "gr-qc",
    doi = "10.1103/PhysRevD.110.084007",
    journal = "Phys. Rev. D",
    volume = "110",
    number = "8",
    pages = "084007",
    year = "2024"
}

@article{Wardell:2024yoi,
    author = "Wardell, Barry and Kavanagh, Chris and Dolan, Sam R.",
    title = "{Sourced metric perturbations of Kerr spacetime in Lorenz gauge}",
    eprint = "2406.12510",
    archivePrefix = "arXiv",
    primaryClass = "gr-qc",
    doi = "10.1088/1361-6382/ae0918",
    journal = "Class. Quant. Grav.",
    volume = "42",
    number = "20",
    pages = "205007",
    year = "2025"
}

@article{Dolan:2023enf,
    author = "Dolan, Sam R. and Durkan, Leanne and Kavanagh, Chris and Wardell, Barry",
    title = "{Metric perturbations of Kerr spacetime in Lorenz gauge: circular equatorial orbits}",
    eprint = "2306.16459",
    archivePrefix = "arXiv",
    primaryClass = "gr-qc",
    doi = "10.1088/1361-6382/ad52e3",
    journal = "Class. Quant. Grav.",
    volume = "41",
    number = "15",
    pages = "155011",
    year = "2024"
}

@article{Dolan:2021ijg,
    author = "Dolan, Sam R. and Kavanagh, Chris and Wardell, Barry",
    title = "{Gravitational Perturbations of Rotating Black Holes in Lorenz Gauge}",
    eprint = "2108.06344",
    archivePrefix = "arXiv",
    primaryClass = "gr-qc",
    doi = "10.1103/PhysRevLett.128.151101",
    journal = "Phys. Rev. Lett.",
    volume = "128",
    number = "15",
    pages = "151101",
    year = "2022"
}

@article{Heffernan:2012su,
    author = "Heffernan, Anna and Ottewill, Adrian and Wardell, Barry",
    title = "{High-order expansions of the Detweiler-Whiting singular field in Schwarzschild spacetime}",
    eprint = "1204.0794",
    archivePrefix = "arXiv",
    primaryClass = "gr-qc",
    doi = "10.1103/PhysRevD.86.104023",
    journal = "Phys. Rev. D",
    volume = "86",
    pages = "104023",
    year = "2012"
}

@article{Pound:2014xva,
    author = "Pound, Adam and Miller, Jeremy",
    title = "{Practical, covariant puncture for second-order self-force calculations}",
    eprint = "1403.1843",
    archivePrefix = "arXiv",
    primaryClass = "gr-qc",
    doi = "10.1103/PhysRevD.89.104020",
    journal = "Phys. Rev. D",
    volume = "89",
    number = "10",
    pages = "104020",
    year = "2014"
}

@article{Ottewill:2008uu,
    author = "Ottewill, Adrian C. and Wardell, Barry",
    title = "{Quasi-local contribution to the scalar self-force: Non-geodesic Motion}",
    eprint = "0810.1961",
    archivePrefix = "arXiv",
    primaryClass = "gr-qc",
    doi = "10.1103/PhysRevD.79.024031",
    journal = "Phys. Rev. D",
    volume = "79",
    pages = "024031",
    year = "2009"
}

@Book{Grad14,
  author       = {I. S. Gradshteyn and I. M. Ryzhik},
  editor       = {Daniel Zwillinger and Victor Moll},
  publisher    = {Elsevier},
  title        = {Table of Integrals, Series, and Products},
  year         = {2014},
  edition      = {8},
  isbn         = {9780123849335},
  doi          = {10.1016/c2010-0-64839-5},
  translator   = {{Scripta Technica, Inc}},
}

@article{Vega:2009qb,
    author = "Vega, Ian and Diener, Peter and Tichy, Wolfgang and Detweiler, Steven L.",
    title = "{Self-force with $(3+1)$ codes: A Primer for numerical relativists}",
    eprint = "0908.2138",
    archivePrefix = "arXiv",
    primaryClass = "gr-qc",
    doi = "10.1103/PhysRevD.80.084021",
    journal = "Phys. Rev. D",
    volume = "80",
    pages = "084021",
    year = "2009"
}

@article{Toomani:2021jlo,
	archiveprefix = {arXiv},
	author = {Toomani, Vahid and Zimmerman, Peter and Spiers, Andrew and Hollands, Stefan and Pound, Adam and Green, Stephen R.},
	doi = {10.1088/1361-6382/ac37a5},
	eprint = {2108.04273},
	journal = {Class. Quant. Grav.},
	number = {1},
	pages = {015019},
	primaryclass = {gr-qc},
	title = {{New metric reconstruction scheme for gravitational self-force calculations}},
	volume = {39},
	year = {2022}}

@article{Dolan:2012jg,
	archiveprefix = {arXiv},
	author = {Dolan, Sam R. and Barack, Leor},
	doi = {10.1103/PhysRevD.87.084066},
	eprint = {1211.4586},
	journal = {Phys. Rev. D},
	pages = {084066},
	primaryclass = {gr-qc},
	title = {{Self-force via $m$-mode regularization and 2+1D evolution: III. Gravitational field on Schwarzschild spacetime}},
	volume = {87},
	year = {2013}}

@article{Dolan:2014pja,
	archiveprefix = {arXiv},
	author = {Dolan, Sam R. and Nolan, Patrick and Ottewill, Adrian C. and Warburton, Niels and Wardell, Barry},
	doi = {10.1103/PhysRevD.91.023009},
	eprint = {1406.4890},
	journal = {Phys. Rev. D},
	number = {2},
	pages = {023009},
	primaryclass = {gr-qc},
	title = {{Tidal invariants for compact binaries on quasicircular orbits}},
	volume = {91},
	year = {2015}}

@article{Sago:2008id,
	archiveprefix = {arXiv},
	author = {Sago, Norichika and Barack, Leor and Detweiler, Steven L.},
	doi = {10.1103/PhysRevD.78.124024},
	eprint = {0810.2530},
	journal = {Phys. Rev. D},
	pages = {124024},
	primaryclass = {gr-qc},
	title = {{Two approaches for the gravitational self force in black hole spacetime: Comparison of numerical results}},
	volume = {78},
	year = {2008}}

@article{Akcay:2012ea,
	archiveprefix = {arXiv},
	author = {Akcay, Sarp and Barack, Leor and Damour, Thibault and Sago, Norichika},
	doi = {10.1103/PhysRevD.86.104041},
	eprint = {1209.0964},
	journal = {Phys. Rev. D},
	pages = {104041},
	primaryclass = {gr-qc},
	title = {{Gravitational self-force and the effective-one-body formalism between the innermost stable circular orbit and the light ring}},
	volume = {86},
	year = {2012}}

@article{Detweiler:2008ft,
	archiveprefix = {arXiv},
	author = {Detweiler, Steven L.},
	doi = {10.1103/PhysRevD.77.124026},
	eprint = {0804.3529},
	journal = {Phys. Rev. D},
	pages = {124026},
	primaryclass = {gr-qc},
	title = {{A Consequence of the gravitational self-force for circular orbits of the Schwarzschild geometry}},
	volume = {77},
	year = {2008}}

@article{Shah:2012gu,
	archiveprefix = {arXiv},
	author = {Shah, Abhay G. and Friedman, John L. and Keidl, Tobias S.},
	doi = {10.1103/PhysRevD.86.084059},
	eprint = {1207.5595},
	journal = {Phys. Rev. D},
	pages = {084059},
	primaryclass = {gr-qc},
	title = {{EMRI corrections to the angular velocity and redshift factor of a mass in circular orbit about a Kerr black hole}},
	volume = {86},
	year = {2012}}

@article{Mino:1996nk,
	archiveprefix = {arXiv},
	author = {Mino, Yasushi and Sasaki, Misao and Tanaka, Takahiro},
	doi = {10.1103/PhysRevD.55.3457},
	eprint = {gr-qc/9606018},
	journal = {Phys. Rev. D},
	pages = {3457--3476},
	reportnumber = {OU-TAP-38, KUNS-1394},
	title = {{Gravitational radiation reaction to a particle motion}},
	volume = {55},
	year = {1997}}

@article{Detweiler:2002mi,
	archiveprefix = {arXiv},
	author = {Detweiler, Steven L. and Whiting, Bernard F.},
	doi = {10.1103/PhysRevD.67.024025},
	eprint = {gr-qc/0202086},
	journal = {Phys. Rev. D},
	pages = {024025},
	title = {{Selfforce via a Green's function decomposition}},
	volume = {67},
	year = {2003}}

@phdthesis{Berndtson:2007gsc,
	archiveprefix = {arXiv},
	author = {Berndtson, Mark V.},
	eprint = {0904.0033},
	primaryclass = {gr-qc},
	school = {University of Colorado},
	title = {{Harmonic gauge perturbations of the Schwarzschild metric}},
	year = {2007}}

@article{Durkan:2022fvm,
	archiveprefix = {arXiv},
	author = {Durkan, Leanne and Warburton, Niels},
	doi = {10.1103/PhysRevD.106.084023},
	eprint = {2206.08179},
	journal = {Phys. Rev. D},
	number = {8},
	pages = {084023},
	primaryclass = {gr-qc},
	title = {{Slow evolution of the metric perturbation due to a quasicircular inspiral into a Schwarzschild black hole}},
	volume = {106},
	year = {2022}}

@article{Barack:2007tm,
	archiveprefix = {arXiv},
	author = {Barack, Leor and Sago, Norichika},
	doi = {10.1103/PhysRevD.75.064021},
	eprint = {gr-qc/0701069},
	journal = {Phys. Rev. D},
	pages = {064021},
	title = {{Gravitational self force on a particle in circular orbit around a Schwarzschild black hole}},
	volume = {75},
	year = {2007}}

@article{Warburton:2021kwk,
	archiveprefix = {arXiv},
	author = {Warburton, Niels and Pound, Adam and Wardell, Barry and Miller, Jeremy and Durkan, Leanne},
	doi = {10.1103/PhysRevLett.127.151102},
	eprint = {2107.01298},
	journal = {Phys. Rev. Lett.},
	number = {15},
	pages = {151102},
	primaryclass = {gr-qc},
	title = {{Gravitational-Wave Energy Flux for Compact Binaries through Second Order in the Mass Ratio}},
	volume = {127},
	year = {2021}}

@article{Wardell:2021fyy,
	archiveprefix = {arXiv},
	author = {Wardell, Barry and Pound, Adam and Warburton, Niels and Miller, Jeremy and Durkan, Leanne and Le Tiec, Alexandre},
	doi = {10.1103/PhysRevLett.130.241402},
	eprint = {2112.12265},
	journal = {Phys. Rev. Lett.},
	number = {24},
	pages = {241402},
	primaryclass = {gr-qc},
	title = {{Gravitational Waveforms for Compact Binaries from Second-Order Self-Force Theory}},
	volume = {130},
	year = {2023}}

@article{Pound:2019lzj,
	archiveprefix = {arXiv},
	author = {Pound, Adam and Wardell, Barry and Warburton, Niels and Miller, Jeremy},
	doi = {10.1103/PhysRevLett.124.021101},
	eprint = {1908.07419},
	journal = {Phys. Rev. Lett.},
	number = {2},
	pages = {021101},
	primaryclass = {gr-qc},
	title = {{Second-Order Self-Force Calculation of Gravitational Binding Energy in Compact Binaries}},
	volume = {124},
	year = {2020}}

@article{Campanelli:1998jv,
    author = "Campanelli, Manuela and Lousto, Carlos O.",
    title = "{Second order gauge invariant gravitational perturbations of a Kerr black hole}",
    eprint = "gr-qc/9811019",
    archivePrefix = "arXiv",
    reportNumber = "AEI-096",
    doi = "10.1103/PhysRevD.59.124022",
    journal = "Phys. Rev. D",
    volume = "59",
    pages = "124022",
    year = "1999"
}

@article{Pound:2015wva,
	archiveprefix = {arXiv},
	author = {Pound, Adam},
	doi = {10.1103/PhysRevD.92.104047},
	eprint = {1510.05172},
	journal = {Phys. Rev. D},
	number = {10},
	pages = {104047},
	primaryclass = {gr-qc},
	title = {{Second-order perturbation theory: problems on large scales}},
	volume = {92},
	year = {2015}}

@misc{BHPToolkit,
	howpublished = {(\href{https://bhptoolkit.org/}{bhptoolkit.org})},
	title = {{Black Hole Perturbation Toolkit}}}

@article{Hopper:2012ty,
	archiveprefix = {arXiv},
	author = {Hopper, Seth and Evans, Charles R.},
	doi = {10.1103/PhysRevD.87.064008},
	eprint = {1210.7969},
	journal = {Phys. Rev. D},
	number = {6},
	pages = {064008},
	primaryclass = {gr-qc},
	reportnumber = {AEI-2012-190},
	title = {{Metric perturbations from eccentric orbits on a Schwarzschild black hole: I. Odd-parity Regge-Wheeler to Lorenz gauge transformation and two new methods to circumvent the Gibbs phenomenon}},
	volume = {87},
	year = {2013}}

@article{Akcay:2010dx,
	archiveprefix = {arXiv},
	author = {Akcay, Sarp},
	doi = {10.1103/PhysRevD.83.124026},
	eprint = {1012.5860},
	journal = {Phys. Rev. D},
	pages = {124026},
	primaryclass = {gr-qc},
	title = {{A Fast Frequency-Domain Algorithm for Gravitational Self-Force: I. Circular Orbits in Schwarzschild Spacetime}},
	volume = {83},
	year = {2011}}

@article{Leather:2024mls,
	archiveprefix = {arXiv},
	author = {Leather, Benjamin},
	eprint = {2411.14976},
	journal = {Gen. Rel. Grav.},
	pages = {112},
	primaryclass = {gr-qc},
	title = {{Gravitational self-force with hyperboloidal slicing and spectral methods}},
	volume = {57},
	year = {2025}}

@article{Teukolsky:1973ha,
	author = {Teukolsky, S. A.},
	doi = {10.1086/152444},
	journal = {Astrophys. J.},
	pages = {635--647},
	title = {{Perturbations of a rotating black hole. 1. Fundamental equations for gravitational, electromagnetic, and neutrino-field perturbations}},
	volume = {185},
	year = {1973}}

@article{vandeMeent:2015lxa,
  archiveprefix = {arXiv},
  author        = {van de Meent, Maarten and Shah, Abhay G.},
  doi           = {10.1103/PhysRevD.92.064025},
  eprint        = {1506.04755},
  journal       = {Phys. Rev. D},
  number        = {6},
  pages         = {064025},
  primaryclass  = {gr-qc},
  title         = {{Metric perturbations produced by eccentric equatorial orbits around a Kerr black hole}},
  volume        = {92},
  year          = {2015}
}

@InCollection{Pound:2021qin,
  author        = {Adam Pound and Barry Wardell},
  booktitle     = {Handbook of Gravitational Wave Astronomy},
  publisher     = {Springer},
  title         = {Black hole perturbation theory and gravitational self-force},
  year          = {2022},
  address       = {Singapore},
  editor        = {Cosimo Bambi and Stavros Katsanevas and Konstantinos D. Kokkotas},
  month         = mar,
  archiveprefix = {arXiv},
  doi           = {10.1007/978-981-15-4702-7_38-1},
  eprint        = {2101.04592},
  primaryclass  = {gr-qc},
}

@article{vandeMeent:2016pee,
  archiveprefix = {arXiv},
  author        = {van de Meent, Maarten},
  doi           = {10.1103/PhysRevD.94.044034},
  eprint        = {1606.06297},
  journal       = {Phys. Rev. D},
  number        = {4},
  pages         = {044034},
  primaryclass  = {gr-qc},
  title         = {{Gravitational self-force on eccentric equatorial orbits around a Kerr black hole}},
  volume        = {94},
  year          = {2016}
}

@misc{Nasipak_pybhpt_2026,
    author = {Nasipak, Zachary},
    license = {GPL-3.0},
    month = sep,
    title = {{pybhpt}},
    url = {https://github.com/znasipak/pybhpt},
    doi = {10.5281/zenodo.18317623},
    version = {v1.0.0},
    year = {2026}
}

@article{Nasipak:2025tby,
    author = "Nasipak, Zachary",
    title = "{Metric reconstruction and the Hamiltonian for eccentric, precessing binaries in the small-mass-ratio limit}",
    eprint = "2507.07746",
    archivePrefix = "arXiv",
    primaryClass = "gr-qc",
    doi = "10.1103/7zkx-vbg9",
    journal = "Phys. Rev. D",
    volume = "113",
    number = "12",
    pages = "124051",
    year = "2026"
}

@article{Kinnersley:1969zza,
    author = "Kinnersley, William",
    title = "{Type D Vacuum Metrics}",
    doi = "10.1063/1.1664958",
    journal = "J. Math. Phys.",
    volume = "10",
    pages = "1195--1203",
    year = "1969"
}

@article{Dolan:2010mt,
    author = "Dolan, Sam R. and Barack, Leor",
    title = "{Self force via m-mode regularization and 2+1D evolution: Foundations and a scalar-field implementation on Schwarzschild}",
    eprint = "1010.5255",
    archivePrefix = "arXiv",
    primaryClass = "gr-qc",
    doi = "10.1103/PhysRevD.83.024019",
    journal = "Phys. Rev. D",
    volume = "83",
    pages = "024019",
    year = "2011"
}

@article{Barsanti:2026ulr,
    author = "Barsanti, Susanna and Burke, Ollie and Maselli, Andrea and Sotiriou, Thomas P. and Spiers, Andrew and Warburton, Niels",
    title = "{The significance of first post-adiabatic contributions for scalar charge measurements with intermediate and extreme mass ratio inspirals}",
    eprint = "2607.09310",
    archivePrefix = "arXiv",
    primaryClass = "gr-qc",
    month = "7",
    year = "2026",
    journal = ""
}

@article{Spiers:2023cva,
    author = "Spiers, Andrew and Maselli, Andrea and Sotiriou, Thomas P.",
    title = "{Measuring scalar charge with compact binaries: High accuracy modeling with self-force}",
    eprint = "2310.02315",
    archivePrefix = "arXiv",
    primaryClass = "gr-qc",
    doi = "10.1103/PhysRevD.109.064022",
    journal = "Phys. Rev. D",
    volume = "109",
    number = "6",
    pages = "064022",
    year = "2024"
}

@article{Dolan:2011dx,
    author = "Dolan, Sam R. and Barack, Leor and Wardell, Barry",
    title = "{Self force via m-mode regularization and 2+1D evolution: II. Scalar-field implementation on Kerr spacetime}",
    eprint = "1107.0012",
    archivePrefix = "arXiv",
    primaryClass = "gr-qc",
    doi = "10.1103/PhysRevD.84.084001",
    journal = "Phys. Rev. D",
    volume = "84",
    pages = "084001",
    year = "2011"
}

@article{Spiers:2023cip,
    author = "Spiers, Andrew and Pound, Adam and Wardell, Barry",
    title = "{Second-order perturbations of Kerr black holes: Formalism and reconstruction of the first-order metric}",
    eprint = "2306.17847",
    archivePrefix = "arXiv",
    primaryClass = "gr-qc",
    doi = "10.1103/PhysRevD.110.064030",
    journal = "Phys. Rev. D",
    volume = "110",
    number = "6",
    pages = "064030",
    year = "2024"
}

@article{bondi1962gravitational,
    author = "Bondi, H. and van der Burg, M. G. J. and Metzner, A. W. K.",
    title = "{Gravitational waves in general relativity. 7. Waves from axisymmetric isolated systems}",
    doi = "10.1098/rspa.1962.0161",
    journal = "Proc. Roy. Soc. Lond. A",
    volume = "269",
    pages = "21--52",
    year = "1962"
}

@article{kegeles1979constructive,
    author = "Kegeles, L. S. and Cohen, J. M.",
    title = "{Constructive procedure for perturbations of space-times}",
    doi = "10.1103/PhysRevD.19.1641",
    journal = "Phys. Rev. D",
    volume = "19",
    pages = "1641--1664",
    year = "1979"
}

@article{cohen1975space,
  title={Space-time perturbations},
  author={J M Cohen and L S Kegeles},
  journal={Physics Letters A},
  volume={54},
  number={1},
  pages={5--7},
  year={1975},
  publisher={Elsevier},
  doi = "10.1016/0375-9601(75)90583-6"
}

@article{chrzanowski1975vector,
    author = "Chrzanowski, P. L.",
    title = "{Vector Potential and Metric Perturbations of a Rotating Black Hole}",
    doi = "10.1103/PhysRevD.11.2042",
    journal = "Phys. Rev. D",
    volume = "11",
    pages = "2042--2062",
    year = "1975"
}

@article{Bucciotti:2024zyp,
    author = "Bucciotti, Bruno and Juliano, Leonardo and Kuntz, Adrien and Trincherini, Enrico",
    title = "{Quadratic quasinormal modes of a Schwarzschild black hole}",
    eprint = "2405.06012",
    archivePrefix = "arXiv",
    primaryClass = "gr-qc",
    doi = "10.1103/PhysRevD.110.104048",
    journal = "Phys. Rev. D",
    volume = "110",
    number = "10",
    pages = "104048",
    year = "2024"
}

@article{newman1962behavior,
    author = "Newman, Ezra T. and Unti, Theodore W. J.",
    title = "{Behavior of Asymptotically Flat Empty Spaces}",
    doi = "10.1063/1.1724303",
    journal = "J. Math. Phys.",
    volume = "3",
    number = "5",
    pages = "891",
    year = "1962"
}

@article{Leather:2026zhl,
    author = "Leather, Benjamin and Spiers, Andrew and Pound, Adam and Upton, Samuel D. and Wardell, Barry and Durkan, Leanne and Warburton, Niels",
    title = "{Second-order Teukolsky calculations for nonspinning, quasicircular binaries}",
    eprint = "2609.08995",
    archivePrefix = "arXiv",
    primaryClass = "gr-qc",
    month = "9",
    year = "2026",
    journal = ""
}

@article{maselli2020detecting,
  title={Detecting scalar fields with extreme mass ratio inspirals},
  author={Maselli, Andrea and Franchini, Nicola and Gualtieri, Leonardo and Sotiriou, Thomas P},
  journal={Physical Review Letters},
  volume={125},
  number={14},
  pages={141101},
  year={2020},
  publisher={APS}
}

@article{Hollands:2024iqp,
    author = "Hollands, Stefan and Toomani, Vahid",
    title = "{Metric reconstruction in Kerr spacetime}",
    eprint = "2405.18604",
    archivePrefix = "arXiv",
    primaryClass = "gr-qc",
    doi = "10.1088/1361-6382/ad87a1",
    journal = "Class. Quant. Grav.",
    volume = "43",
    number = "5",
    pages = "055001",
    year = "2026"
}

@inproceedings{barnich2013note,
  title={A Note on the {Newman-Unti} group and the {BMS} charge algebra in terms of {Newman-Penrose} coefficients},
  author={Barnich, Glenn and Lambert, Pierre-Henry},
  booktitle={Journal of Physics: Conference Series},
  volume={410},
  number={1},
  pages={012142},
  year={2013}
}

@article{Spiers:2026yqx,
    author = "Spiers, Andrew and Pound, Adam and Moxon, Jordan",
    title = "{The Bondi--Sachs gauge, BMS frames, and memory in black hole perturbation theory}",
    eprint = "2606.24816",
    archivePrefix = "arXiv",
    primaryClass = "gr-qc",
    month = "6",
    year = "2026",
    journal = ""
}

@article{sachs1962BMS-fav,
    author = "Sachs, R.",
    title = "{Asymptotic symmetries in gravitational theory}",
    doi = "10.1103/PhysRev.128.2851",
    journal = "Phys. Rev.",
    volume = "128",
    pages = "2851--2864",
    year = "1962"
}

@article{spiers2023second,
    author = "Spiers, Andrew and Pound, Adam and Moxon, Jordan",
    title = "{Second-order Teukolsky formalism in Kerr spacetime: Formulation and nonlinear source}",
    eprint = "2305.19332",
    archivePrefix = "arXiv",
    primaryClass = "gr-qc",
    doi = "10.1103/PhysRevD.108.064002",
    journal = "Phys. Rev. D",
    volume = "108",
    number = "6",
    pages = "064002",
    year = "2023"
}

@article{Kavanagh:2016idg,
    author = "Kavanagh, Chris and Ottewill, Adrian C. and Wardell, Barry",
    title = "{Analytical high-order post-Newtonian expansions for spinning extreme mass ratio binaries}",
    eprint = "1601.03394",
    archivePrefix = "arXiv",
    primaryClass = "gr-qc",
    doi = "10.1103/PhysRevD.93.124038",
    journal = "Phys. Rev. D",
    volume = "93",
    number = "12",
    pages = "124038",
    year = "2016"
}

@article{Vega:2007mc,
    author = "Vega, Ian and Detweiler, Steven L.",
    title = "{Regularization of fields for self-force problems in curved spacetime: Foundations and a time-domain application}",
    eprint = "0712.4405",
    archivePrefix = "arXiv",
    primaryClass = "gr-qc",
    doi = "10.1103/PhysRevD.77.084008",
    journal = "Phys. Rev. D",
    volume = "77",
    pages = "084008",
    year = "2008"
}

@article{Barack:2018yvs,
    author = "Barack, Leor and Pound, Adam",
    title = "{Self-force and radiation reaction in general relativity}",
    eprint = "1805.10385",
    archivePrefix = "arXiv",
    primaryClass = "gr-qc",
    doi = "10.1088/1361-6633/aae552",
    journal = "Rept. Prog. Phys.",
    volume = "82",
    number = "1",
    pages = "016904",
    year = "2019"
}

@article{Hinderer:2008dm,
  archiveprefix = {arXiv},
  author        = {Hinderer, Tanja and Flanagan, Eanna E.},
  doi           = {10.1103/PhysRevD.78.064028},
  eprint        = {0805.3337},
  journal       = {Phys.Rev.},
  pages         = {064028},
  primaryclass  = {gr-qc},
  slaccitation  = {%%CITATION = ARXIV:0805.3337;%%},
  title         = {{Two timescale analysis of extreme mass ratio inspirals
                   in Kerr. I. Orbital Motion}},
  volume        = {D78},
  year          = {2008}
}

@article{Barack:2007we,
    author = "Barack, Leor and Golbourn, Darren A. and Sago, Norichika",
    title = "{m-Mode Regularization Scheme for the Self Force in Kerr Spacetime}",
    eprint = "0709.4588",
    archivePrefix = "arXiv",
    primaryClass = "gr-qc",
    doi = "10.1103/PhysRevD.76.124036",
    journal = "Phys. Rev. D",
    volume = "76",
    pages = "124036",
    year = "2007"
}

@article{Osburn:2022bby,
    author = "Osburn, Thomas and Nishimura, Nami",
    title = "{New self-force method via elliptic partial differential equations for Kerr inspiral models}",
    eprint = "2206.07031",
    archivePrefix = "arXiv",
    primaryClass = "gr-qc",
    doi = "10.1103/PhysRevD.106.044056",
    journal = "Phys. Rev. D",
    volume = "106",
    number = "4",
    pages = "044056",
    year = "2022"
}

@article{Osburn:2026uct,
    author = "Osburn, Thomas and Wardell, Barry and Battaglia, Erin",
    title = "{Schwarzschild perturbations in Lorenz gauge via elliptic differential equations}",
    eprint = "2608.07934",
    archivePrefix = "arXiv",
    primaryClass = "gr-qc",
    month = "8",
    year = "2026",
    journal = ""
}

@article{Vu:2026ypc,
    author = "Vu, Nils L. and Nishimura, Nami and Osburn, Thomas and Thompson, Jonathan E. and Kidder, Lawrence E. and Upton, Samuel D. and Wardell, Barry",
    title = "{Self-force calculations with numerical relativity methods}",
    eprint = "2606.04998",
    archivePrefix = "arXiv",
    primaryClass = "gr-qc",
    doi = "10.1103/lbyk-26gh",
    journal = "Phys. Rev. D",
    volume = "114",
    number = "6",
    pages = "064077",
    year = "2026"
}

@article{Upton:2025bja,
    author = "Upton, Samuel D. and Wardell, Barry and Pound, Adam and Warburton, Niels and Barack, Leor",
    title = "{Effective source for second-order self-force calculations: Quasicircular orbits in Schwarzschild spacetime}",
    eprint = "2508.00087",
    archivePrefix = "arXiv",
    primaryClass = "gr-qc",
    doi = "10.1103/f9j2-k64r",
    journal = "Phys. Rev. D",
    volume = "113",
    number = "6",
    pages = "064013",
    year = "2026"
}

@article{Heffernan:2022zgc,
    author = "Heffernan, Anna",
    title = "{Self-force regularization of a point particle for generic orbits in Kerr spacetime: electromagnetic and gravitational cases}",
    eprint = "2209.05450",
    archivePrefix = "arXiv",
    primaryClass = "gr-qc",
    month = "9",
    year = "2022",
    journal = ""
}

@article{Upton:2023tcv,
    author = "Upton, Samuel D.",
    title = "{Second-order gravitational self-force in a highly regular gauge: Covariant and coordinate punctures}",
    eprint = "2309.03778",
    archivePrefix = "arXiv",
    primaryClass = "gr-qc",
    doi = "10.1103/PhysRevD.109.044021",
    journal = "Phys. Rev. D",
    volume = "109",
    number = "4",
    pages = "044021",
    year = "2024"
}

@article{Warburton:2010eq,
    author = "Warburton, Niels and Barack, Leor",
    title = "{Self force on a scalar charge in Kerr spacetime: circular equatorial orbits}",
    eprint = "1003.1860",
    archivePrefix = "arXiv",
    primaryClass = "gr-qc",
    doi = "10.1103/PhysRevD.81.084039",
    journal = "Phys. Rev. D",
    volume = "81",
    pages = "084039",
    year = "2010"
}

@phdthesis{Cunningham_2026, title={Contributions to Gravitational Waveform Models for Intermediate and Extreme Mass-Ratio Inspirals}, publisher={University College Dublin. School of Mathematics and Statistics},
author={Cunningham, Kevin Oisín}, year={2026},
school = "{University College Dublin}",
 url = {http://hdl.handle.net/10197/32257}
}

@article{Hughes:1999bq,
    author = "Hughes, Scott A.",
    title = "{The Evolution of circular, nonequatorial orbits of Kerr black holes due to gravitational wave emission}",
    eprint = "gr-qc/9910091",
    archivePrefix = "arXiv",
    doi = "10.1103/PhysRevD.61.084004",
    journal = "Phys. Rev. D",
    volume = "61",
    number = "8",
    pages = "084004",
    year = "2000",
    note = "[Erratum: Phys. Rev. D 63, 049902 (2001); ibid. 65, 069902 (2002); ibid. 67, 089901 (2003); ibid. 78, 109902 (2008); ibid. 90, 109904 (2014)]"
}

@misc{xAct,
  author       = {Jos{\'{e}} M. Mart{\'{i}}n-Garc{\'{i}}a},
  howpublished = {\url{https://xact.es/}},
  title        = {{xAct}: Efficient tensor computer algebra for the {Wolfram Language}},
  year  = "2026"
}

@misc{xCoba,
  author       = {David Yllanes and Jos{\'{e}} M. Mart{\'{i}}n-Garc{\'{i}}a},
  howpublished = {\url{https://xact.es/xCoba/}},
  title        = {{xCoba}: General component tensor computer algebra},
  year  = "2026"
}

@article{Martin-Garcia:2008ysv,
    author = "Mart{\'\i}n-Garc{\'\i}a, Jos{\'e} M.",
    title = "{xPerm: fast index canonicalization for tensor computer algebra}",
    eprint = "0803.0862",
    archivePrefix = "arXiv",
    primaryClass = "cs.SC",
    doi = "10.1016/j.cpc.2008.05.009",
    journal = "Comput. Phys. Commun.",
    volume = "179",
    number = "8",
    pages = "597--603",
    year = "2008"
}

@misc{Mathematica,
  author = "{Wolfram Research{,} Inc.}",
  title = {Mathematica, {V}ersion 15.0},
  url = {https://www.wolfram.com/mathematica},
  note = "{Champaign, IL}",
  year = "2026"
}

@misc{DLMF,
  key          = "{\relax DLMF}",
  title        = "{\it NIST Digital Library of Mathematical Functions}",
  howpublished = "\url{https://dlmf.nist.gov/}, Release 1.2.8 of 2026-09-15",
  url          = "https://dlmf.nist.gov/",
  note         = "{F}.~W.~J. Olver, A.~B. {Olde Daalhuis}, D.~W. Lozier, B.~I. Schneider,
                  R.~F. Boisvert, C.~W. Clark, B.~R. Miller, B.~V. Saunders,
                  H.~S. Cohl, and M.~A. McClain, eds.",
    year = "2026"
}

@unpublished{SpectreGR,
    author = "Nishimura, Nami and Vu, Nils L. and Gispert-Fernández, Adrià and Kelly, Cillian and Leather, Benjamin and Osburn, Thomas and Thompson, Jonathan E. and Upton, Samuel D. and Wardell, Barry",
    title = "{Kerr metric perturbations in Lorenz gauge via numerical relativity methods}",
    note = {(in preparation)}
}

@article{LISA:2024hlh,
    author = "Colpi, Monica and others",
    collaboration = "LISA",
    title = "{LISA Definition Study Report}",
    eprint = "2402.07571",
    archivePrefix = "arXiv",
    primaryClass = "astro-ph.CO",
    month = "2",
    year = "2024",
    journal = ""
}

@article{ET:2025xjr,
    author = "Abac, Adrian and others",
    collaboration = "ET",
    title = "{The Science of the Einstein Telescope}",
    eprint = "2503.12263",
    archivePrefix = "arXiv",
    primaryClass = "gr-qc",
    reportNumber = "ET-0036C-25",
    doi = "10.1088/1475-7516/2026/03/081",
    journal = "JCAP",
    volume = "2026",
    number = "03",
    pages = "081",
    year = "2026"
}

@article{Hall:2022dik,
    author = "Hall, Evan D.",
    title = "{Cosmic Explorer: A Next-Generation Ground-Based Gravitational-Wave Observatory}",
    doi = "10.3390/galaxies10040090",
    journal = "Galaxies",
    volume = "10",
    number = "4",
    pages = "90",
    year = "2022"
}

@article{LISAConsortiumWaveformWorkingGroup:2023arg,
    author = "Afshordi, Niayesh and others",
    collaboration = "LISA Consortium Waveform Working Group",
    title = "{Waveform modelling for the Laser Interferometer Space Antenna}",
    eprint = "2311.01300",
    archivePrefix = "arXiv",
    primaryClass = "gr-qc",
    doi = "10.1007/s41114-025-00056-1",
    journal = "Living Rev. Rel.",
    volume = "28",
    number = "1",
    pages = "9",
    year = "2025"
}

@article{Gupta:2023lga,
    author = "Gupta, Ish and others",
    title = "{Characterizing gravitational wave detector networks: from A$^\sharp$ to cosmic explorer}",
    eprint = "2307.10421",
    archivePrefix = "arXiv",
    primaryClass = "gr-qc",
    reportNumber = "CE Document No. P2300019, CE Document No. P2300019-v2",
    doi = "10.1088/1361-6382/ad7b99",
    journal = "Class. Quant. Grav.",
    volume = "41",
    number = "24",
    pages = "245001",
    year = "2024"
}

@article{Babak:2017tow,
    author = "Babak, Stanislav and Gair, Jonathan and Sesana, Alberto and Barausse, Enrico and Sopuerta, Carlos F. and Berry, Christopher P. L. and Berti, Emanuele and Amaro-Seoane, Pau and Petiteau, Antoine and Klein, Antoine",
    title = "{Science with the space-based interferometer LISA. V: Extreme mass-ratio inspirals}",
    eprint = "1703.09722",
    archivePrefix = "arXiv",
    primaryClass = "gr-qc",
    doi = "10.1103/PhysRevD.95.103012",
    journal = "Phys. Rev. D",
    volume = "95",
    number = "10",
    pages = "103012",
    year = "2017"
}

@article{Wald:1978vm,
    author = "Wald, Robert M.",
    title = "{Construction of Solutions of Gravitational, Electromagnetic, Or Other Perturbation Equations from Solutions of Decoupled Equations}",
    doi = "10.1103/PhysRevLett.41.203",
    journal = "Phys. Rev. Lett.",
    volume = "41",
    pages = "203--206",
    year = "1978"
}

@article{Thornburg:2026eeb,
    author = "Thornburg, Jonathan",
    title = "{Time-domain evolution of Lorenz-gauge metric perturbations: taming the $\ell=m=1$ gauge instability}",
    eprint = "2606.27016",
    archivePrefix = "arXiv",
    primaryClass = "gr-qc",
    month = "6",
    year = "2026",
    journal = ""
}

@article{Miller:2016hjv,
    author = "Miller, Jeremy and Wardell, Barry and Pound, Adam",
    title = "{Second-order perturbation theory: the problem of infinite mode coupling}",
    eprint = "1608.06783",
    archivePrefix = "arXiv",
    primaryClass = "gr-qc",
    doi = "10.1103/PhysRevD.94.104018",
    journal = "Phys. Rev. D",
    volume = "94",
    number = "10",
    pages = "104018",
    year = "2016"
}

@article{Poisson:2011nh,
    author = "Poisson, Eric and Pound, Adam and Vega, Ian",
    title = "{The Motion of point particles in curved spacetime}",
    eprint = "1102.0529",
    archivePrefix = "arXiv",
    primaryClass = "gr-qc",
    doi = "10.12942/lrr-2011-7",
    journal = "Living Rev. Rel.",
    volume = "14",
    pages = "7",
    year = "2011"
}

@article{Goldberg:1966uu,
    author = "Goldberg, J. N. and MacFarlane, A. J. and Newman, E. T. and Rohrlich, F. and Sudarshan, E. C. G.",
    title = "{Spin-$s$ spherical harmonics and $\eth$}",
    doi = "10.1063/1.1705135",
    journal = "J. Math. Phys.",
    volume = "8",
    pages = "2155",
    year = "1967"
}

@article{Quinn:1996am,
    author = "Quinn, Theodore C. and Wald, Robert M.",
    title = "{An Axiomatic approach to electromagnetic and gravitational radiation reaction of particles in curved space-time}",
    eprint = "gr-qc/9610053",
    archivePrefix = "arXiv",
    doi = "10.1103/PhysRevD.56.3381",
    journal = "Phys. Rev. D",
    volume = "56",
    pages = "3381--3394",
    year = "1997"
}

@article{Li:2026foy,
    author = "Li, Dongjun and Kuan, Hao-Jui and Weller, Colin and Yang, Chaoyi and Lam, Kelvin Ka-Ho and Bourg, Patrick and Chung, Adrian Ka-Wai and Chen, Yanbei and Yunes, Nicol{\'a}s and Yang, Huan",
    title = "{Modeling extreme mass-ratio inspirals in astrophysical and dark-matter environments}",
    eprint = "2610.22480",
    archivePrefix = "arXiv",
    primaryClass = "gr-qc",
    month = "10",
    year = "2026",
    journal = ""
}

@article{Berti:2025hly,
    author = "Berti, Emanuele and others",
    editor = "Berti, Emanuele and Cardoso, Vitor and Carullo, Gregorio",
    title = "{Black hole spectroscopy: from theory to experiment}",
    eprint = "2505.23895",
    archivePrefix = "arXiv",
    primaryClass = "gr-qc",
    reportNumber = "RUP-25-10; YITP-25-64; RIKEN-iTHEMS-Report-25",
    doi = "10.1088/1361-6382/ae59e2",
    journal = "Class. Quant. Grav.",
    volume = "43",
    number = "12",
    pages = "123001",
    year = "2026"
}

@article{Casals:2024ynr,
    author = "Casals, Marc and Hollands, Stefan and Pound, Adam and Toomani, Vahid",
    title = "{Spin-2 Green{\textquoteright}s functions on Kerr in radiation gauge}",
    eprint = "2402.15468",
    archivePrefix = "arXiv",
    primaryClass = "gr-qc",
    doi = "10.1088/1361-6382/ad7cbc",
    journal = "Class. Quant. Grav.",
    volume = "41",
    number = "21",
    pages = "215010",
    year = "2024"
}

@article{Li:2026rkf,
    author = "Li, Dongjun and Yunes, Nicol{\'a}s",
    title = "{Metric Reconstruction for Generic Black-Hole Perturbations}",
    eprint = "2605.11080",
    archivePrefix = "arXiv",
    primaryClass = "gr-qc",
    month = "5",
    year = "2026",
    journal = ""
}

@article{Green:2019nam,
    author = "Green, Stephen R. and Hollands, Stefan and Zimmerman, Peter",
    title = "{Teukolsky formalism for nonlinear Kerr perturbations}",
    eprint = "1908.09095",
    archivePrefix = "arXiv",
    primaryClass = "gr-qc",
    doi = "10.1088/1361-6382/ab7075",
    journal = "Class. Quant. Grav.",
    volume = "37",
    number = "7",
    pages = "075001",
    year = "2020"
}

@article{Loutrel:2020wbw,
    author = "Loutrel, Nicholas and Ripley, Justin L. and Giorgi, Elena and Pretorius, Frans",
    title = "{Second Order Perturbations of Kerr Black Holes: Reconstruction of the Metric}",
    eprint = "2008.11770",
    archivePrefix = "arXiv",
    primaryClass = "gr-qc",
    doi = "10.1103/PhysRevD.103.104017",
    journal = "Phys. Rev. D",
    volume = "103",
    number = "10",
    pages = "104017",
    year = "2021"
}

@article{Ripley:2020xby,
    author = "Ripley, Justin L. and Loutrel, Nicholas and Giorgi, Elena and Pretorius, Frans",
    title = "{Numerical computation of second order vacuum perturbations of Kerr black holes}",
    eprint = "2010.00162",
    archivePrefix = "arXiv",
    primaryClass = "gr-qc",
    doi = "10.1103/PhysRevD.103.104018",
    journal = "Phys. Rev. D",
    volume = "103",
    pages = "104018",
    year = "2021"
}

@unpublished{Cunningham_inprep,
  author = {Cunningham, Kevin Ois\'in and Wardell, Barry and Dolan, Sam},
  title  = "{Metric perturbations of Kerr spacetime in Lorenz gauge: Precessing eccentric orbits}",
  note   = {(in preparation)},
  year   = {2026}
}

@article{Blanchet:2012at,
    author = "Blanchet, Luc and Buonanno, Alessandra and Le Tiec, Alexandre",
    title = "{First law of mechanics for black hole binaries with spins}",
    eprint = "1211.1060",
    archivePrefix = "arXiv",
    primaryClass = "gr-qc",
    doi = "10.1103/PhysRevD.87.024030",
    journal = "Phys. Rev. D",
    volume = "87",
    number = "2",
    pages = "024030",
    year = "2013"
}

@software{BHPToolkit:Teukolsky,
    author       = {Wardell, Barry and Warburton, Niels and Cunningham, Kevin and Durkan, Leanne and Leather, Benjamin and Nasipak, Zachary and Kavanagh, Chris and Ottewill, Adrian and Casals, Marc and Torres, Theo and Neef, Jakob and Barsanti, Susanna},
    title        = {Teukolsky},
    month        = sep,
    year         = 2026,
    publisher    = {Zenodo},
    version      = {1.2.1},
    doi          = {10.5281/zenodo.22873185},
    url          = {https://doi.org/10.5281/zenodo.22873185},
    swhid        = {swh:1:dir:27715775ff72233751790eea6b97c7a06ed018d7;origin=https://doi.org/10.5281/zenodo.7037850;visit=swh:1:snp:7b793883182b18b32d1f0a7c93bea3c87509a7cf;anchor=swh:1:rel:fbd2b37bd49620ef8c3d242dcc1947f5a737aba4;path=BlackHolePerturbationToolkit-Teukolsky-80a554d},
}

@software{BHPToolkit:SpinWeightedSpheroidalHarmonics,
  author       = {Wardell, Barry and Warburton, Niels and Cunningham, Kevin and Ottewill, Adrian and Casals, Marc and Neef, Jakob and Upton, Samuel D. and Fransen, Kwinten},
  title        = {SpinWeightedSpheroidalHarmonics},
  month        = sep,
  year         = 2026,
  publisher    = {Zenodo},
  version      = {1.1.1},
  doi          = {10.5281/zenodo.22809903},
  url          = {https://doi.org/10.5281/zenodo.22809903},
  swhid        = {swh:1:dir:f09182c98751055ea8450962b95b6aaa5b67d049;origin=https://doi.org/10.5281/zenodo.8090680;visit=swh:1:snp:e4facb38391d33453c83f50c0e6b1c79d7853e17;anchor=swh:1:rel:1f9649ba435521ce327117247e9177ce29266695;path=BlackHolePerturbationToolkit-SpinWeightedSpheroidalHarmonics-48cd355},
}

@Book{Synge1960,
  author    = {J. L. Synge},
  publisher = {North-Holland},
  title     = {Relativity: The General Theory},
  year      = {1960},
  address   = {Amsterdam},
  series    = {Series in Physics},
  pagetotal = {505},
}

@article{Upton:2021oxf,
    author = "Upton, Samuel D. and Pound, Adam",
    title = "{Second-order gravitational self-force in a highly regular gauge}",
    eprint = "2101.11409",
    archivePrefix = "arXiv",
    primaryClass = "gr-qc",
    doi = "10.1103/PhysRevD.103.124016",
    journal = "Phys. Rev. D",
    volume = "103",
    number = "12",
    pages = "124016",
    year = "2021"
}

@article{Detweiler:2011tt,
    author = "Detweiler, Steven",
    title = "{Gravitational radiation reaction and second order perturbation theory}",
    eprint = "1107.2098",
    archivePrefix = "arXiv",
    primaryClass = "gr-qc",
    doi = "10.1103/PhysRevD.85.044048",
    journal = "Phys. Rev. D",
    volume = "85",
    pages = "044048",
    year = "2012"
}

@article{Musaeus:2026kgp,
    author = "Musaeus, J{\o}rgen and Pound, Adam and Upton, Samuel D.",
    title = "{Point particles in general relativity: beyond linear perturbation theory}",
    eprint = "2609.10521",
    archivePrefix = "arXiv",
    primaryClass = "gr-qc",
    month = "9",
    year = "2026",
    journal = ""
}

@misc{Upton2026Capra,
  author       = {Upton, Sam D.},
  title        = {Progress on punctures and effective source for spherical orbits in {Kerr}},
  howpublished = {Talk presented at the 29th Capra Meeting on Radiation Reaction in General Relativity},
  month        = {June},
  year         = {2026},
  address      = {Brussels, Belgium},
  note         = {Held at Université Libre de Bruxelles, June 29--July 3, 2026}
}

@article{Wardell:2011gb,
    author = "Wardell, Barry and Vega, Ian and Thornburg, Jonathan and Diener, Peter",
    title = "{A Generic effective source for scalar self-force calculations}",
    eprint = "1112.6355",
    archivePrefix = "arXiv",
    primaryClass = "gr-qc",
    doi = "10.1103/PhysRevD.85.104044",
    journal = "Phys. Rev. D",
    volume = "85",
    pages = "104044",
    year = "2012"
}

@article{Haas:2006ne,
    author = "Haas, Roland and Poisson, Eric",
    title = "{Mode-sum regularization of the scalar self-force: Formulation in terms of a tetrad decomposition of the singular field}",
    eprint = "gr-qc/0605077",
    archivePrefix = "arXiv",
    doi = "10.1103/PhysRevD.74.044009",
    journal = "Phys. Rev. D",
    volume = "74",
    pages = "044009",
    year = "2006"
}

@article{Mino:2001mq,
    author = "Mino, Yasushi and Nakano, Hiroyuki and Sasaki, Misao",
    title = "{Covariant selfforce regularization of a particle orbiting a Schwarzschild black hole: Mode decomposition regularization}",
    eprint = "gr-qc/0111074",
    archivePrefix = "arXiv",
    reportNumber = "UTAP-386, OUTAP-157",
    doi = "10.1143/PTP.108.1039",
    journal = "Prog. Theor. Phys.",
    volume = "108",
    pages = "1039--1064",
    year = "2003"
}

\end{document}